\documentclass[hidelinks, conference]{IEEEtran}
\IEEEoverridecommandlockouts
\PassOptionsToPackage{bookmarks={false}}{hyperref} 

\usepackage[a-1b]{pdfx}
\usepackage{color}
\usepackage{url}
\usepackage[latin9]{inputenc}
\usepackage{amsthm}
\usepackage{amsmath}
\usepackage{amsthm}
\usepackage{amssymb}
\usepackage{graphicx}
\usepackage{epstopdf}
\usepackage{wrapfig}
\usepackage{algorithmic}
\usepackage{dirtytalk}
\usepackage{algorithm}
\usepackage{mathrsfs}
\usepackage{float}
\usepackage{mathtools}
\usepackage{tabulary}
\usepackage{booktabs}
\usepackage{caption}
\usepackage{subfig}
\usepackage{xprintlen}
\usepackage{multirow}

\graphicspath{{./}{./files/figures_eps/}}

\PassOptionsToPackage{bookmarks={false}}{hyperref}

\def\BibTeX{{\rm B\kern-.05em{\sc i\kern-.025em b}\kern-.08em
   T\kern-.1667em\lower.7ex\hbox{E}\kern-.125emX}}

\newcommand{\comment}[1]{ }
\newcommand{\proposed}{{\tt{EPS}}} 

\newcommand\subparagraph{%
  \@startsection{subparagraph}{0}
  {\parindent}
  {0ex \@plus 0ex \@minus 0ex}
  {-1em}
  {\normalfont\normalsize\bfseries}}
\makeatother

\begin{document}




\title{EPS: Distinguishable IQ Data Representation for Domain-Adaptation Learning of Device Fingerprints}
 
\author{\IEEEauthorblockN{Abdurrahman Elmaghbub}
\IEEEauthorblockA{Oregon State University\\
elmaghba@oregonstate.edu}
\and
\IEEEauthorblockN{Bechir Hamdaoui}
\IEEEauthorblockA{Oregon State University\\
hamdaoui@oregonstate.edu}
\thanks{This work is supported in part by NSF/Intel Award No. 2003273.}
}

\maketitle
\thispagestyle{plain}
\pagestyle{plain}

\begin{abstract}
Deep learning (DL)-based RF fingerprinting (RFFP) technology has emerged as a powerful physical-layer security mechanism, enabling device identification and authentication based on unique device-specific signatures that can be extracted from the received RF signals. However, DL-based RFFP methods face major challenges concerning their  ability to adapt to domain (e.g., day/time, location, channel, etc.) changes and variability. This work proposes a novel IQ data representation and feature design, termed Double-Sided Envelope Power Spectrum or \proposed, that is proven to overcome the domain adaptation problems significantly. By accurately capturing device hardware impairments while suppressing irrelevant domain information, \proposed~offers improved feature selection for DL models in RFFP. Experimental evaluations demonstrate its effectiveness, achieving over 99\% testing accuracy in same-day/channel/location evaluations and 93\% accuracy in cross-day evaluations, outperforming the traditional IQ representation. Additionally, \proposed~excels in cross-location evaluations, achieving a 95\% accuracy. The proposed representation significantly enhances the robustness and generalizability of DL-based RFFP methods, thereby presenting a transformative solution to IQ data-based device fingerprinting.

\end{abstract}


\section{Introduction}
\label{sec:into}
Deep learning (DL)-based RF fingerprinting (RFFP) technology is being recognized as a powerful physical-layer security mechanism \cite{hamdaoui2020deep, sankhe2019oracle, jian2020deep,ding2018specific,basha2022leveraging,shen2021radio}, enabling device identification and authentication through the extraction of unique device fingerprints embedded in the devices' transmitted RF signals. These fingerprints arise as a result of inherent hardware manufacturing imperfections of various RF circuitry components (e.g., local oscillators, mixers, power amplifiers) yielding RF signal distortions~\cite{elmaghbub2021,sankhe2019oracle} that collectively shape distinctive device signatures that can be extracted using DL models.
%
Although DL has eliminated the need for data preprocessing and domain-knowledge to extract features from raw RF data, most of the DL-RFFP approaches rely on the assumption that the training and testing data are drawn from the same distribution, which falls short of the conditions of realistic RF scenarios~\cite{hamdaoui2022uncovering,fu2023deep}. 
In other words, these approaches do not perform well in practical scenarios, in which the testing data is collected under a domain that is different from that used during training, where a {\em domain} here refers to a network condition (e.g., setting, environment) under which data is collected. This includes the collection time, the channel condition, the receiver hardware, the device location, and the protocol configuration, as well as other aspects.
Therefore, any considerable change in the testing versus training settings yields a different domain, and as such, we define robustness to domain changes as the ability of a learning model to maintain its training domain performance when tested under new domains. This is often referred to as {\em domain adaptation} in the machine learning community.



Several experimental studies using LoRa and WiFi devices (e.g., \cite{elmaghbub2021, hamdaoui2022deep, al2020exposing, hanna_wisig_2022}) have revealed the sensitivity of DL-based RFFP approaches to domain changes, thereby limiting their practical adoption to security applications. 
Unraveling such sensitivity issues is a complex task involving two black boxes: the deep learning model and the microelectronic circuitry of the RF devices. However, it is widely believed that the wireless channel, influenced by various confounding factors, plays a significant role in the failure of these approaches to adapt and generalize to different domains. 
To address the impact of the channel and enhance domain generalization, some studies have focused on removing channel dynamics from the raw signal through techniques like channel equalization \cite{hanna_wisig_2022, basha2023channel,2019deepradioid,hamdaoui2023deep} or hardware impairment compensation \cite{shen2021radio}. However, these approaches have drawbacks. Channel equalization can inadvertently remove crucial device-specific features, resulting in discriminative information loss and poor RF fingerprinting performance \cite{rajendran2022rf}. Impairment compensation techniques, on the other hand, target specific channel impairments, limiting their generalizability across different environments and wireless channels. 
Recently, domain adaptation techniques have emerged as a means to reduce the presence of domain information in the feature vector through adversarial learning \cite{li2022radionet}, or to decouple device-specific and domain-specific features through adversarial disentanglement learning \cite{elmaghbub2023adl} and generative adversarial networks \cite{al2023signcrf}. However, these techniques do not scale well and fail to maintain a practically acceptable performance for medium to large numbers of devices. Moreover, they are designed to adapt to the target domain dataset, leading to similar generalization issues when encountering domains that are unseen during training. 

Most DL-based RFFP approaches use the time-domain IQ representation of the received RF signal as the input for the learning models. This is because of the ability of these models to extract relevant features from the raw IQ data without needing preprocessing or prior domain knowledge. However, recent studies \cite{elmaghbub2021, snoap2022robust} have indicated that DL models relying solely on IQ values fail to adapt to domain changes, such as changes in the wireless channel and/or receiver hardware \cite{gaskin2022tweak}. The observation implies that the IQ representation encompasses a substantial amount of information that is not specific to the devices being classified. Consequently, models tend to overfit to particular training settings, such as the channel, receiver hardware, and device location, rather than effectively capturing the device-distinctive hardware impairments. The intricate nature of RF data, coupled with its dissimilarity to other data types used in DL, such as images and text, contributes to the suboptimal performance of imported DL models from vision/text domains when directly applied to raw RF data. This underscores the need for novel RF data representations that effectively capture the hardware impairments of devices. Such representations can enhance the feature selection process of DL models, preventing reliance on incorrect and unreliable features without unnecessarily complicating the models.

To fulfill this need, this paper proposes a novel RF data representation that significantly enhances the accuracy and robustness of DL-based RFFP methods to domain changes. Our motivation stems from the realization that raw IQ data representation contains a significant amount of device-irrelevant information. Consequently, extracting meaningful fingerprints from this raw IQ data becomes akin to finding a needle in a haystack filled with numerous deceptive needle-like objects. We overcome this limitation by proposing a novel RF data representation that vividly captures the device's hardware impairments while suppressing device-irrelevant information. Specifically, the proposed data representation closely mirrors the impaired behavior of a key RF hardware component, the oscillator, whose impairment substantially contributes to the device's unique fingerprint. To generate this representation, we extract the outer shape or envelope of the IQ signal, eliminate the resulting amplitude offset, and calculate the double-sided envelope's power spectrum or simply \proposed, yielding a novel data representation of the IQ signal that serves as an effective input for machine learning classifiers. 

Through extensive evaluation on a testbed of 15 IEEE 802.11b WiFi devices, we demonstrate the effectiveness of our proposed approach in real-world scenarios. 
Our experimental results show that when combined with standard CNN (Convolution Neural Network) models, the proposed \proposed-based device fingerprinting framework achieves outstanding performance. Notably, it achieves a testing accuracy of over 99\% in same-domain (day or location) scenarios, where training and testing are done on same day/location. More importantly, in cross-location scenarios, the proposed framework maintains a testing accuracy of over 95\%, whereas the same model trained with the conventional IQ representation achieves only 55\%. 
Therefore, our \proposed~representation offers a transformative solution  
that significantly advances the RF fingerprinting field by substantially improving the accuracy and generalizability of DL-based RFFP approaches.


Our key contributions can thus be summarized as follows: 
\begin{itemize}
    \item We propose \proposed, a novel RF signal representation input to DL-based RFFP approaches that substantially enhances the accuracy, robustness, and generalizability across various domains and hardware configurations.

    \item We demonstrate through extensive experimentation the distinguishability and reliability of the \proposed~representation across time, channel, and location domains, justifying its applicability to RF fingerprinting applications. 
    
    \item We release massive 8TB IEEE 802.11b WiFi datasets of 15 Pycom devices that include both raw and processed files for more than 5000 packets for each device for four scenarios: Wired Setup, Wireless Setup, Different Locations Setup, and Random Deployment Setup. 
    
    \item We extensively assess the performance of \proposed~when used as an input to standard CNNs for classifying Pycom devices and showcase an exceptional cross-domain performance in real-world scenarios, achieving an average testing accuracy of 93\% and 95\% respectively for the cross-days and cross-location scenarios.

    \item We demonstrate the impact of the local oscillator's frequency instability during transceiver hardware warm-up and stabilization on the performance of DL-based RFFP.

\end{itemize}

The rest of the paper is organized as follows. Sec.~\ref{sec:related} presents the related works. Sec.~\ref{sec:motiv} studies the impact of carrier frequency offset and inaccuracy on the behavior of IQ signals. Sec.~\ref{sec:eps} presents the proposed IQ data representation approach, \proposed. Sec.~\ref{sec:proposed} presents the proposed \proposed-based device fingerprinting framework. Sec.~\ref{sec:setup} describes the testbed and RF datasets used for evaluating the proposed framework, which is presented in Sec.~\ref{sec:eval}. Sec.~\ref{sec:stability} highlights the impact of hardware warm-up and stabilization on RF fingerprinting accuracy. Finally, the paper is concluded in Sec.~\ref{sec:conc}.

\section{Related Work}
\label{sec:related}
Prior works that aimed to address the domain generalizability challenges of DL-based RFFP can be broadly categorized into two approaches: data-centric and architecture-centric.
For the data-centric approaches, various data augmentation techniques have been explored to expose the DL models to a wider range of wireless channel instances, thereby enhancing their robustness against channel variations. For instance, Soltani et al. \cite{soltani2020more} and Al-Shuwaili et al. \cite{al2021deeplora} integrated data augmentation engines into the training process. These engines incorporated WLAN TGn and ITU-R channel models, respectively, along with an additive white noise model. While these techniques demonstrated a marginal improvement in testing accuracy, they do not offer a practical and scalable solution suitable for commercial deployment. Additionally, due to the intrinsic nature of wireless channels, it is challenging and impractical to devise a universal channel-augmenting model capable of significantly improving performance across a wide range of wireless channels. Other data-centric approaches mitigate the impact of the channel through channel equalization \cite{hanna_wisig_2022, basha2023channel, 2019deepradioid} or impairment compensation \cite{shen2021radio}.

For the DL architecture-centric approaches, researchers have formulated the RFFP generalizability challenge as a domain adaptation problem \cite{redko2020survey, gaskin2022tweak} and capitalized on the advancement in transfer learning to address it. The underlying assumption in these frameworks is that the source and target domains exhibit slightly different distributions. One notable domain adaptation framework, ADL-ID \cite{elmaghbub2023adl}, integrates disentangled representation learning with adversarial learning to tackle the challenge of short-term temporal generalization in RFFP. ADL-ID involves segregating the feature vector into two distinct components: device-specific (fingerprints) and domain-specific features. During the inference stage, only the device-specific features, which encompass characteristics that remain invariant across the source and target domains, are utilized. Similarly, SignCRF \cite{al2023signcrf} leverages a cycle-consistent generative adversarial network to construct an environment translator that effectively decouples hardware impairments from channel and environmental conditions. Adversarial domain adaptation techniques have also been employed in this context. For instance, RadioNet \cite{li2022radionet} adopts an adversarial learning scheme, utilizing a domain discriminator and a reversal gradient layer to minimize the domain-related information in the feature vector. The trained feature extractor is subsequently connected to a KNN classifier fine-tuned using the target data. Following a similar calibration approach, the Tweak approach, proposed by Gaskin et al. \cite{gaskin2022tweak}, combines metric learning with lightweight calibration using the target data to enhance generalization across hardware, channel, and configuration dimensions. 

While these domain adaptation methods provide valuable insights in enhancing RFFP generalizability, they exhibited limitations in providing satisfactory performance for medium-scale testbeds. Furthermore, they encounter challenges when faced with significant distribution gaps between the enrollment and deployment datasets or when confronted with unseen environmental conditions that differ from the target domain employed during adaptation. 

\section{Understanding the Impact of Carrier Frequency Inaccuracy on IQ Signal Behavior}
\label{sec:motiv}
Local oscillators are the transceiver hardware component that are responsible for producing oscillating signals needed for signal up-conversion at the sender side and for signal down-conversion at the receiver side. The inaccuracy and instability of the oscillating signal's frequency, typically caused by external factors like temperature, vibration and electromagnetic interference, impact the overall system performance behavior. 
As such, in an effort to improve their robustness to these external factors, various types of different crystal oscillators have been developed over the years, including temperature-controlled crystal oscillators (TCXOs), which feature temperature compensation, and oven-controlled crystal oscillators (OCXOs), which place the crystal in a temperature-controlled environment to keep their temperatures at a constant level, thereby improving the accuracy of their oscillating frequencies \cite{zhou2008frequency}.

\subsection{The Carrier Frequency Offset (CFO) Impairment}
One crucial hardware impairment resulting from the oscillating frequency inaccuracy is the inevitable mismatch between the receiver's local oscillating frequency and that of the sender's. 
This mismatch in frequency is known as the {\em Carrier Frequency Offset or CFO} and often leads to signal distortion. This CFO impairment has been leveraged to provide unique device fingerprints that could serve as inputs to DL-based RFFP techniques, as in \cite{vo2016fingerprinting}, where the mean and probability distribution of CFO values are used to differentiate between WiFi devices. Because CFO information is also embedded in the time-domain IQ representation of RF signals, most DL-based RFFP frameworks use raw IQ data as the input to their learning models, with the expectation that the models will capture and extract such CFO values on their own.

As mentioned earlier, the problem, however, is that recent studies \cite{elmaghbub2021, al2020exposing} reveal that DL models relying solely on IQ samples fail to adapt to different domains. For instance, when the DL models are trained on data collected on one day but tested on data collected on a different day, the testing accuracy drops significantly \cite{gaskin2022tweak}. As a result, new RF signal representations that effectively capture device-specific impairments are needed to improve the feature selection process of DL models and to enhance their robustness to domain changes. In this work, we propose robust RF signal representations that accurately capture the impairments of oscillators and their CFO manifestation to serve as distinctive features that improve RFFP's adaptation to domain variations.

To be able to design IQ data representations that effectively capture CFO impairments, it is important to first study and understand the impact of the carrier frequency offset and inaccuracy on the behavior of the received IQ signals.

\subsection{The Impact of Carrier Frequency Inaccuracy}\label{subsec:impact}
To investigate, study and acquire some good understanding of the impact of the oscillating frequency inaccuracy on the IQ signal behavior, we leveraged our experimental testbed of 15 Pycom/IoT devices to observe, analyze, and compare the IQ signals collected from multiple different (but identical in hardware) off-the-shelf devices.
This is done by having each of the 15 Pycom devices transmit multiple IEEE 802.11b WiFi packets after being powered on for 20 minutes to ensure hardware stabilization \cite{elmaghbub2023impact}. We want to emphasize here the importance of waiting until the end of the warm-up period of the devices' hardware before starting the data collection process to ensure robust and consistent measurements; we provide further explanation and illustration on this in Sec.~\ref{sec:stability}.
The transmitted signals are then captured and sampled at 45MSps using a USRP B210 receiver. More description and details on the testbed are provided later in Sec.~\ref{sec:setup}.

\begin{figure}
\centering
    \begin{minipage}[t]{\linewidth}  
    \centering 
    \subfloat[The I component of the received IQ signal]{%
    \includegraphics[width=\linewidth,height=.4\linewidth]{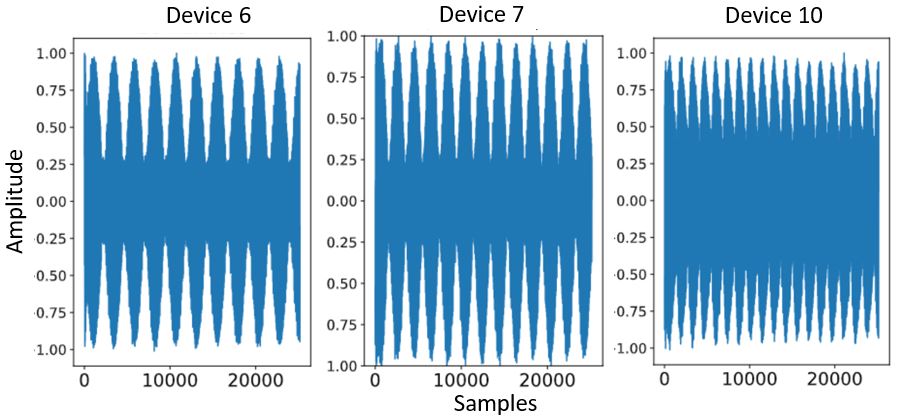} 
     \label{all-devices-I}}\\ 
    \subfloat[The Q component of the IQ signal]{%
    \includegraphics[width=\linewidth,height=.4\linewidth]{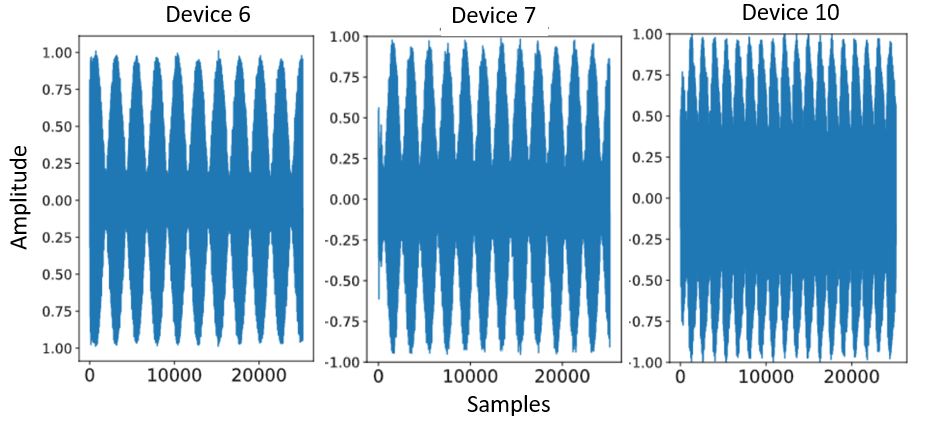}
    \label{all-devices-Q}}
    \caption{The time-domain IQ signal behavior across three different Pycom devices. The number of 'humps' are: 11 for Device 6; 13 for Device 7; and 19 for Device 10.} 
    \label{all-devices-abcd} 
    \end{minipage}
\end{figure}

We show in Fig.~\ref{all-devices-abcd} the time behavior of both the I (in-phase) and Q (quadrature) signal components collected from three testbed devices: Devices 6, 7 and 10. 
A couple of key observations we want to draw from this experiment. First, observe the `sinusoidal' behavior of the Envelopes\footnote{The Envelope of an oscillating signal is the smooth boundary function that outlines the extremes of the signal (e.g., see \cite{johnson2011software}, Appendix C).} of both the I and Q signals. 
More importantly, note that the number of `humps' of the Envelope is different across the devices; it is 11 for Device 6, 13 for Device 7, and 19 for Device 10. 
Another observation we make here is that the Envelopes of the I and Q signals of a given device vary in the opposite direction of one another---i.e., shifted by 180 degrees, though still exhibiting the same number of `humps'.
It is worth mentioning that although shown for only three devices here, these reported sinusoidal behaviors of the IQ signals' Envelopes are observed across all of the 15 tested Pycom devices, but with each device exhibiting a slightly different number of humps.

The questions that arise now are: (i) what is the cause of the observed sinusoidal behavior of the IQ signal Envelope? and (ii) why does the number of `humps' differ from one device to another?
We will show that the main cause behind such a behavior is the CFO (carrier frequency offset) between the Pycom device's oscillating frequency and that of the USRP receiver that exists due to the instability and inaccuracy of the device's local oscillator. 
Specifically, we will next show that the number of humps in the sinusoidal Envelope depends on the CFO value. This explains that the reason why different devices exhibit different numbers of humps is because each device presents a different CFO, which varies across devices due to the device's oscillator hardware imperfections incurred during manufacturing. Later in Sec.~\ref{sec:stability}, we will also demonstrate that the CFO value (and hence the number of humps) of a given device keeps changing over time until the device hardware is stabilized; i.e., the CFO value keeps varying over time until the end of the hardware warm-up period.
This is due to the instability that the carrier frequency exhibits when the oscillator hardware of the device is still warming up.

\subsection{The Cause of the Observed Envelope Behavior} \label{subsec:dem}
We now demonstrate and affirm that the CFO impairment is what is behind the sinusoidal behavior of the IQ signal's Envelope illustrated in Sec.~\ref{subsec:impact}. 
We do so here both through Matlab simulation and analytically; we also confirmed this via hardware experimentation in Section IV.C of~\cite{elmaghbub2023impact}.

\subsubsection{Simulated Affirmation}
\begin{figure}
\centering
    \begin{minipage}[t]{\linewidth} 
    \centering 
    \subfloat[CFO = 0 (ideal scenario)]{
    \includegraphics[width=.45\linewidth,height=0.43\linewidth]{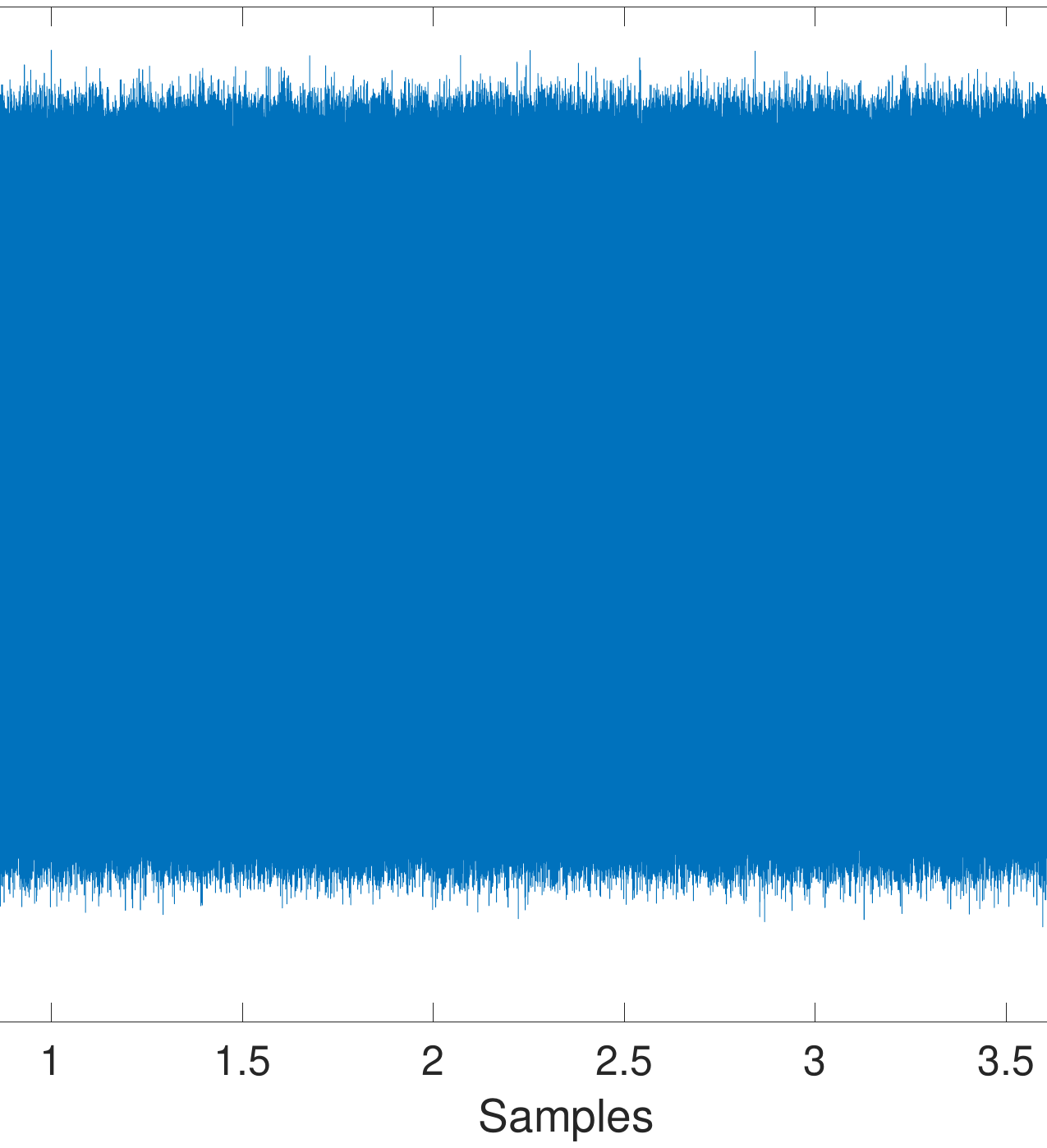} 
     \label{cfo0}}
     \hfill
    \subfloat[CFO = 50 Hz]{
    \includegraphics[width=.45\linewidth,height=0.43\linewidth]{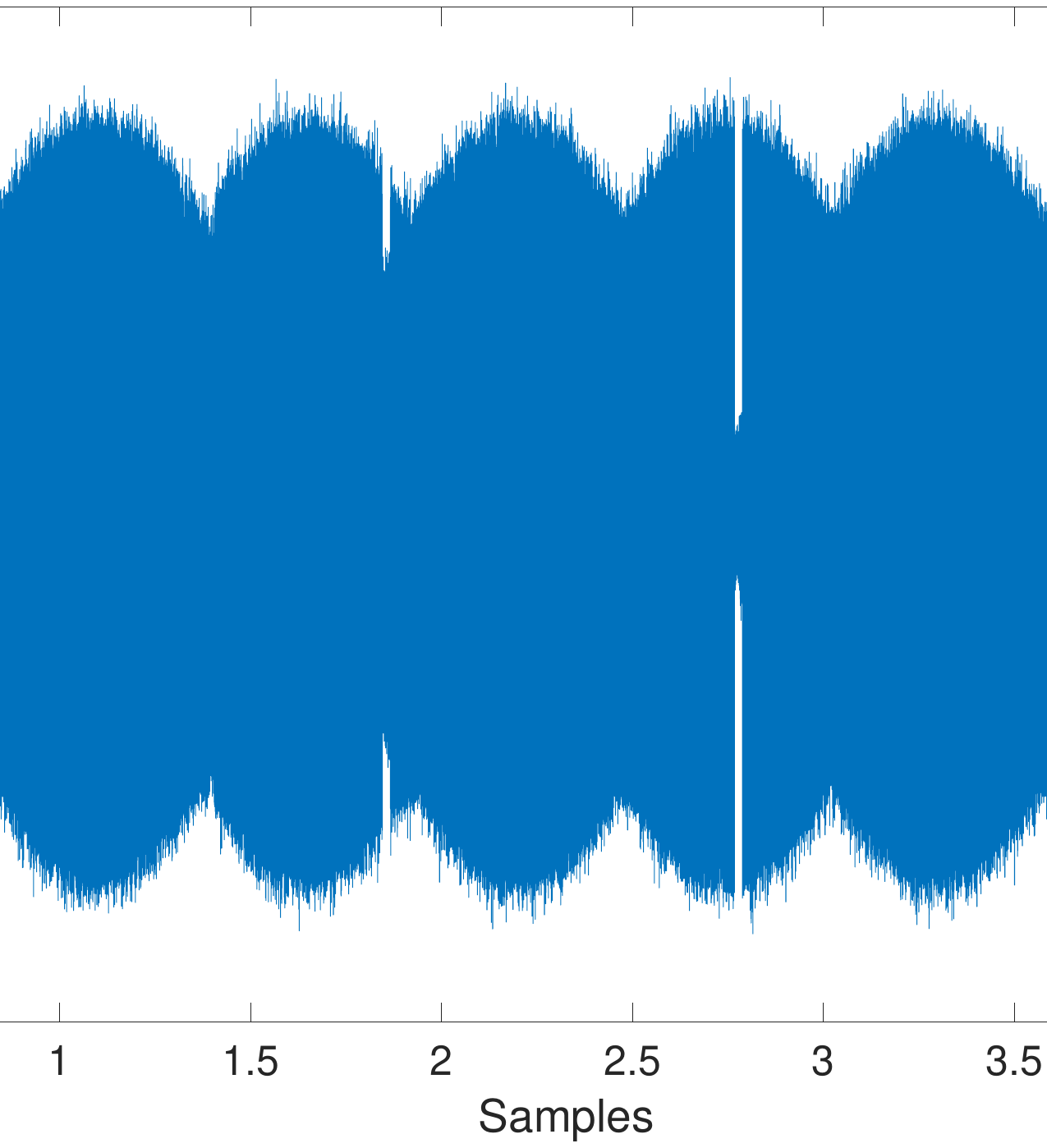}
    \label{cfo50}}\\
    \subfloat[CFO = 100 Hz]{
    \includegraphics[width=.45\linewidth,height=0.43\linewidth]{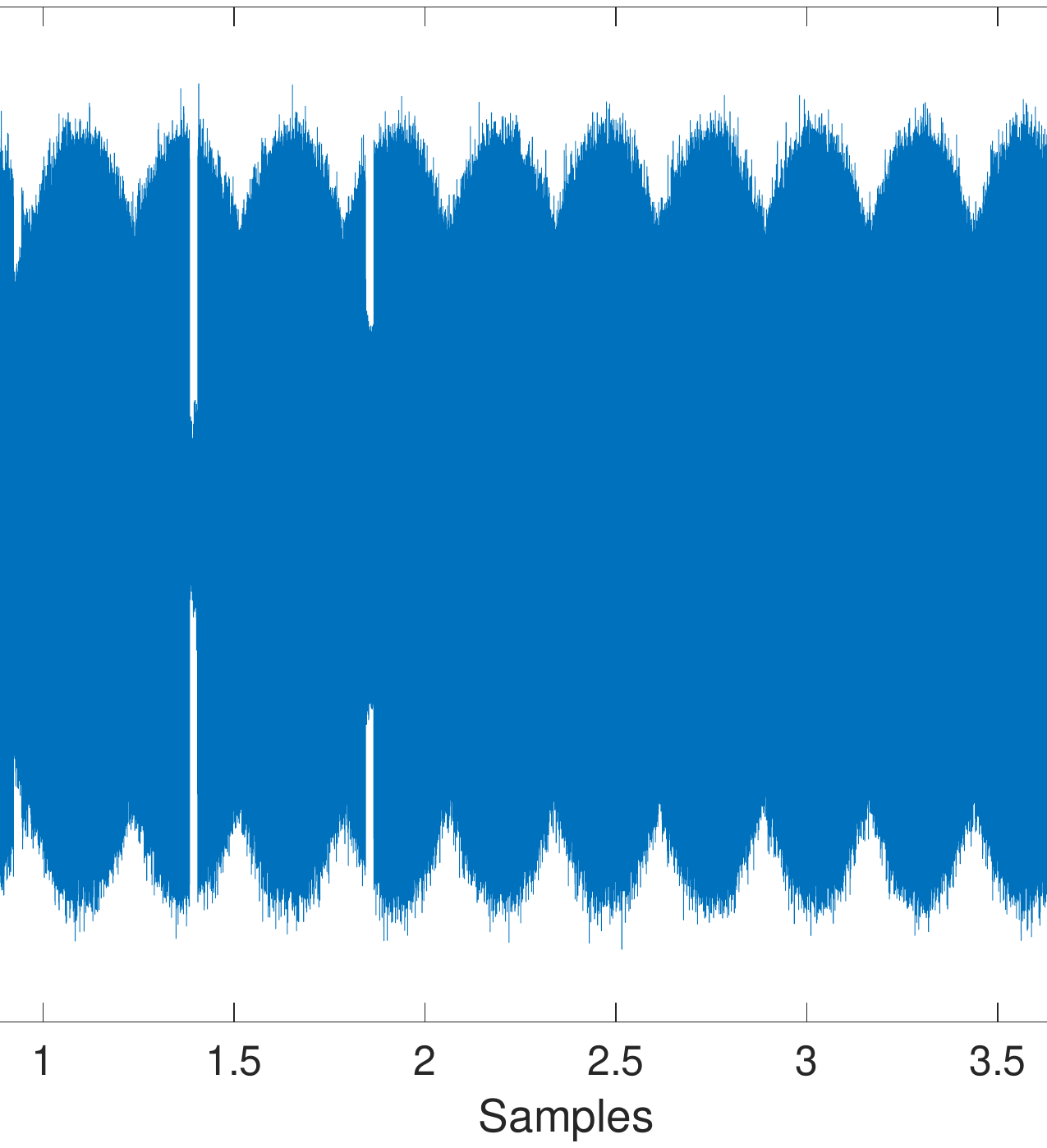}
    \label{cfo100}}
     \hfill
\subfloat[CFO = 200 Hz]{
    \includegraphics[width=.45\linewidth,height=0.43\linewidth]{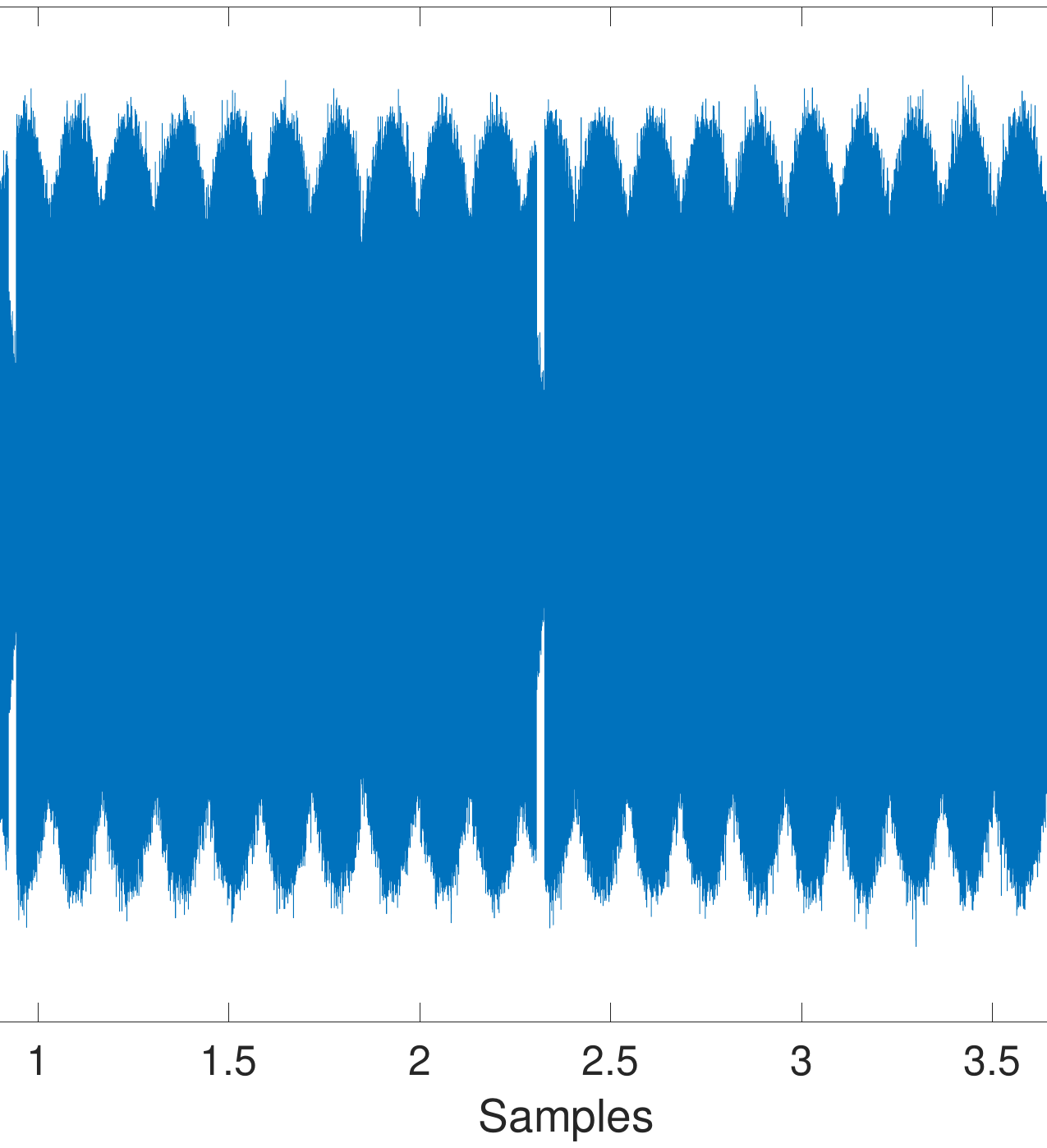}
    \label{cfo200}}
    \caption{Simulated time-domain I signal component. Y-axis is the amplitude of the I values; X-axis is the time in samples.} 
    \label{cfo-matlab} 
    \end{minipage}
\end{figure}

We used MATLAB R2023b to build our wireless communication model, allowing to vary the CFO value between the sending device and the receiving device. We used MATLAB's WLAN toolbox to generate multiple IEEE 802.11b WiFi DSSS waveforms impaired with the following CFO values: 0 Hz (ideal device), 50 Hz, 100 Hz, and 200 Hz.
The CFO-impaired transmitted signal is then first passed through an AWGN channel, and then down-converted and sampled by the receiver to generate IQ data samples.
For each case, we collected $10$ WiFi frames, with each frame having a size of $1000$ bits. 
Then, we extracted the real (I) components of the signals and plotted them separately for CFO = 0 in Fig.~\ref{cfo0}, CFO = 50Hz in Fig.~\ref{cfo50}, CFO = 100Hz in Fig.~\ref{cfo100}, and CFO = 200Hz in Fig.~\ref{cfo200}.
The simulated results clearly show the dependency between the CFO values and the number of observed `humps' in the I signal's Envelope, and that the CFO is what causes the observed Envelope shape. The same trends were observed for the Q signal components as well, but we did not include them here to limit redundancy.

We want to mention that we also experimented with varying other hardware impairments, including IQ imbalance, Phase Noise, and DC offset, but have not noticed any `sinusoidal' behavior of the Envelopes. This confirms that other transceiver hardware impairments, though do manifest themselves in other types of distortions, do not yield the Envelope behavior we observed with the CFO impairment.

\subsubsection{Analytic Affirmation}
Now that we've demonstrated and affirmed through Matlab modeling and simulation that the cause of the sinusoidal shape of the IQ envelope is the CFO impairments, we also do so here but analytically.
%
To illustrate, consider a baseband signal $a(t)\exp{(j\phi(t))}$ modulated by an oscillating signal with a carrier frequency $f_c$ and a CFO, $\Delta f$. The passband transmitted signal can be expressed as:
\begin{equation}\nonumber
    s(t) = a(t) \cos(2\pi (f_c + \Delta f) t + \phi(t))
\end{equation}

At the receiver side, the In-phase (I) component, $r_I(t)$, of the received signal can be recovered (demodulated) as:
\begin{equation}\nonumber
    r_I(t) = s(t) \cos(2\pi f_ct)
\end{equation}
Due to the CFO (i.e., the carrier frequency at the receiver is slightly different from the carrier frequency used by the sender), the phase of the wave shifts over time, causing the amplitude of the received signal to vary sinusoidally over time. This can be seen by rewriting $r_I(t)$ as
\begin{equation}\nonumber
    r_I(t)\! =\! \frac{a(t)}{2}[\cos(2\pi (2f_c  + \Delta f) t + \phi(t) ) + \cos(2\pi \Delta ft + \phi(t))]
    \label{env}
\end{equation}

\begin{figure}
\centering
    \begin{minipage}[t]{\linewidth} 
    \centering 
    \subfloat[CFO = $\Delta f$ = 0]{
    \includegraphics[width=.47\linewidth,height=0.16\linewidth]{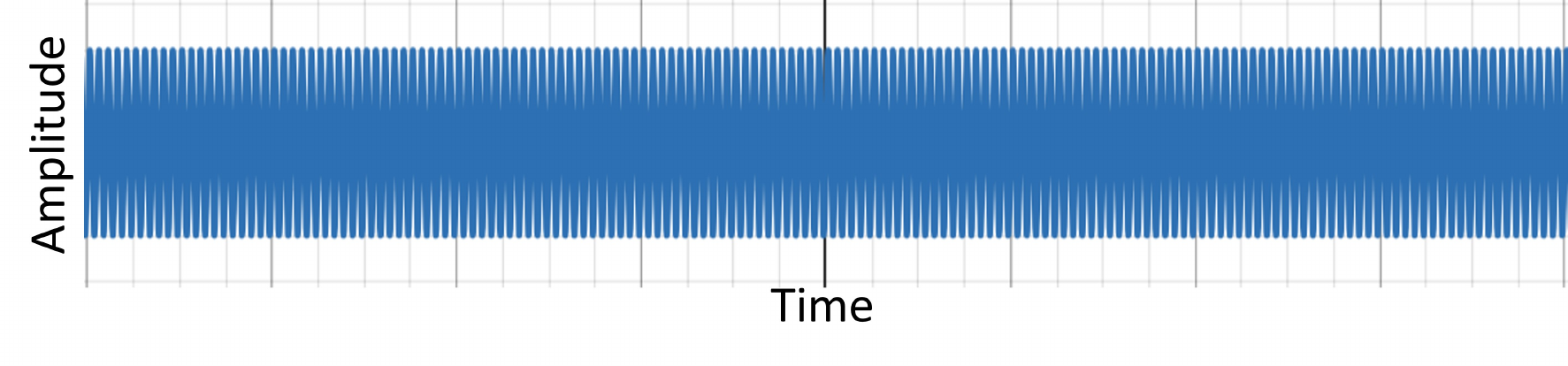} 
     \label{cfo-a0}}
     \hfill
    \subfloat[CFO = $\Delta f$ = 0.1Hz]{
    \includegraphics[width=.47\linewidth,height=0.16\linewidth]{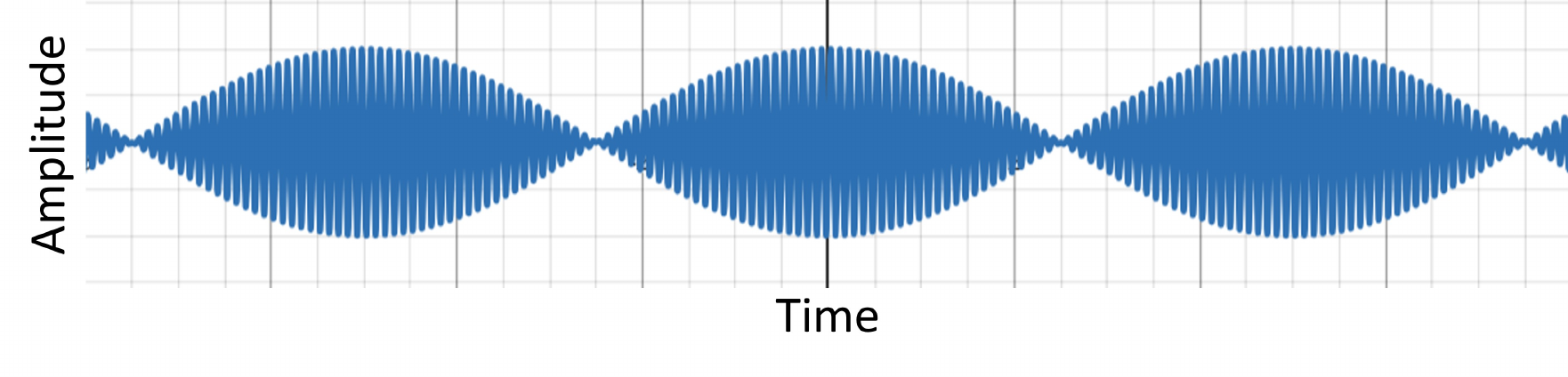}
    \label{cfo-a01}}\\
    \subfloat[CFO = $\Delta f$ = 0.2Hz]{
    \includegraphics[width=.47\linewidth,height=0.16\linewidth]{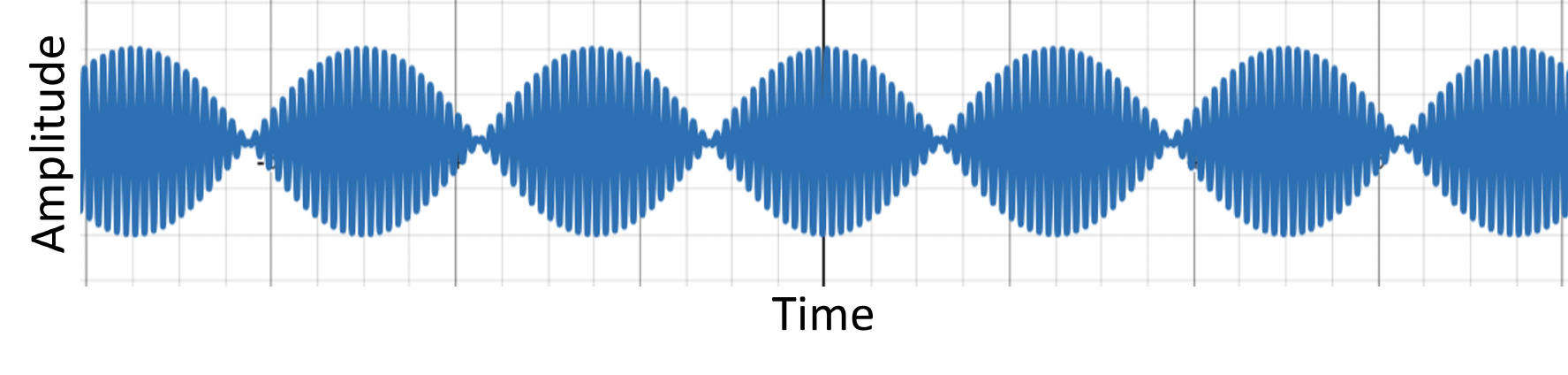}
    \label{cfo-a02}}
     \hfill
\subfloat[CFO = $\Delta f$ = 0.5Hz]{
    \includegraphics[width=.47\linewidth,height=0.16\linewidth]{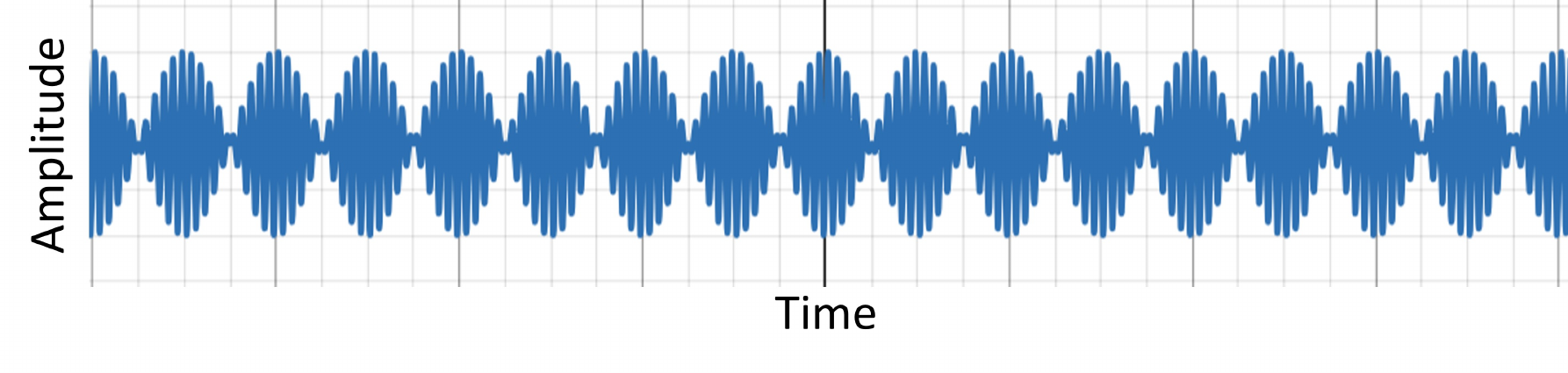}
    \label{cfo-a05}}
    \caption{Simple examples illustrating the impact of CFO on the Envelope behavior when $a(t)=\sin(2\pi10t)$.} 
    \label{cfo-analytic} 
    \end{minipage}
\end{figure}

While the left term in the equation above represents the frequency sum of the two frequencies, typically filtered out by a bandpass filter, the right term represents the frequency difference which is, in this case, the carrier frequency mismatch or CFO, $\Delta f$, between the local oscillators of the transmitter and the receiver. When CFO $=\Delta f = 0$, the second term becomes $1$ and the signal maintains a scaled version of its original amplitude, $\frac{a(t)}{2}$ (e.g. corresponding to Fig.~\ref{cfo0} in the simulation case). However, a frequency mismatch will cause the received signal to be modulated at the frequency $\Delta f$, resulting in an amplitude of $\frac{a(t)}{2}\cos(2\pi \Delta f t + \phi(t))$ (e.g. corresponding to Figs.~\ref{cfo50},~\ref{cfo100},~\ref{cfo200} in the simulation case). 
For illustration purposes, we show in Fig.~\ref{cfo-analytic} the impact of CFO (i.e., $\Delta f$) on the Envelope behavior of a simple/toy example with $a(t)=\sin(2\pi10t)$.
The Envelope clearly exhibits a `sinusoidal' shape when CFO is nonzero, and that the number of humps in the Envelope increases with the CFO.

In conclusion, we confirm that these observed device-dependent, sinusoidal Envelope behaviors of the IQ signals collected from the different off-the-shelf PyCom devices are indeed attributed to the CFO impairments.
%
These CFOs exist because the carrier frequency generated by the local oscillator at the USRP receiver is (slightly) different from that generated by the local oscillator at the Pycom device.
%
%
These demonstrations also confirm that devices with (even slightly) different oscillating frequencies yield different number of `humps' in the received signals' Envelope. And this work leverages such a difference in the number of humps across different devices to propose efficient IQ representations that are shown to significantly improve the accuracy and robustness of DL-based RF fingerprinting to domain changes.

\section{Novel IQ Data Representation for Distinguishable Neural Network Features}
\label{sec:eps}
In this section, we begin by presenting a novel IQ signal representation/feature extracted from the oscillator's envelope shape (observed and explained in the previous section) that substantially improves the robustness of device fingerprinting to domain changes and variations. 
We then evaluate the effectiveness of the proposed feature design vis-a-vis of its fingerprinting ability to (i) {\em distinguish between devices} and (ii) {\em adapt to domain changes} by maintaining high accuracy performance under varying domains.

\subsection{Capturing the Oscillator's Envelope Behavior}
\label{EPS-extraction}
In order to extract the CFO value resulting from the mismatch between the sender's and receiver's oscillating frequencies, which is embedded in the signal's envelope shape as observed and explained in Sec.~\ref{sec:motiv}, we first create the analytic signal, $z(t)$, of the time-domain representation of the receiver packet, $r(t)$. The analytic signal $z(t)$ is a complex-valued signal, comprising the original signal, $r(t)$, as its real part and the Hilbert transform (HT) of $r(t)$ as its imaginary part, and can hence be written as 
    $z(t) = r(t) + j HT(r(t))$
where $HT(r(t)) = \frac{1}{\pi} \int_{-\infty}^{\infty} \frac{r(t- \tau)}{\tau}d\tau$.
Formally, the envelope $e(t)$ of the signal $r(t)$ is the magnitude of its analytic signal; i.e., 
\begin{equation}
    e(t) \triangleq |z(t)| = \sqrt{r(t)^2 + HT(r(t))^2}
\end{equation}

\subsection{The Proposed IQ Data Representation: The Double-Sided Envelope's Power Spectrum (\proposed)}
\begin{figure}
\begin{center}
   \includegraphics[width=1\columnwidth]{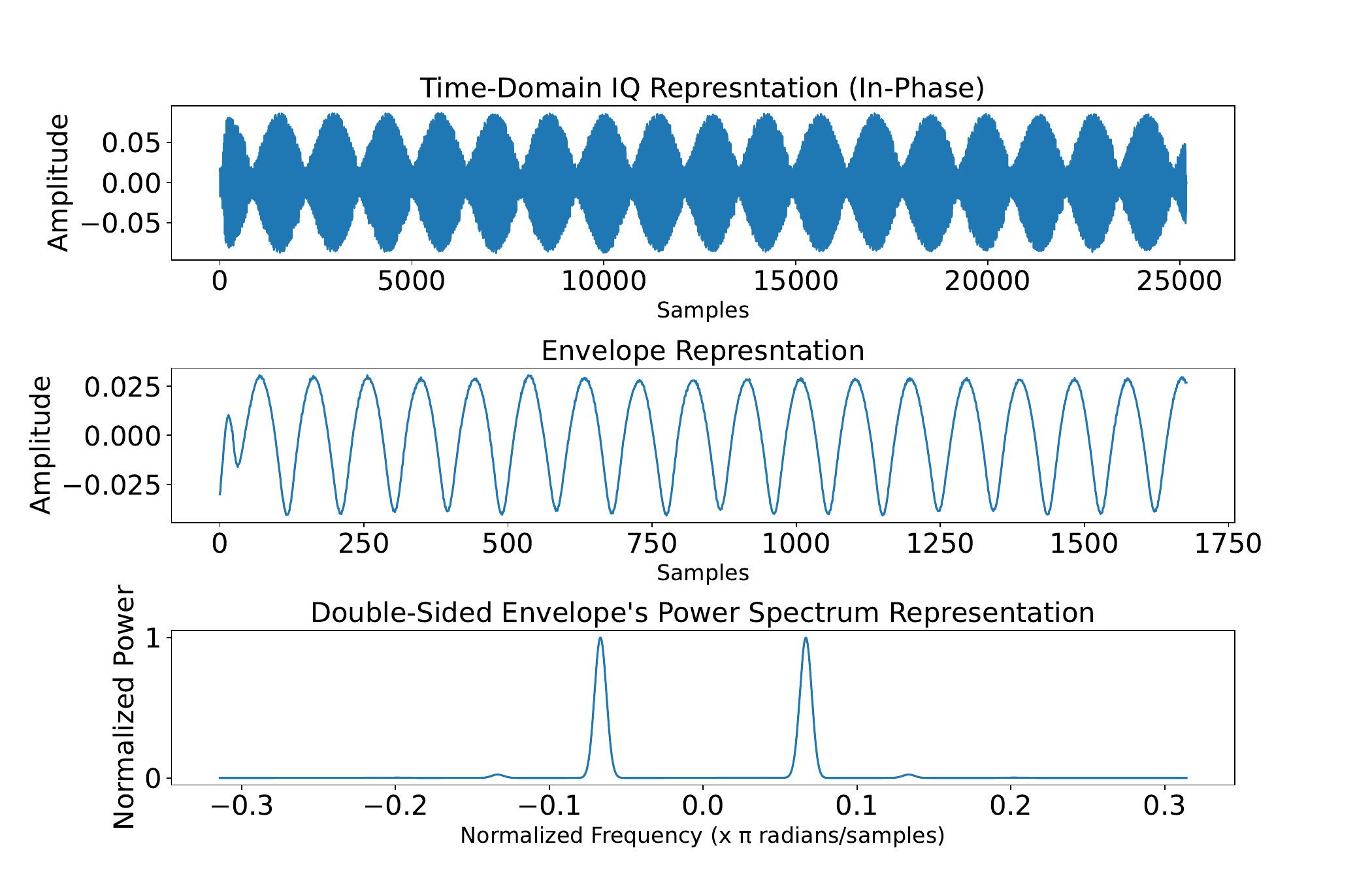}
\end{center}
\caption{Extracting the \proposed~feature from a WiFi Frame.}  
\label{env_repr}
\end{figure}

After extracting the envelope of the IQ signal using the analytic signal presentation as described in Sec.~\ref{EPS-extraction}, we remove the DC offset of the envelope and compute its normalized double-sided power spectrum, which results in one main sideband and its harmonics on each side. We propose this double-sided envelope's power spectrum, termed \proposed~for short, as the new IQ data representation  to use as input to the deep learning models. As we show later, this improves the models' accuracy significantly and makes them highly robust to the domain adaptation challenges we described in Sec.~\ref{sec:into}.
Fig. \ref{env_repr} shows the three stages involved in extracting \proposed~from a WiFi frame sent by one of the Pycom devices. The figure at the top displays the time-domain I component values of the WiFi frame, which exhibits sinusoidal variations in amplitude due to the impairments of the crystal oscillator. The figure in the middle depicts the extracted envelope of the frame using the analytic signal representation. The figure at the bottom shows the double-sided envelope's power spectrum, \proposed.

\subsection{\proposed~Feature Distinguishability Across Different Devices}

 \begin{figure}[t] 
    \centering 
    \includegraphics[width=\linewidth]{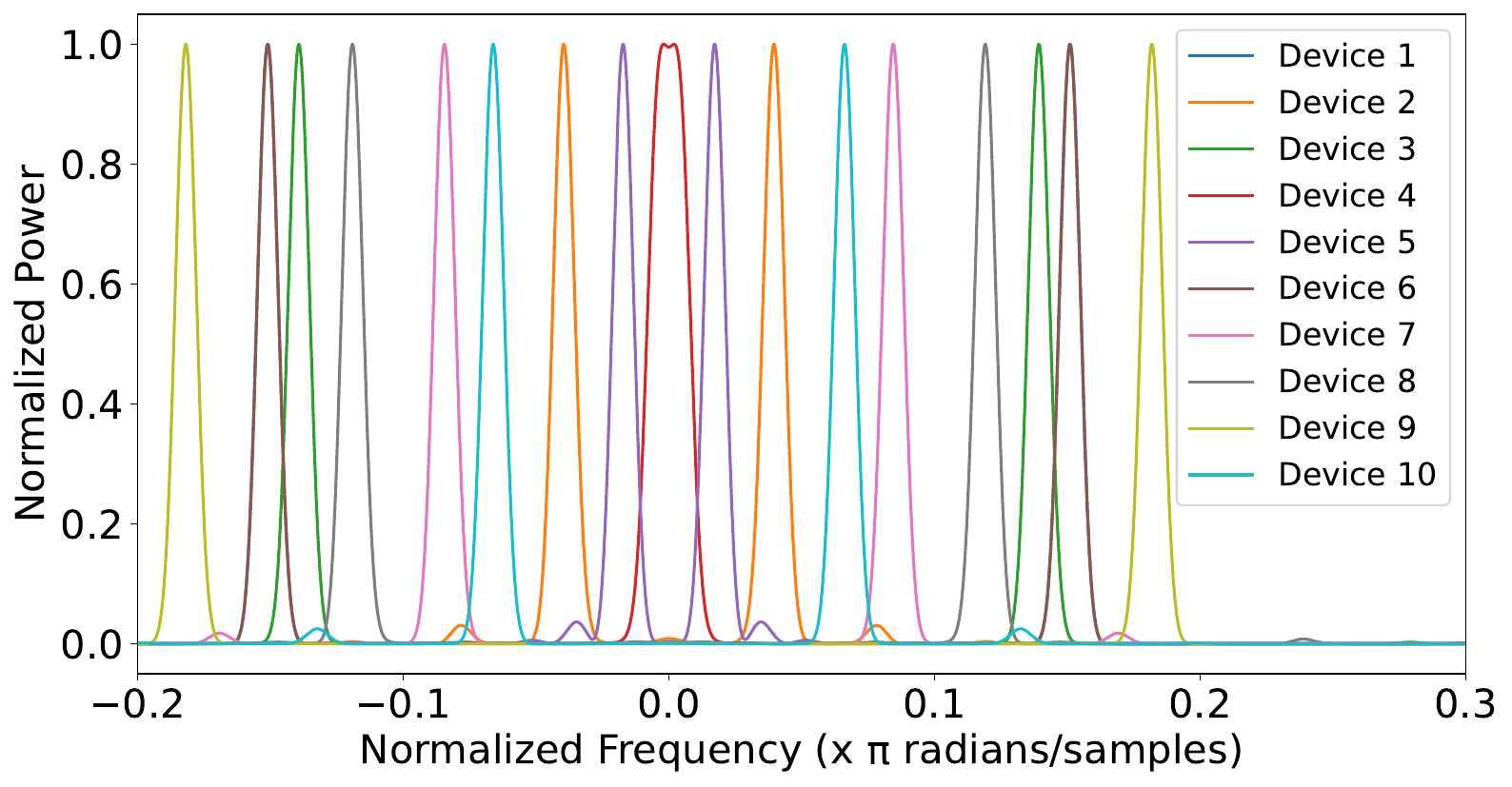} 
    \caption{The \proposed~representation of 10 devices.} 
    \label{fig:unique_spec} 

    \vspace{1em} 

    \includegraphics[width=\linewidth]{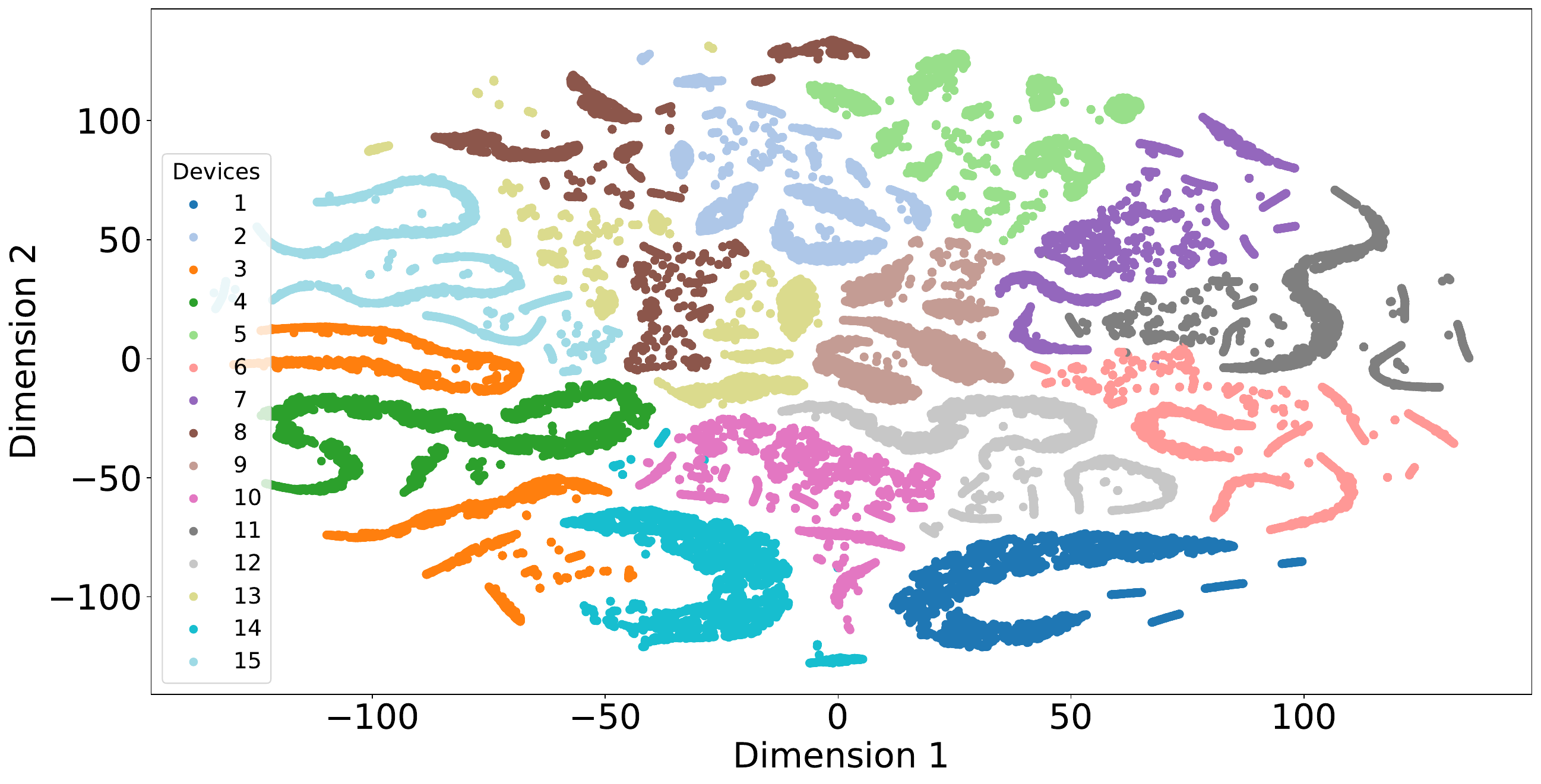} %
    \caption{The t-SNE visualization of the complete testbed.}
    \label{fig:t-sne} 
\end{figure}

In the context of RF fingerprinting, a signal representation that exhibits distinctive device-specific characteristics is critical. The proposed \proposed~feature possesses this property, as it captures the local oscillator's behavior, which is affected by the oscillator's unique hardware impairments. To validate this hypothesis, we conducted an experimental evaluation using our testbed consisting again of $15$ Pycom devices, running the IEEE802.11b protocol and a USRP B210 receiver (more testbed details are provided later in Sec.~\ref{sec:setup}).
Our results depicted in Fig. \ref{fig:unique_spec} reveal that the \proposed~representation is indeed unique for each device, as evidenced by the discernible differences, across the 10 studied devices, in the shape and location of the main sideband and its harmonics. Moreover, this representation or feature contains thousands of samples, making it a high-dimensional data type that can benefit from non-linear dimensionality reduction techniques such as t-distributed stochastic neighbor embedding (t-SNE)~\cite{hinton2002stochastic}. 
By visualizing the representation of $3000$ packets from each device using t-SNE, as shown in Fig. \ref{fig:t-sne}, we demonstrate that the \proposed~representation extracted from more than $40,000$ packets is well-separated and suitable as an effective input for device classification. This will be further validated through experimental results that are presented later in Sec.~\ref{sec:eval}.

 \subsection{\proposed~Feature Robustness to Domain Changes}

\begin{figure} 
    
    \begin{minipage}{\linewidth} 
        \centering 
        \includegraphics[width=\linewidth]{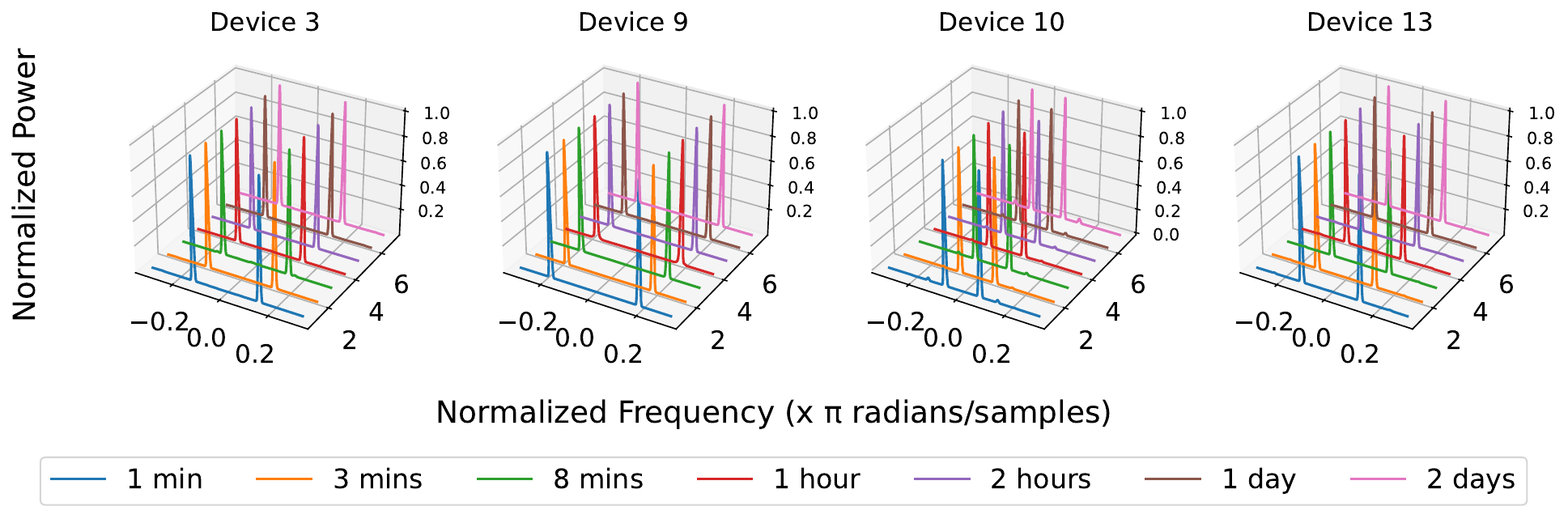} 
        \caption{Time-domain scenario showing the \proposed~representations for 4 devices extracted at 7 different time intervals.} 
        \label{time} 
    \end{minipage}
\end{figure}

Exhibiting unique device-specific features is necessary but insufficient for a representation to serve as a good fingerprint for device classification. If a representation of a device varies randomly each time the signal is captured, it cannot offer a reliable fingerprint and therefore cannot be utilized as an input for the device classification system. Hence, after we showcased the distinguishability property of the \proposed~representation using our testbed, we now test its domain-adaptation ability by examining its robustness to maintaining device separability when there is a change across domains. And we do so by considering three domains: time, channel, and location.


\subsubsection{Time-Agnostic Fingerprinting}
The establishment of stability in any proposed representation is of paramount importance for the practical implementation of RF fingerprinting systems, particularly in dynamic real-world environments where temporal changes are expected. To ascertain the robustness and long-term reliability of our proposed Double-Sided Envelope Power Spectrum (\proposed) representation, a comprehensive set of experiments was performed on our testbed, comprising 15 devices transmitting 802.11b packets. To capture the stability of the devices over time, we employed a wired connection as described in Sec.~\ref{wired_Setup} and initiated the data-capturing process 12 minutes after the devices were activated, ensuring an initial hardware settling and warm-up period \cite{elmaghbub2023impact,shen2021radio}. Precisely, packet captures were obtained at specific intervals, namely 1 minute, 3 minutes, 8 minutes, 1 hour, 2 hours, 1 day, and 2 days, spanning three consecutive days. For each individual device, the \proposed~representation was extracted from the recorded packets at the aforementioned time points over the three-day period. Fig.~\ref{time} depicts the plotted results for four representative devices, clearly showcasing that for each device, all the \proposed~representations extracted at the different time intervals overlap. This demonstrates that the proposed \proposed~representation is time agnostic and remains unchanged over time.
This uniformity in the \proposed~representation was consistently observed across all 15 devices (though shown only for 4 devices in the paper), thus providing compelling evidence of the stability and reliability of our proposed \proposed~representation over time. Furthermore, these findings underscore the efficacy of \proposed~in mitigating the sensitivity to temporal variations encountered in DL-based RF fingerprinting techniques.

\subsubsection{Channel-Agnostic Fingerprinting}
\begin{figure}
    \begin{minipage}{\linewidth} 
        \centering 
        \includegraphics[width=\linewidth]{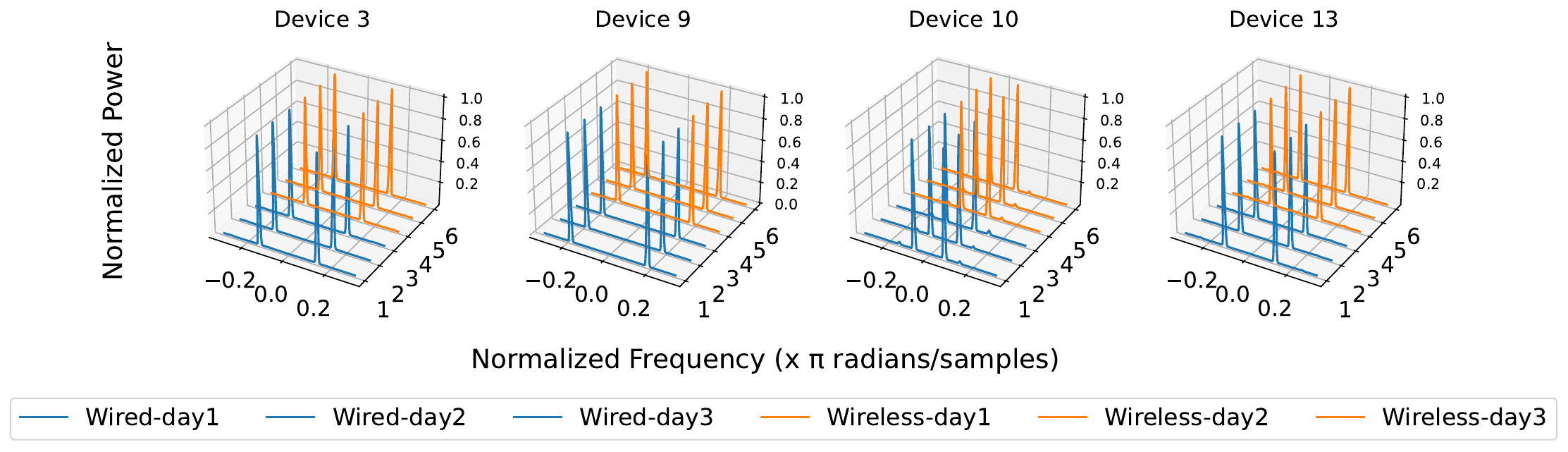} 
        \caption{Channel-domain scenario showing the \proposed~representations for 4 different devices extracted under both wired and wireless setups, each over three days.} 
        \label{channel} 
        \end{minipage}
\end{figure}

To investigate the impact of the wireless channel on the stability and consistency of \proposed, we conducted the following experiment in an indoor environment. The devices were positioned at a fixed distance of $1$ meter from the receiver, and packet captures were performed over a duration of three consecutive days, as detailed later in Sec.~\ref{wireless_Setup}. The objective of this investigation was to compare the \proposed~representations of packets corresponding to each individual device across both the wired and wireless channels over time, thereby discerning the influence of channel variations over time. Fig.~\ref{channel} presents the graphical representations of the \proposed~features obtained from four distinct devices under both wired and wireless channel conditions. Notably, the figures effectively demonstrate that the \proposed~representation of each device remains unaltered regardless of the underlying channel characteristics. This observation unequivocally establishes the inherent stability and reliability of our proposed \proposed~representation, even in the presence of wireless channel effects, over the course of three consecutive days. Importantly, this behavior was consistently observed across all 15 devices, thereby fortifying the empirical evidence supporting the robustness and efficacy of our proposed \proposed. Consequently, these findings firmly substantiate the ability of the \proposed~representation to effectively mitigate the potential sensitivity to channel variations encountered in DL-based RF fingerprinting methods. 

\subsubsection{Location-Agnostic Fingerprinting}
\begin{figure}
        \begin{minipage}{\linewidth}
         \centering 
        \includegraphics[width=\linewidth]{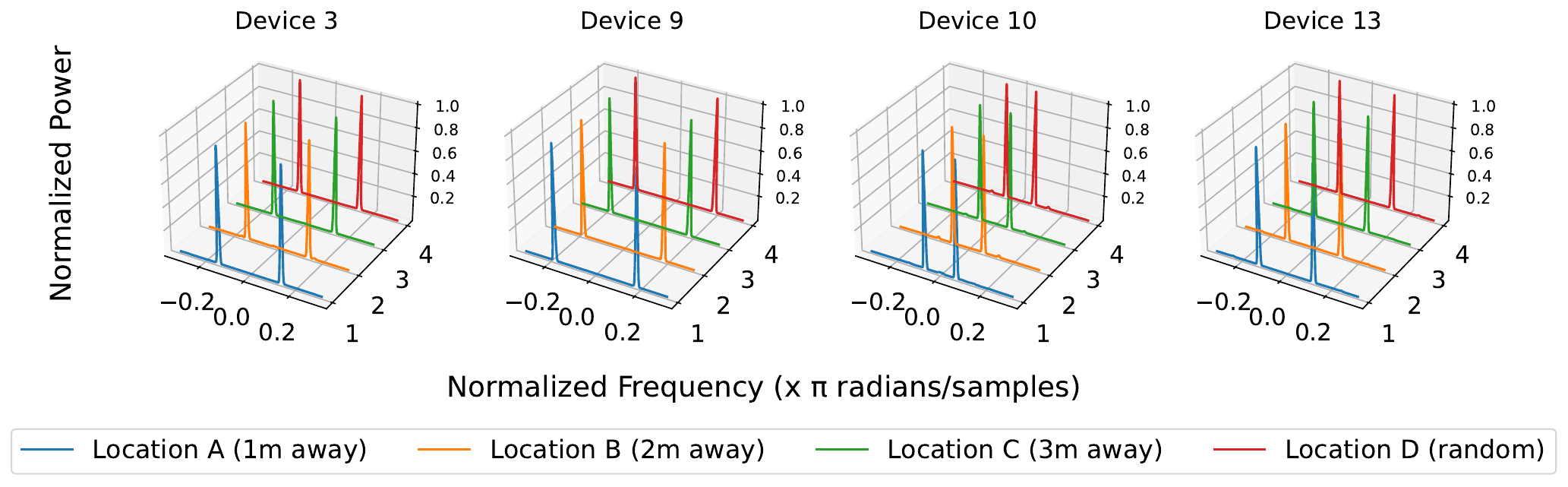} 
        \caption{Location-domain scenario showing the \proposed~representations for 4 devices extracted from 4 different locations. Devices in Locations A, B and C are all placed at fixed distances; devices in Location D are placed at random distances.} 
        \label{locations} 
    \end{minipage}
\end{figure}

Changing the distance between the transmitting devices and the receiver after training can also lead to a drastic drop in performance. To evaluate the robustness of the \proposed~representation to such distance changes, we captured data at three different locations with the devices being placed $1$m-away (Location A), $2$m-away (Location B), and $3$m-away (Location C) from the USRP receiver; this setup is shown in Fig.~\ref{diff_loc} and discussed in more detail in Sec.~\ref{location_Setup}. Fig.~\ref{locations} shows the \proposed~representation of WiFi packets from four devices over the three different locations. 
To extend the reliability test with regard to location and distance, we also considered another realistic scenario in which the devices were randomly deployed within a radius of $3$m from the receiver as shown in Fig.~\ref{rand_loc} (refer to Sec.~\ref{random_Setup} for more details). The corresponding \proposed~representations are depicted as Location D (random) in Fig.~\ref{locations}. The plots in Fig.~\ref{locations} manifest the stability of the \proposed~feature representation over the four studied location scenarios as the signal representations of the four locations completely overlap. 
Our findings confirm the stability of the \proposed~representation in scenarios in which the location, distance, and time of the training and testing sets are different, again making the proposed \proposed~a more reliable and robust input for DL-based RF fingerprinting methods.

\section{\proposed-Based Fingerprinting Framework for Domain-Agnostic Device Classification}
\label{sec:proposed}
Maintaining good performances of RF fingerprinting when faced with domain shifts due, for instance, to changes in channel conditions and/or device location/distance has proven to be very challenging, hindering the widespread adoption of RF fingerprinting technology in real-world applications. Our proposed fingerprinting framework, based on the proposed \proposed~feature representation, has demonstrated stable behavior across various settings and increased resiliency to domain changes, thereby overcoming the aforementioned challenges.

In this section, we evaluate the effectiveness of the proposed \proposed~feature representation vis-a-vis of its ability to adapt to domain (channel, time, and location) changes when the \proposed~data is used as an input to a typical Convolutional Neural Network (CNN) device classifier \cite{2019deepradioid, elmaghbub2020widescan, sankhe2019oracle,liu2017deep}.


\subsection{An Overview of the Proposed EPS-CNN Framework}
\begin{figure*}[t]
    \centering
    \includegraphics[width=\textwidth]{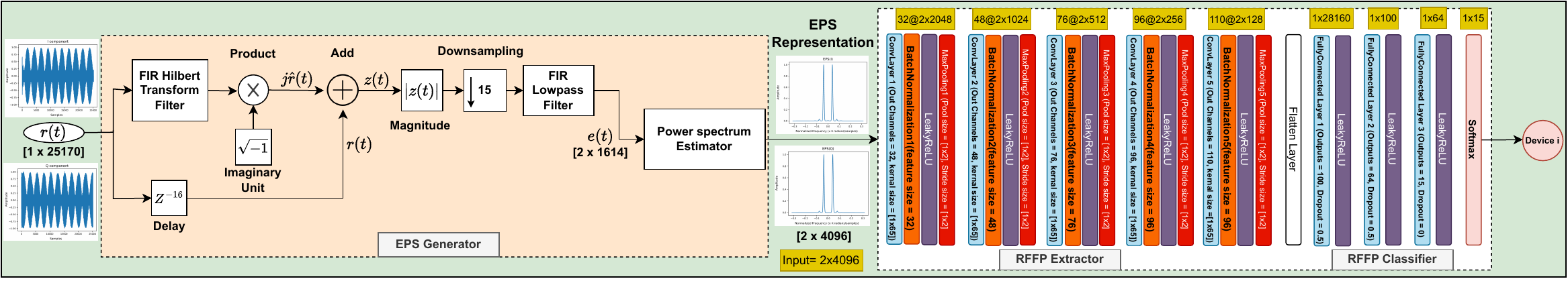}
    \caption{EPS-CNN Framework Overview}
    \label{EPS-CNN}
\end{figure*}


At its high level, the proposed EPS-CNN framework, shown in Fig.~\ref{EPS-CNN}, consists of an EPS generator, which takes the complex-valued IQ representation of a received frame, $r(t)$, as an input 
and then processes the I (In-phase) and Q (Quadrature) components separately. For each frame, the EPS generator first extracts the envelope of the signal, $e(t)$, and then generates the EPS representation of the two components: EPS(I) and EPS(Q). Refer to Sec.~\ref{subsec:eps-extract} for details about the EPS representation generation. The two EPS representations are then concatenated into a tensor (e.g., of size 2x4096) and passed to the CNN network that extracts the suitable features using the six convolution blocks followed by three fully connected layers and a Softmax layer. The CNN is also responsible for learning a classifier from the extracted features to accurately predict the corresponding device of the incoming frame (Device i). 

\subsection{Extraction of \proposed~Feature}\label{subsec:eps-extract}
The implementation we used to extract the signal's envelope, $e(t)$, from the received signal, $r(t)$, is shown in Fig.~\ref{EPS-CNN}; refer back to Sec.~\ref{EPS-extraction} for the derived $e(t)$ expression. We construct the analytic signal by first passing the IQ values of the received frame through an FIR Hilbert transform filter based on the Parks-McClellan algorithm \cite{Park_McClellan} implemented in MATLAB Signal Processing Toolbox. The output of the filter is then multiplied by $\sqrt{-1}$ (the imaginary unit) and added to the time-delayed original signal. It is important to introduce a delay in the input signal because the FIR filter implementation of the Hilbert transform introduces a delay equivalent to half the length of the filter.

The signal's envelope, $e(t)$, is calculated by taking the absolute value of the analytic signal, which is characterized by a lower frequency compared to the original signal. Hence, we first reduce the sampling frequency by a factor of $15$ and then pass the resulting signal through a lowpass filter to effectively eliminate ringing and smooth the envelope. Once the envelope is extracted, we center the envelope's amplitude around zero before generating the corresponding normalized double-sided envelope's power spectrum, i.e., \proposed, representation using the power spectrum estimator.

\subsection{CNN Architecture}
 We train our CNN, using the PyTorch library, 
on an NVIDIA Cuda-enabled NVIDIA GeForce RTX 2080 Ti GPU system for 30 epochs. The input to the model is the EPS representation with a dimension of 2x4096. The first layer applies a 2D convolution with a filter size of 1x65, followed by batch normalization, and LeakyReLU activation. Then, a max-pooling layer with kernel size 1x2 and stride 1x2 is applied to reduce the dimensions of the feature maps. The same pattern is repeated five more times, with increasing numbers of output channels (32, 48, 64, 76, 96, and 110) and decreasing feature map sizes, until a final feature map size of 2x64 is obtained. Then, two fully connected layers with output sizes equal to 100 and 64 are applied, each followed by a dropout layer and LeakyReLU activation. Finally, the output is passed to another fully connected layer that maps it to the 15 output classes, which is passed to a softmax layer to produce the predicted label. We experimentally chose a learning rate of 3e-4 which decays by decaying factor after every 3000 steps. Finally, we use the stochastic gradient descent optimizer with momentum  and a weight decay parameter that adds an L2 regularization on the weights to avoid overfitting. All codes will be released on our GitHub page.

\section{Testbed Setup, Dataset Collection and Experimental Scenarios}
\label{sec:setup}

\begin{figure}
     \subfloat[15 Pycom transmitting devices.\label{subfig-1:tx}]{%
       \includegraphics[width=0.23\textwidth, height = 0.237\textwidth]{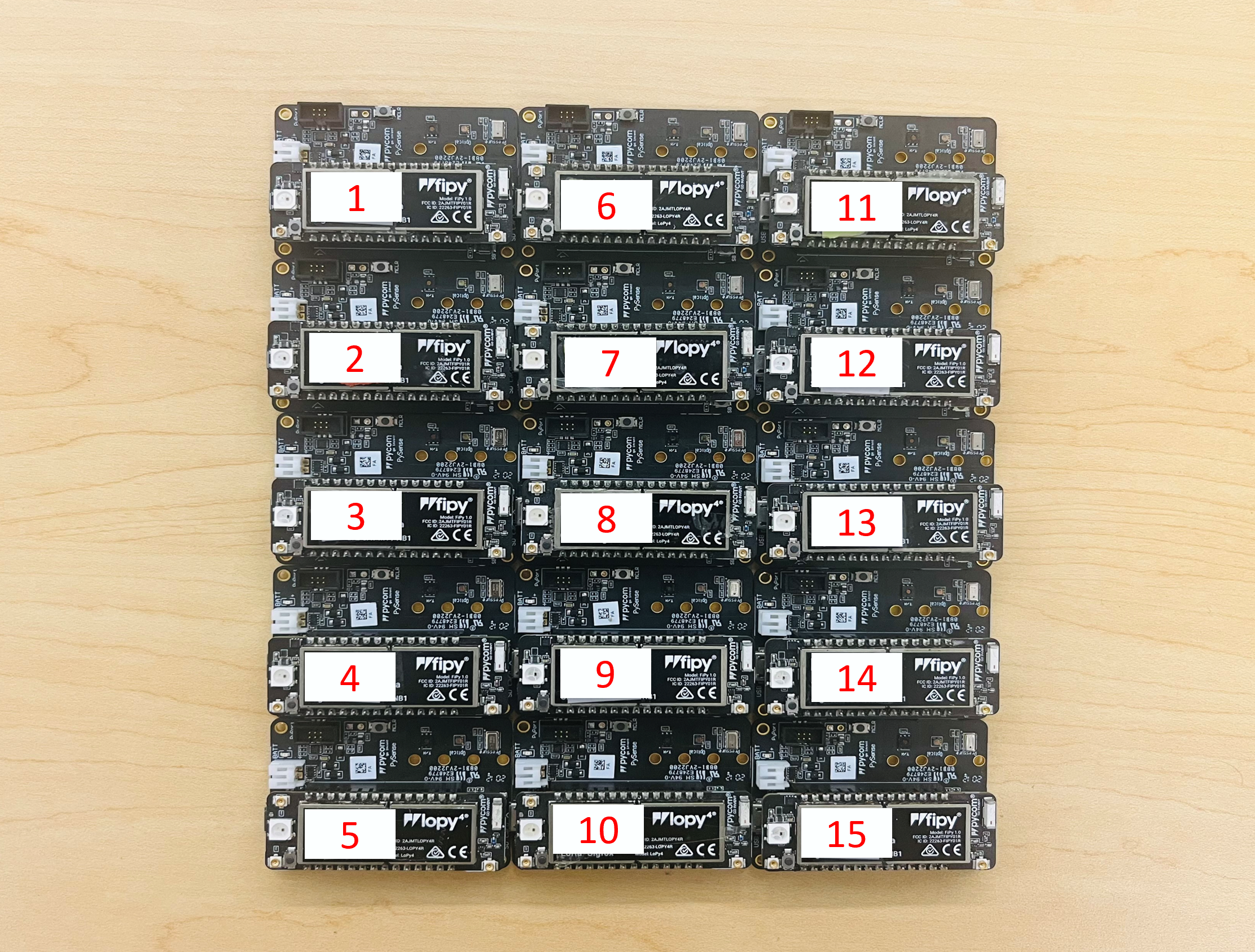}
     }
    \hspace{0.00001cm}
     \subfloat[Wired-WiFi vs. wireless-WiFi.\label{subfig-2:recv}]{%
       \includegraphics[width=0.23\textwidth, height = 0.237\textwidth]{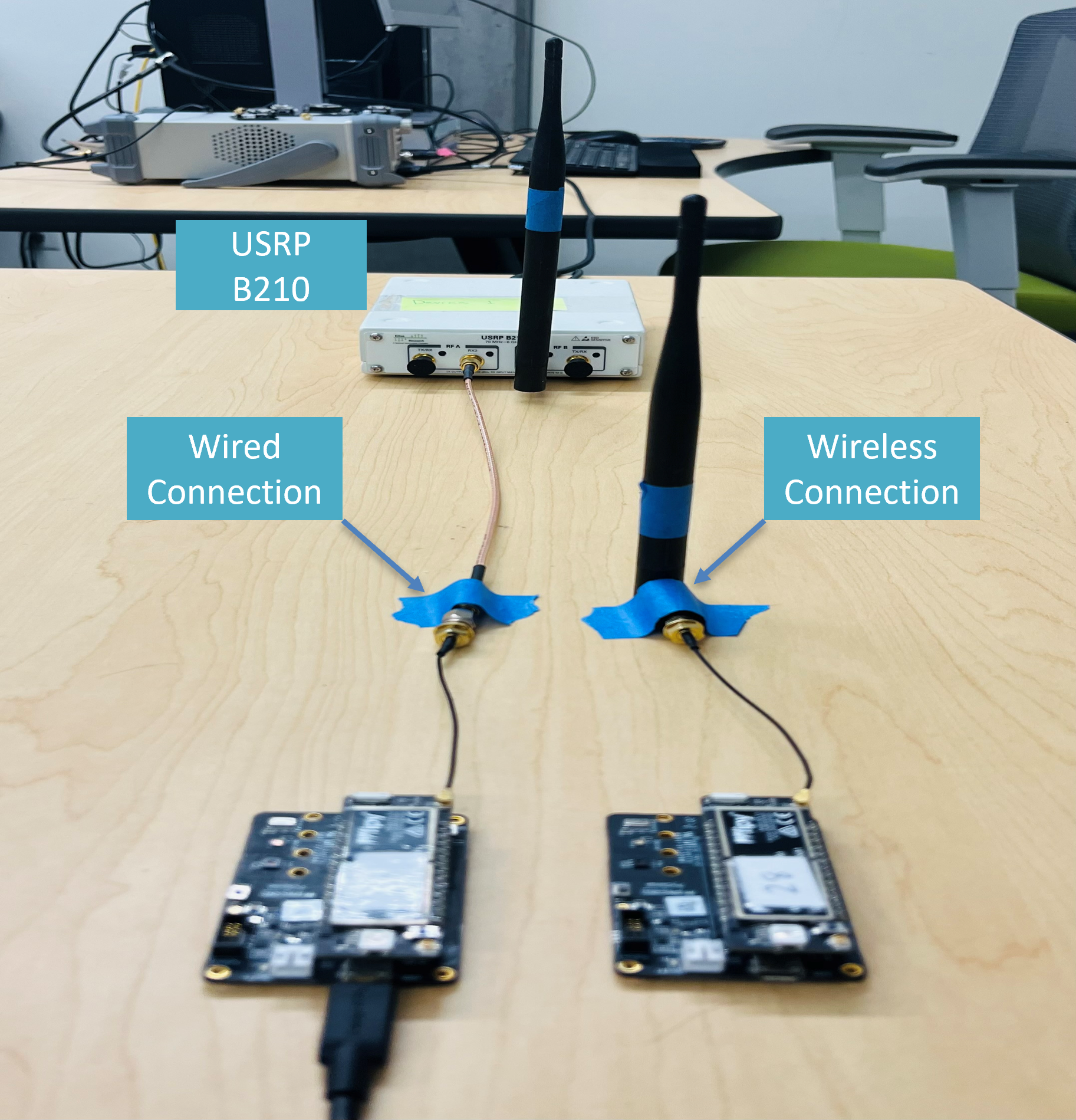}
     }
     \caption{IoT Testbed consisting of 15 Pycom transmitting devices and a USRP B210 receiving device}
     \label{fig:testbed}
\end{figure}

We now describe the testbed setup, the experimental scenarios, and the collected WiFi datasets used for evaluating the effectiveness of the proposed techniques. The WiFi datasets, their description and their download information can be found at
\href{https://research.engr.oregonstate.edu/hamdaoui/datasets/}{\color{blue}{http://research.engr.oregonstate.edu/hamdaoui/datasets}}.

\begin{figure*}
  \centering
  \begin{minipage}{0.48\textwidth}
    \centering
    \includegraphics[width=\textwidth]{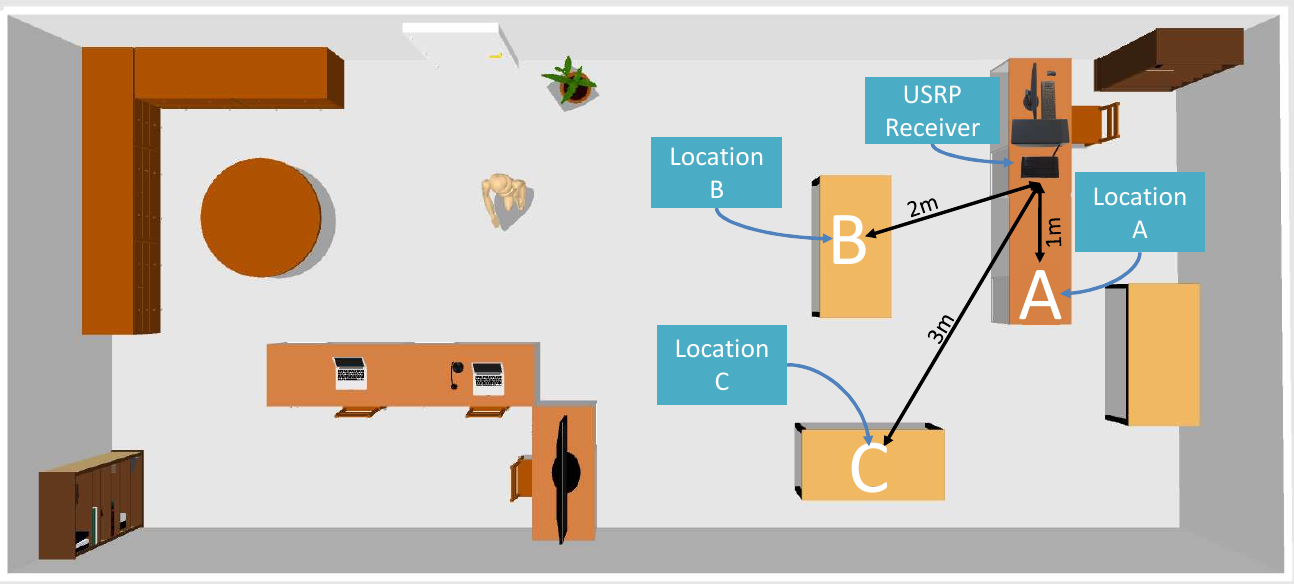}
    \caption{Different-Location Setup}
    \label{diff_loc}
  \end{minipage}\hfill
  \begin{minipage}{0.48\textwidth}
    \centering
    \includegraphics[width=\textwidth]{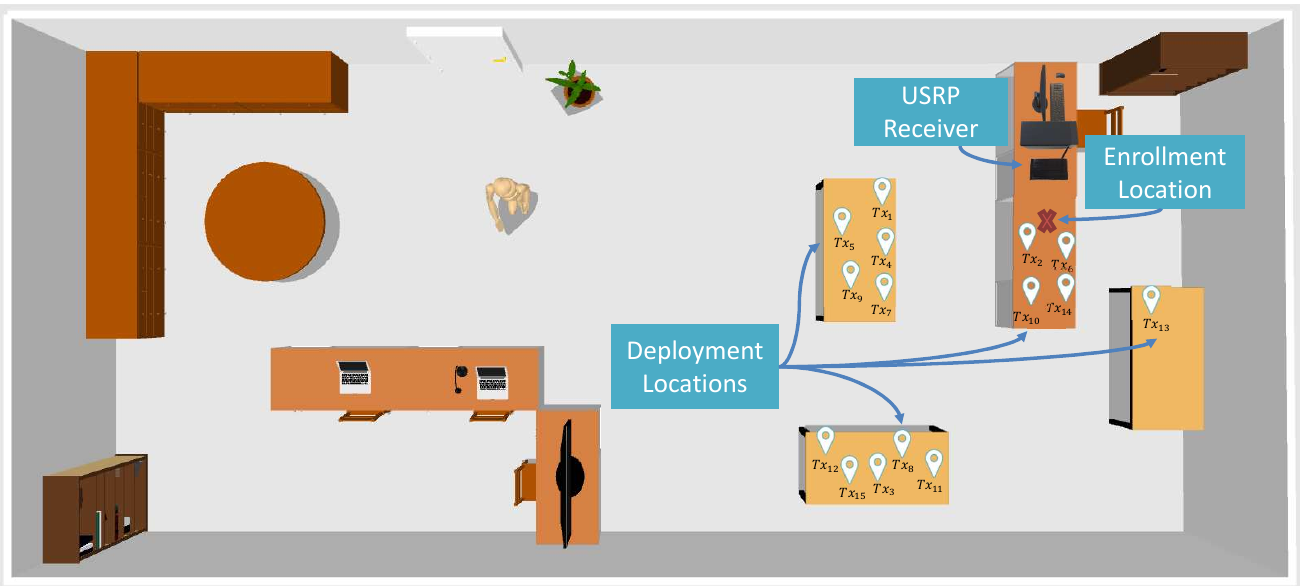}
    \caption{Random-Location Setup}
    \label{rand_loc}
  \end{minipage}
\end{figure*}

\subsection{Testbed Setup}
The testbed used for our evaluation, depicted in Fig.~\ref{fig:testbed}, comprises $15$ Pycom devices, including $10$ FiPy and $5$ LoPy boards, equipped with ESP$32$, Semtech SX$1276$, and Sequans Monarch chips that support WiFi, Bluetooth, LoRa, Sigfox, and LTE network protocols. Data acquisition was performed using an Ettus USRP B$210$ receiver, which was synchronized with an external OCXO for improved sampling accuracy and stability. All devices were powered via USB from an HP laptop and configured to transmit IEEE802.11b WiFi packets using the high-rate direct-sequence spread-spectrum (HR/DSSS) physical layer in the 2GHz spectrum. The transmitting devices transmitted at a rate of 1Mbps with a carrier frequency of 2.412GHz and a bandwidth of 20MHz, while connected to the same 1/2 Wave Whip antenna.

\subsection{Dataset Collection}
We initiated the data-capturing process 12 minutes after the devices were activated, thereby ensuring an initial settling and warm-up period \cite{elmaghbub2023impact}. Each device was configured to operate over WiFi Channel 1 with a center frequency of 2412MHz and a bandwidth of 20MHz. The transmitters were programmed to transmit identical IEEE 802.11b frames with a duration of 559us back to back, separated by a small gap. We captured the first two minutes of transmissions using the USRP B210 at a sample rate of 45MSps. The captured signals were then digitally down-converted to the baseband and stored as IQ samples on our computer. To avoid any data dependency on the identity of the WiFi transmitter, all transmitters were configured to broadcast the same packets, which include the same spoofed MAC address and a payload of zero-bytes.

Finally, we extracted the WiFi packets from the raw IQ sample files and stored them in HDF5 formatted files in the same order they were received. This method allowed us to maintain the integrity of the captured signals and ensured that they were accurately represented in the final dataset.

\subsection{Experimental Scenarios} 
\label{dataset_Section}
Our WiFi dataset contains more than $8$TB of WiFi transmissions from $15$ Pycom devices captured in four different setups/scenarios: Wired, Fixed-Location Wireless, Different-locations Wireless, and Random-Locations Wireless.

\subsubsection{Wired Setup} \label{wired_Setup}
To rule out the impact of the wireless channel, we connected our transmitters directly to the USRP receiver via SMA cabling, and collected data over three days, generating more than $5000$ WiFi frames per device every day.

\subsubsection{Indoor Wireless Setup}\label{wireless_Setup}
Instead of wiring the transmitters to the USRP receiver as done in the Wired Setup, we placed them at a fixed location, $1$m away from the USRP receiver which uses a VERT$900$ antenna to capture the signal. We repeated this experiment over three days to assess the generalizability of the proposed technique over time. This setup generated more than $5000$ WiFi frames per device every day. 

\subsubsection{Different-Locations Wireless Setup}\label{location_Setup}
The location from where the transmitter sends its data impacts the characteristics of the received signal, as signals transmitted from different distances/locations usually experience different channel conditions, which is considered in this work as another varying domain. For each transmitter, we then collected data at three different locations, A, B, and C, which are $1$m, $2$m, and $3$m away from the USRP receiver, respectively, as shown in Fig.~\ref{diff_loc}. This was carried out in one day and generated more than $5000$ WiFi frames per device at each location. 

\subsubsection{Random-Location Wireless Setup}\label{random_Setup}
From a practical viewpoint, when the fingerprinting framework is used for device authentication, the messages that are sent by the devices and are to be used for authentication are likely to come from different random locations, and these random locations are also likely to be different from the locations used in the enrollment (training) stage. 
Therefore, we considered collecting datasets for two random-location scenarios on two different days, each consisting of an enrolment phase (data used for training) and a deployment phase (data used for testing). In both enrolment phases, all the transmitters transmitted from the same location, $1$m away from the receiver, and in both deployment phases, the transmitters were located randomly within a radius of $3$m away from the receiver as shown in the floor plan in Fig.~\ref{rand_loc}. The enrollment datasets were collected in the morning while the random deployment datasets were collected on the night of that same day, generating more than $5000$ WiFi frames per device for each dataset.

\section{Device Identification Results}
\label{sec:eval}

\begin{figure*}
\centering
    \subfloat[Testing accuracy]{
   \includegraphics[width=.48\columnwidth]{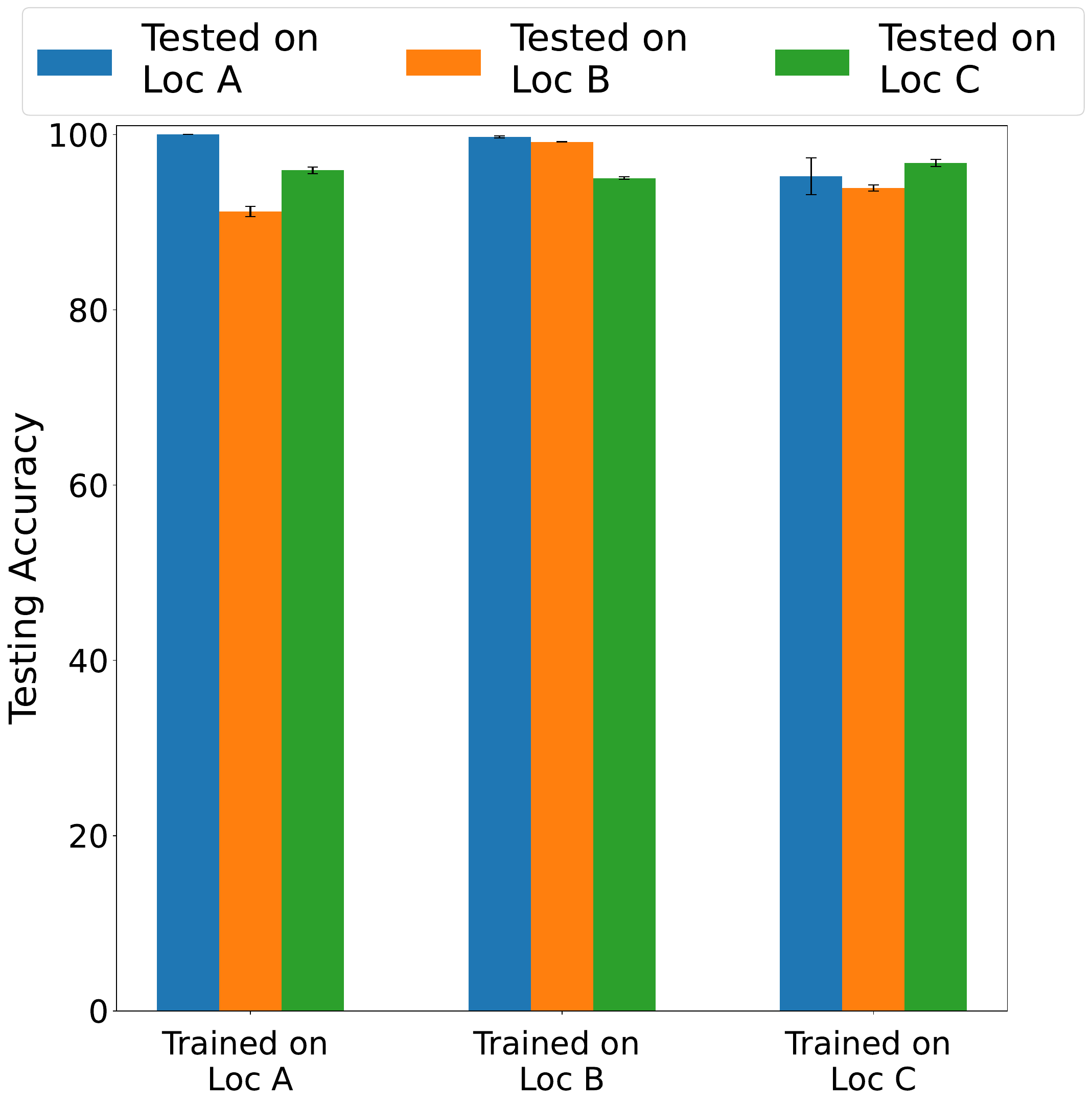}
   \label{eps-trained-locations}}
    \subfloat[ Train-locA / Test-locC]{
   \includegraphics[width=.48\columnwidth]{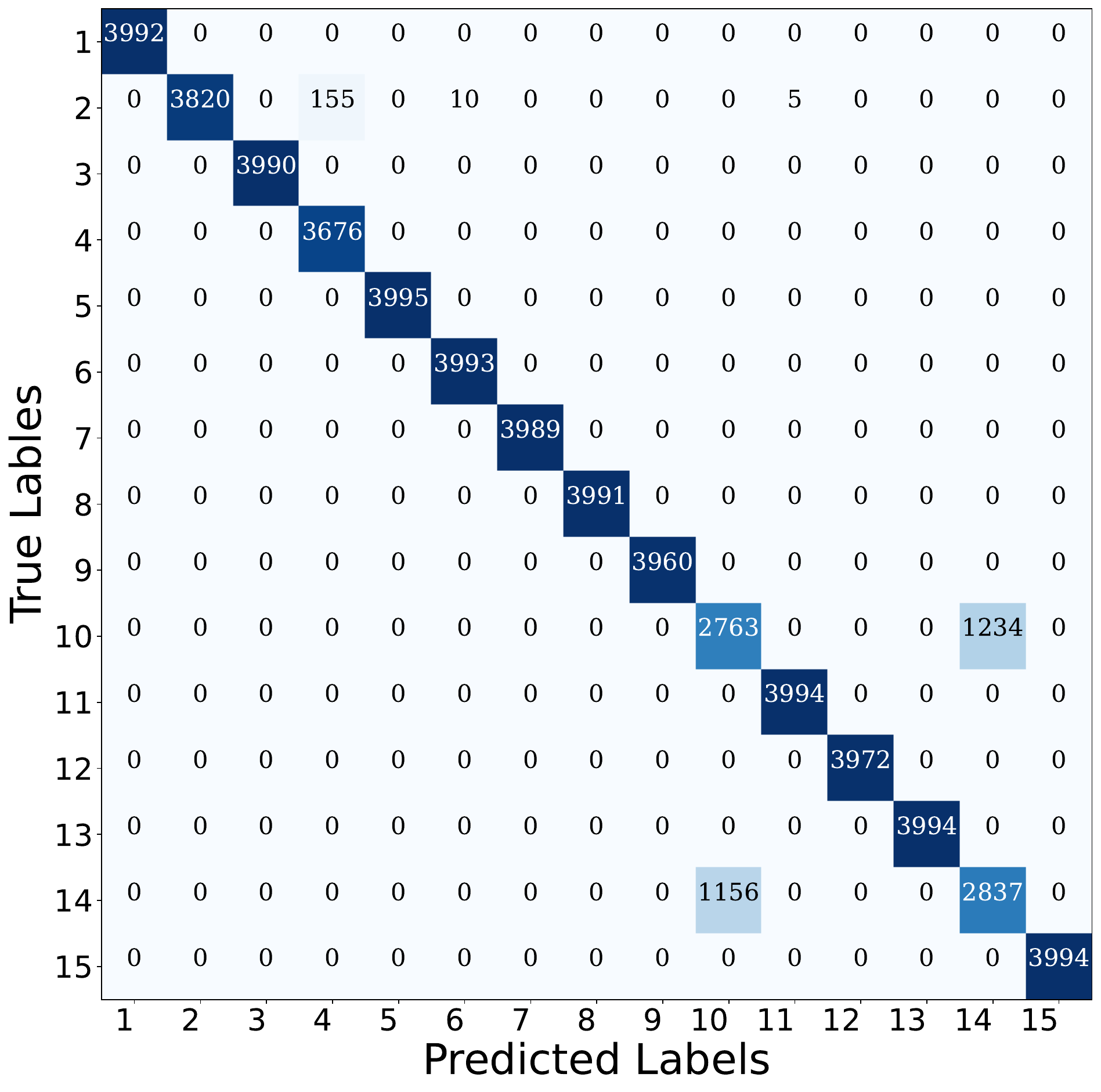}
   \label{eps-loc1}}
    \subfloat[Train-locB / Test-locA]{
   \includegraphics[width=.48\columnwidth]{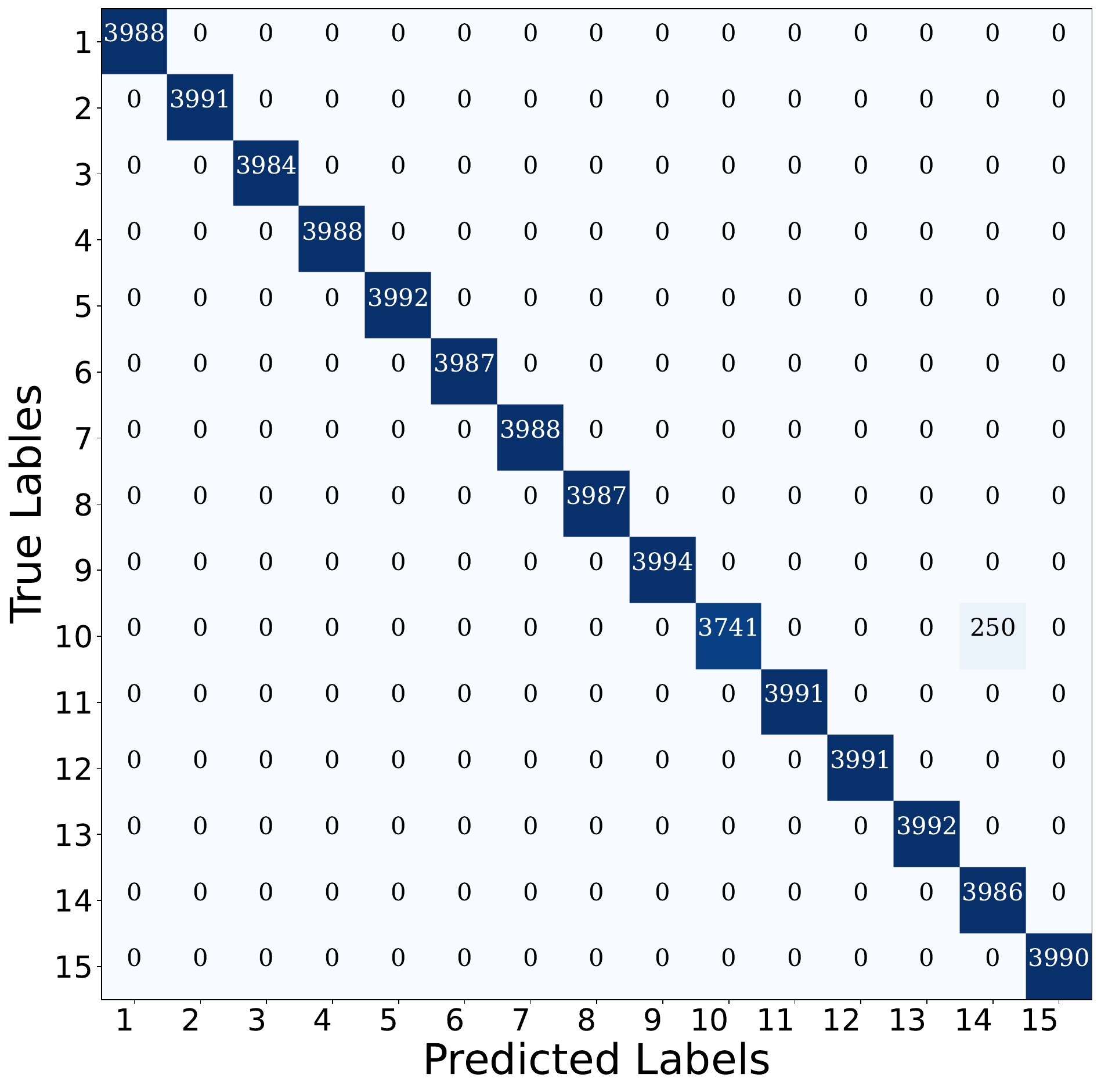}\label{subfig:ocxo3}
   \label{eps-loc2}}
   \subfloat[Train-locC / Test-locA]{
   \includegraphics[width=.48\columnwidth]{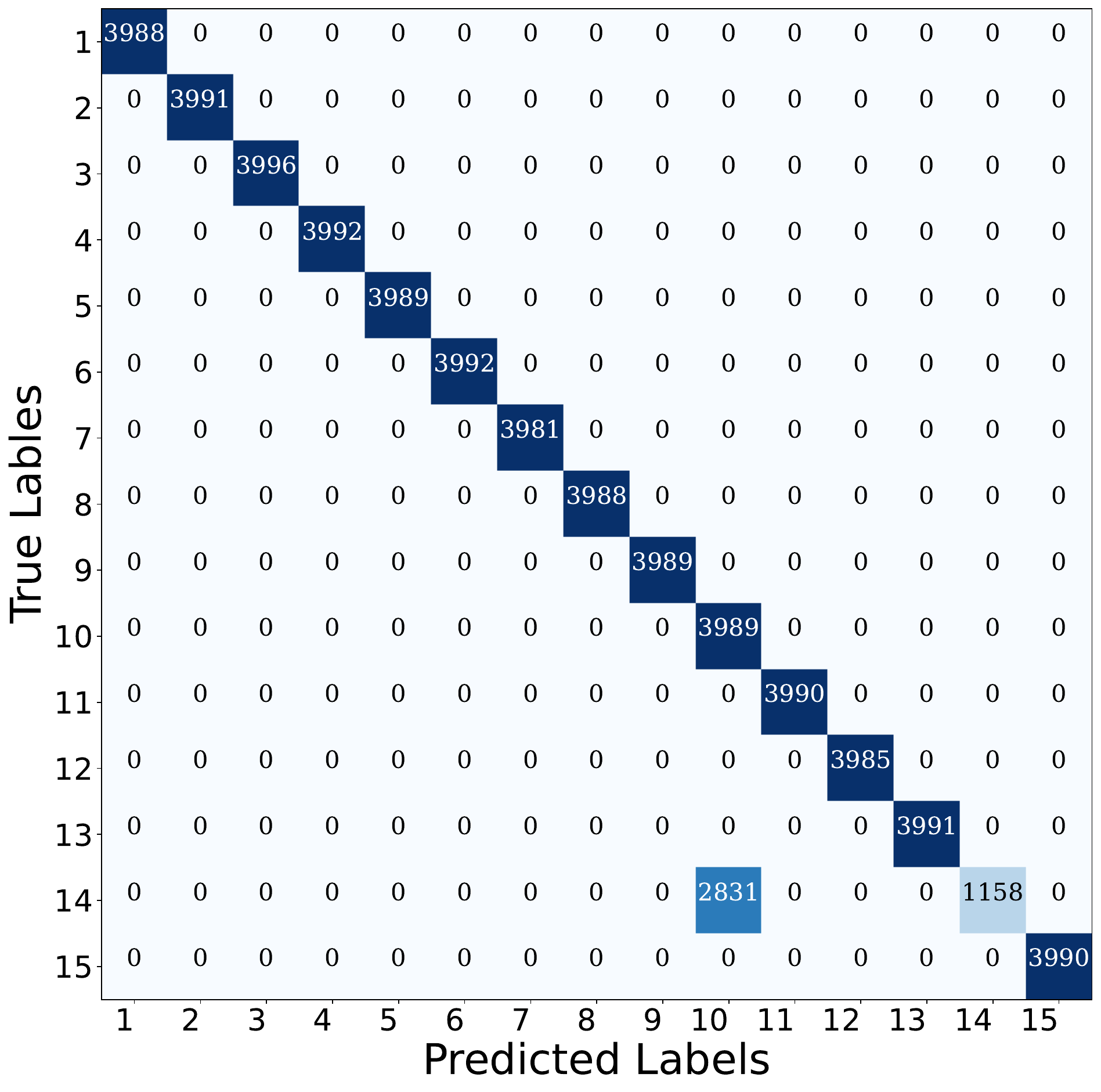}
   \label{eps-loc3}}
\caption{EPS-CNN's performance: (a) Testing accuracy; (b)-(d) confusion matrices for different Train-Location/Test-Location combinations. LocA, LocB, and LocC correspond to when the transmitters are 1m, 2m and 3m away from the receiver.}
\label{eps-locations}
\end{figure*}

\begin{figure*}
\centering
\subfloat[Testing accuracy]{
   \includegraphics[width=.48\columnwidth]{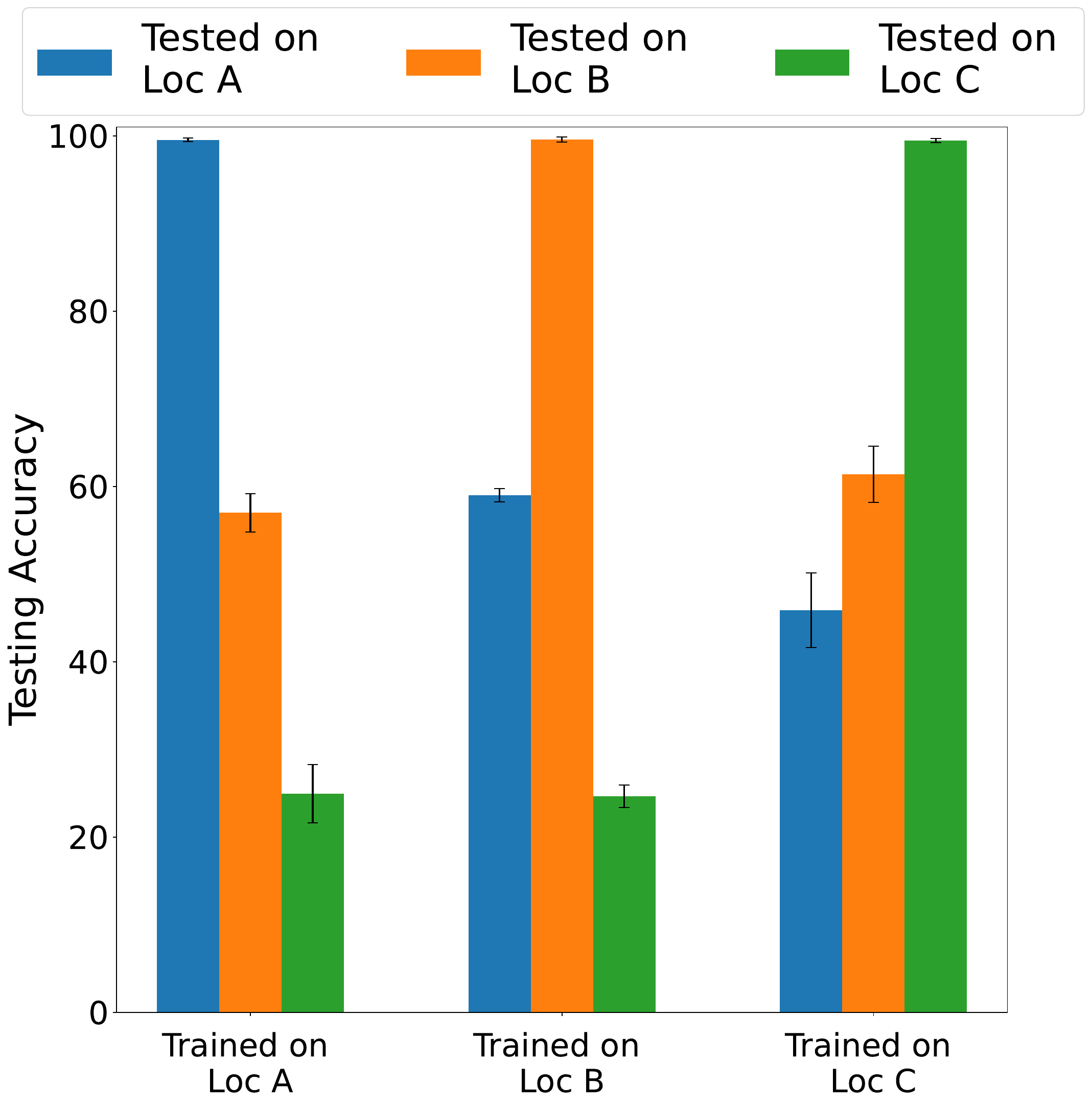}
   \label{IQ-trained-locations}}
\subfloat[Train-locA / Test-locB]{
   \includegraphics[width=.48\columnwidth]{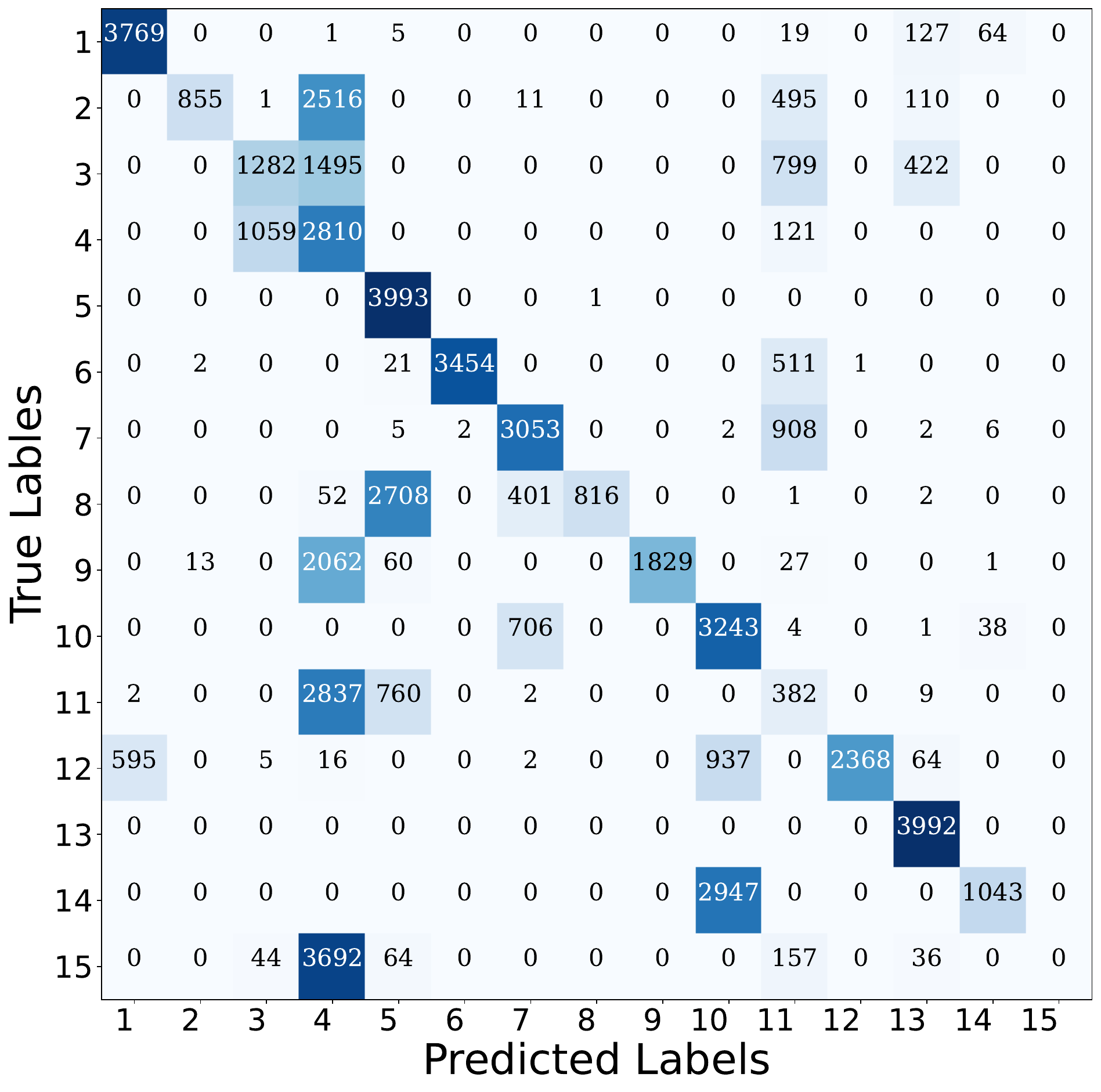}
   \label{IQ-loc1}}
\subfloat[Train-locB / Test-locC]{
   \includegraphics[width=.48\columnwidth]{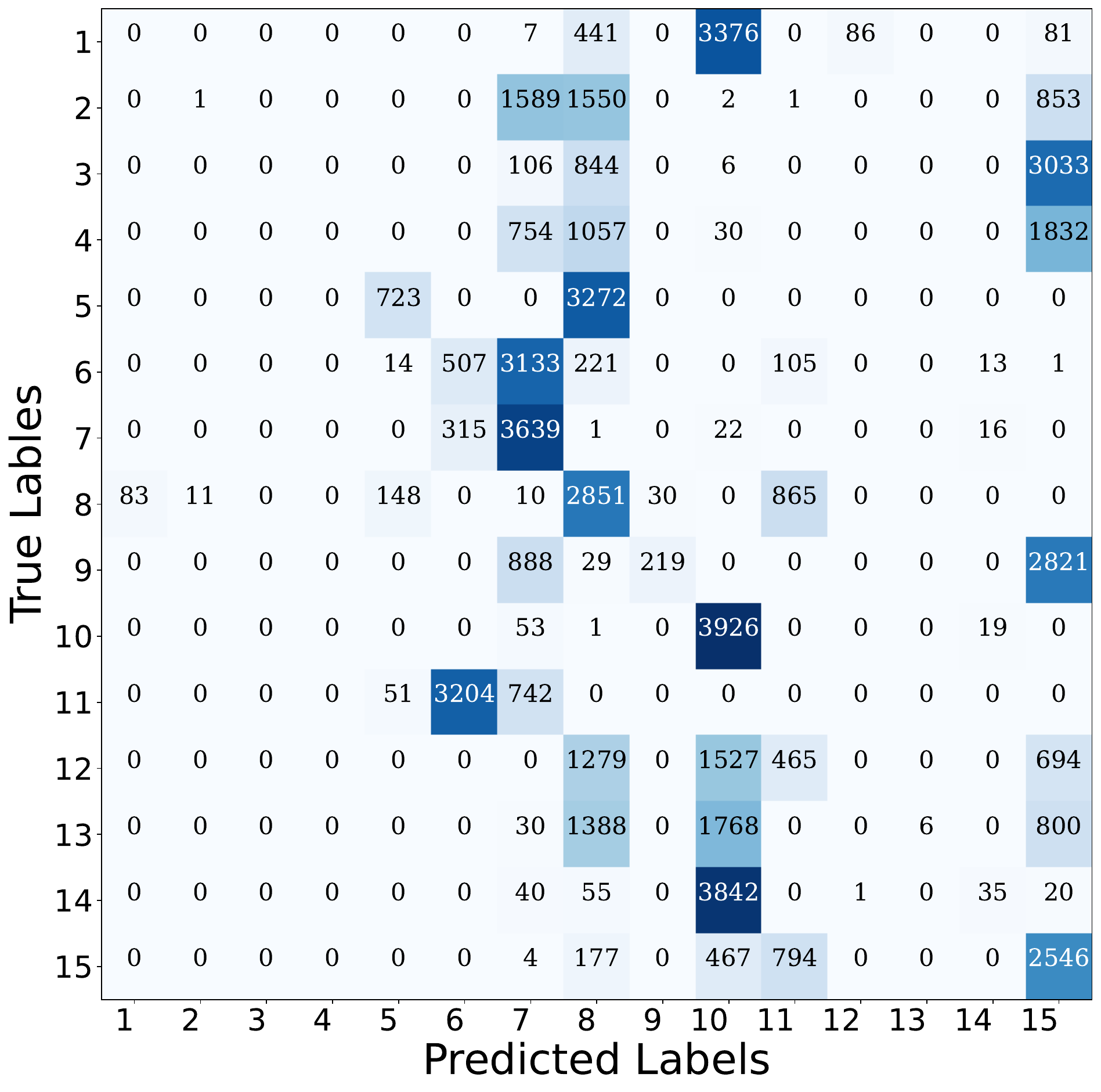}
   \label{IQ-loc2}}
   \subfloat[Train-locC / Test-locA]{
   \includegraphics[width=.48\columnwidth]{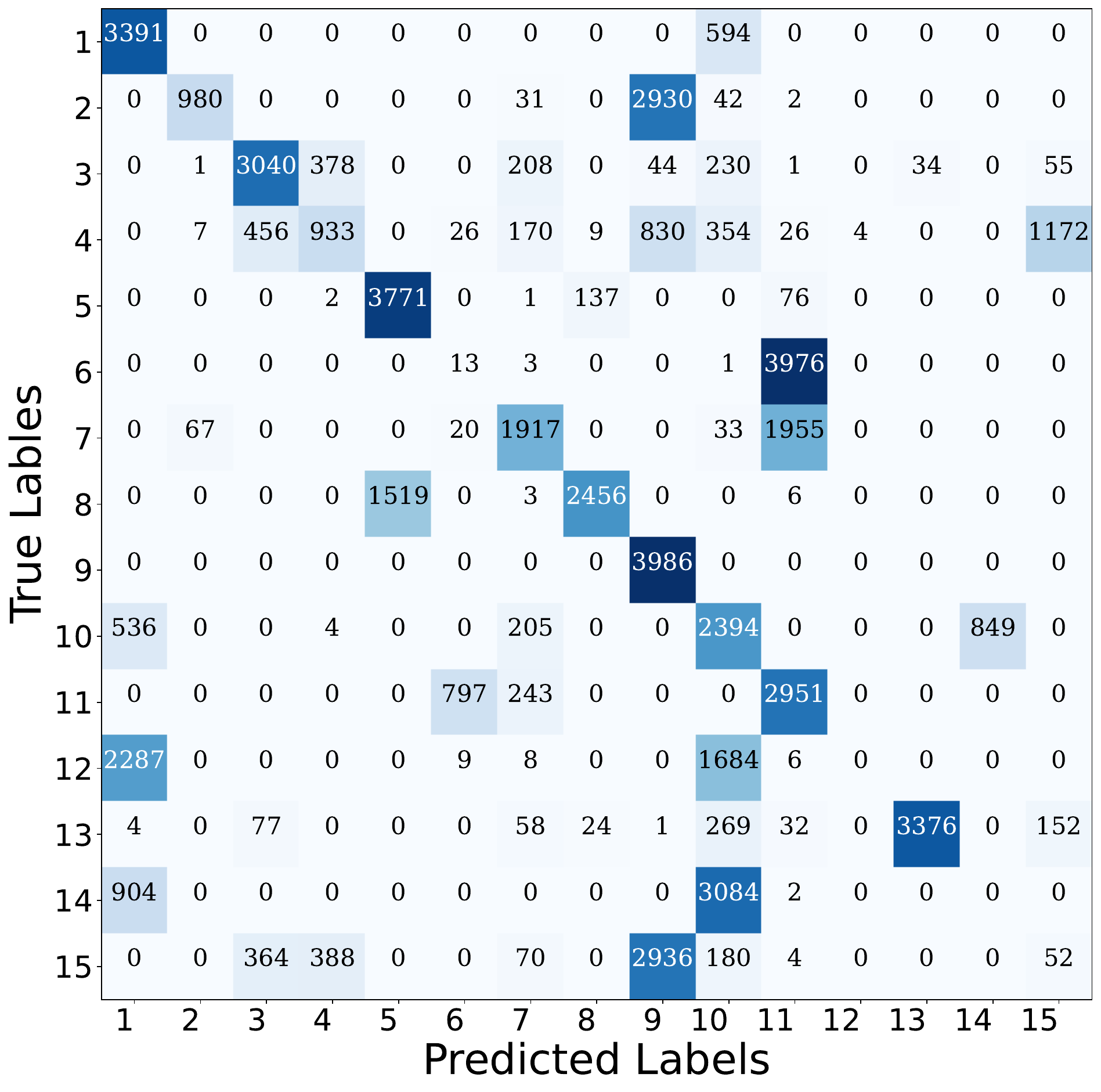}
   \label{IQ-loc3}}
\caption{IQ-CNN's performance: (a) Testing accuracy; (b)-(d) confusion matrices for different Train-Location/Test-Location combinations. LocA, LocB, and LocC correspond to when the transmitters are 1m, 2m and 3m away from the receiver.}
\label{IQ-locations}
\end{figure*}

To assess the effectiveness of our proposed \proposed~feature representation in improving the performance of DL-Based RFFP methods across domains, we considered two performance metrics: same-domain accuracy and cross-domain accuracy. Same-domain accuracy measures the ability of the DL models to identify devices accurately when the testing data/packets (unseen in the training phase) are drawn from the same training domain. On the other hand, cross-domain accuracy evaluates the models' ability to generalize across different domains, such as different locations, channels or days. We evaluated the performance of a standard CNN framework when fed with our proposed \proposed~representation as an input (referred to as EPS-CNN) and compared it with the same CNN framework but when fed with a typical IQ representation as an input (referred to as IQ-CNN). We employed the 5-fold cross-validation method, where each device's data is divided into five non-overlapping, equally-sized partitions. In each fold of cross-validation, we used four partitions for training (3200 packets) and the remaining partition for testing (800 packets). We then averaged the results obtained from each fold to produce a final estimate of the model's performance. For the \proposed~input, we represented each packet by a 2x4096 tensor, which encapsulates the \proposed~representation of both the I and Q components. In contrast, for IQ input, we represented each packet using a 2x8192 tensor, comprising the first 8192 samples of both the I and Q components as this window size provides the best performance for the IQ representation input.

\subsection{Robustness to Location Changes with Fixed Placement}\label{subsub:fix-loc}

First, we begin by evaluating the proposed EPS-CNN framework using the WiFi dataset captured in the three different locations, as described in Sec.~\ref{location_Setup}, to assess its robustness to changes in device locations. The results are shown in Fig.~\ref{eps-locations} for EPS-CNN and Fig.~\ref{IQ-locations} for IQ-CNN, where again LocA, LocB, and LocC correspond to when the transmitters are placed 1m, 2m and 3m away from the USRP receiver.
The results shown in Fig.~\ref{eps-locations} demonstrate that our EPS-CNN framework is highly effective in device fingerprinting, achieving exceptional same-domain testing accuracies at all locations. Specifically, the average testing accuracies at Locations A, B, and C are 100\%, 99.6\%, and 96.7\%, respectively, as shown in Fig.~\ref{eps-trained-locations}. Even more impressive is the performance of our EPS-CNN framework in cross-domain evaluation, where the model is trained on one location and tested on datasets captured in different locations. The results show that EPS-CNN maintains high performances, with average testing accuracies of 91.3\% and 95.5\% when trained on Loc A and tested on Loc B and Loc C, respectively. 
Similarly, EPS-CNN achieves a testing accuracy of 99.7\% and 95.04\% when trained on Loc B and tested on Loc A and Loc C, and an accuracy of 95.3\% and 93.9\% when trained on Loc C and tested on Loc A and Loc B. To the best of our knowledge, this is the highest performance achieved by a DL-based device fingerprinting method when the learning models are tested and trained on different domains. The cross-location confusion matrices are shown in Figs.~\ref{eps-loc1}, \ref{eps-loc2}, \ref{eps-loc3}, further highlighting the effectiveness of EPS-CNN. 

\begin{figure*}
\centering
\subfloat[EPS-CNN]{
   \includegraphics[width=.48\columnwidth]{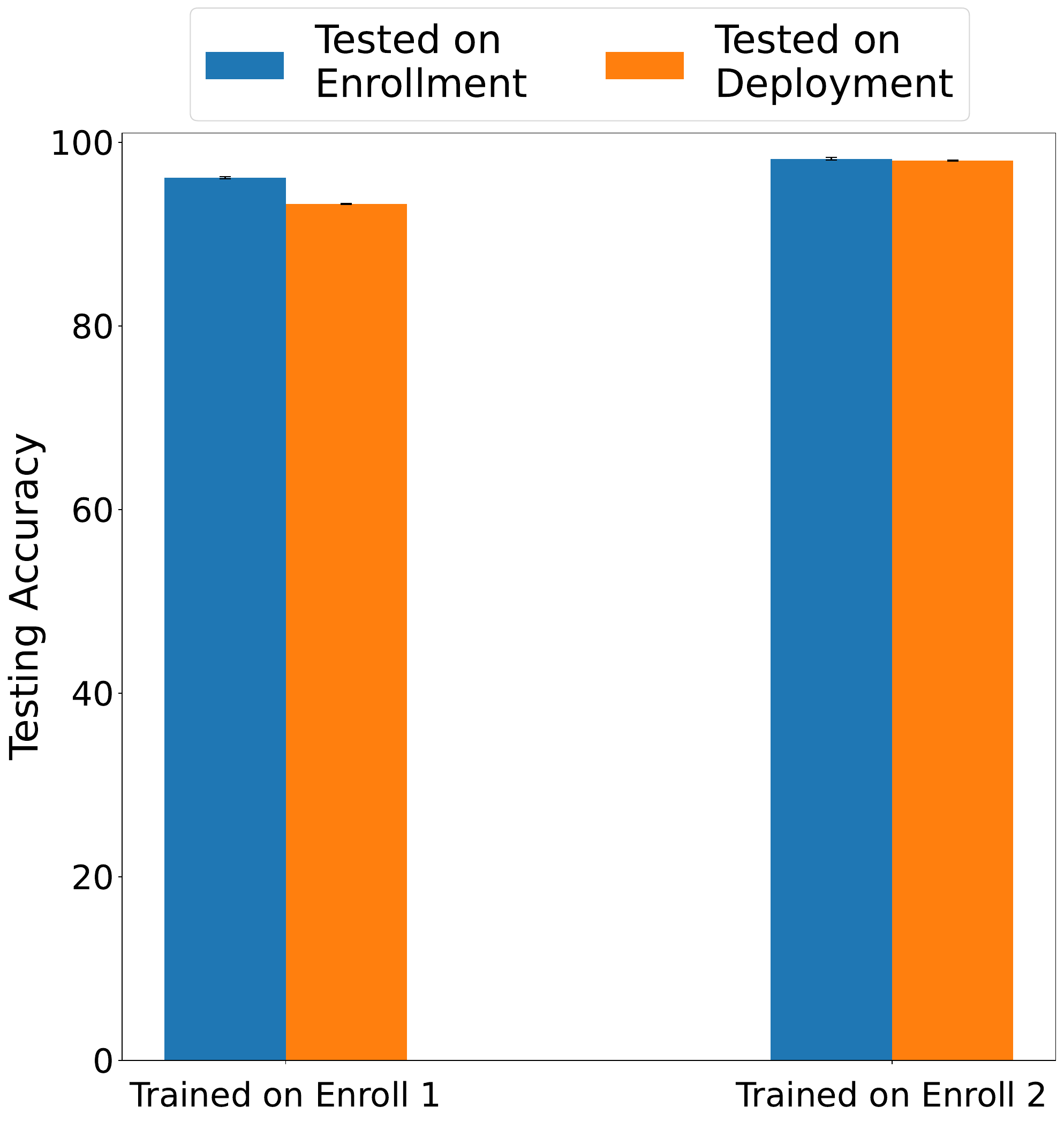}
   \label{eps-rand}}
  \subfloat[IQ-CNN]{
   \includegraphics[width=.48\columnwidth]{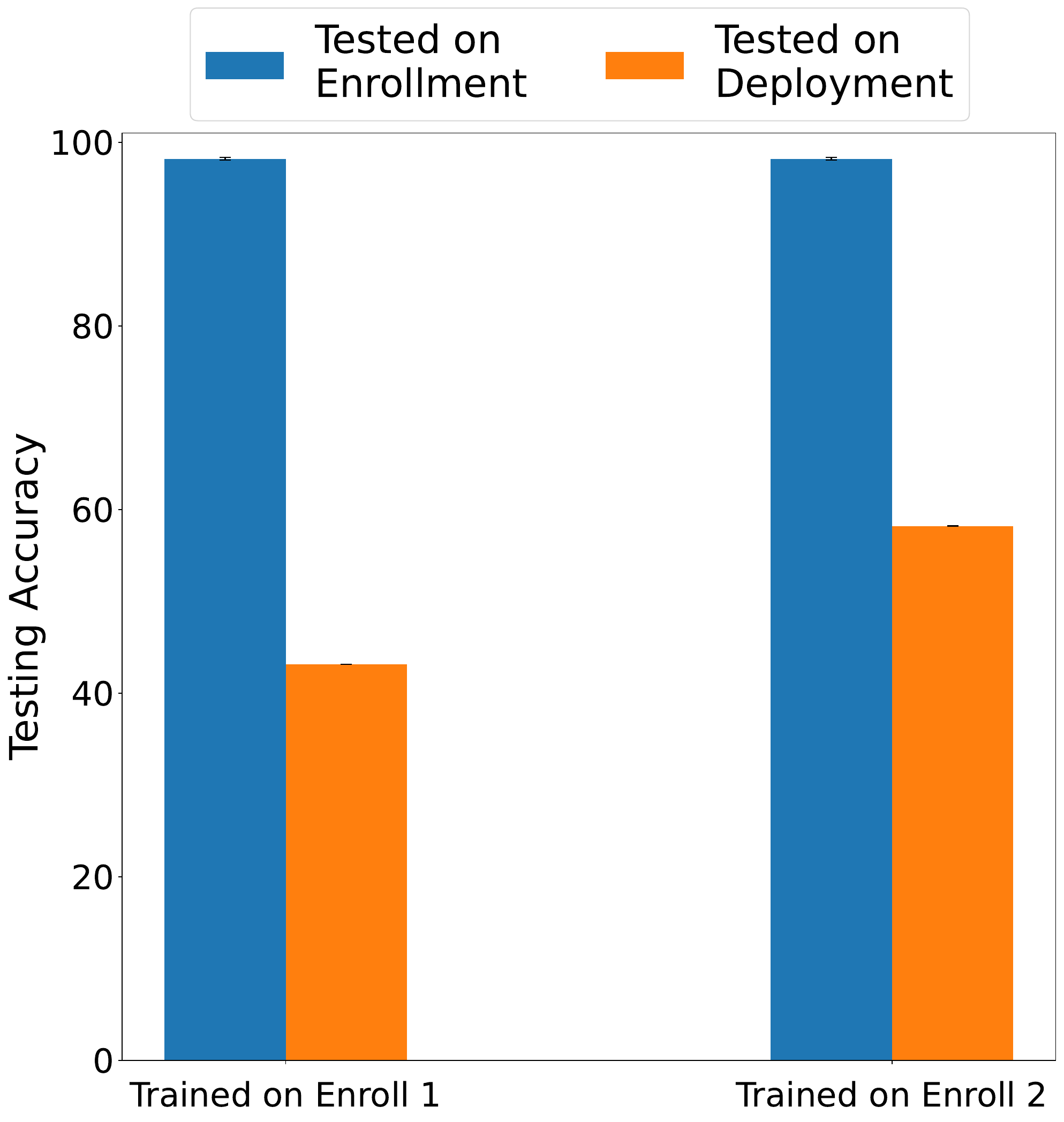}
   \label{IQ-random}}
\subfloat[EPS-CNN]{
   \includegraphics[width=.48\columnwidth]{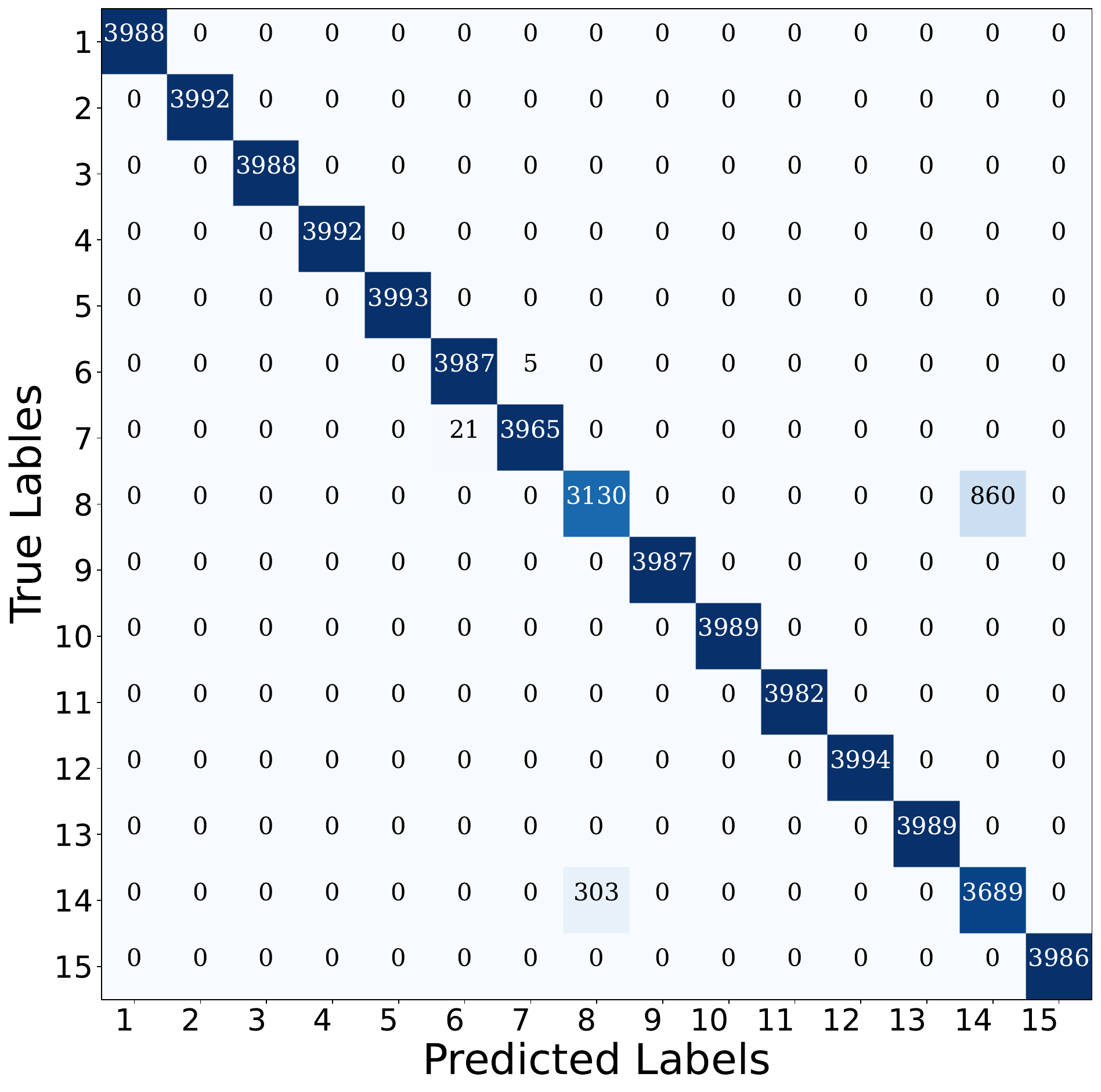}
   \label{eps-rand1}}
    \subfloat[IQ-CNN]{
   \includegraphics[width=.48\columnwidth]{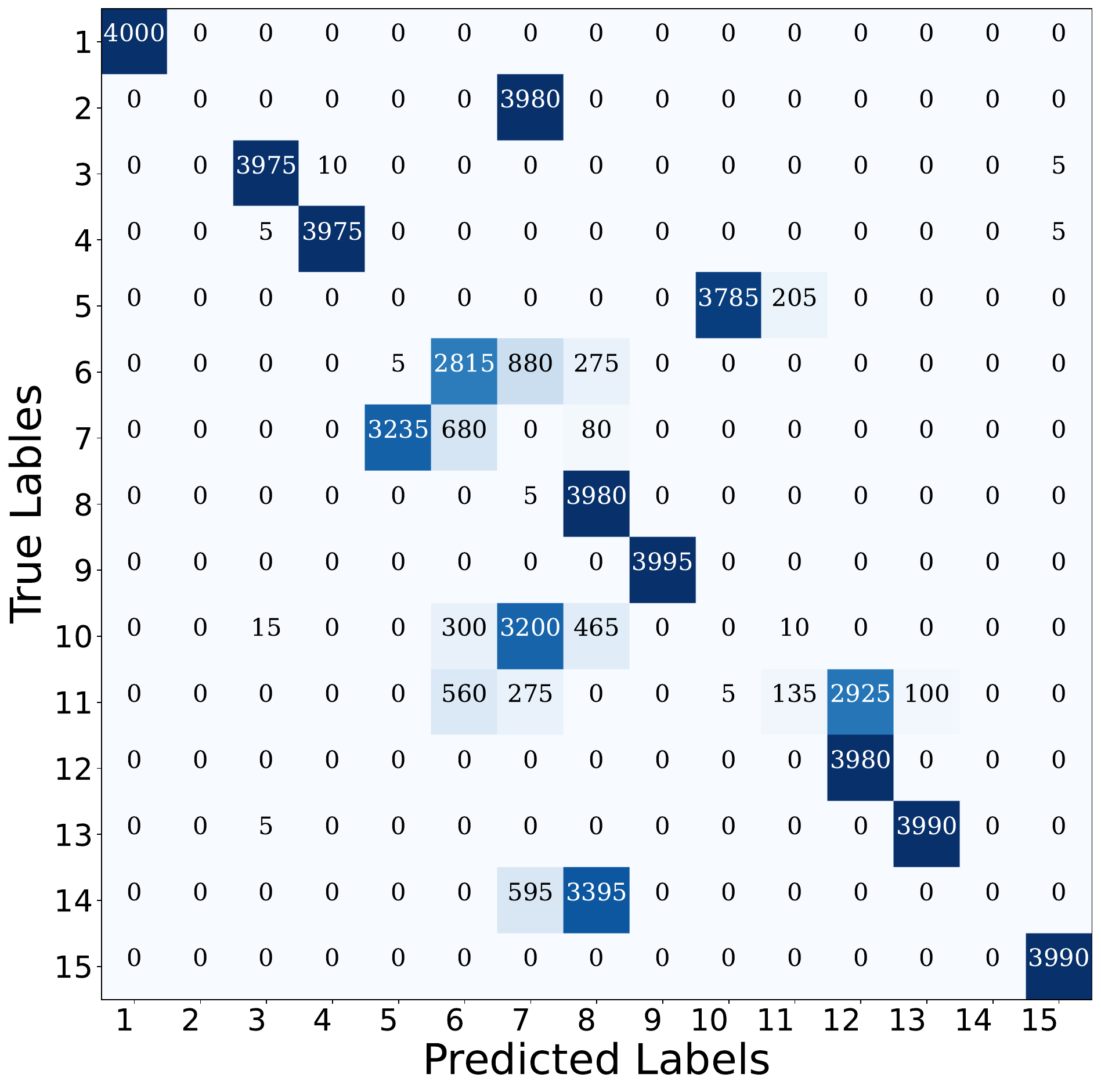}\label{subfig:ocxo4}
   \label{iq-rand1}}
\caption{EPS-CNN/IQ-CNN's performance under random deployments: (a)-(b): testing accuracy; (c)-(d): confusion matrices obtained under random-location setup 2; that is, when training is done under enrolment setup 2 and testing is done under the corresponding deployment setup.}
\label{rand_deployment}
\end{figure*}

In contrast, the conventional IQ-CNN framework, which uses raw IQ representation instead of the \proposed~representation and whose results are shown in Fig.~\ref{IQ-locations}, struggles to maintain its performance in cross-domain evaluation. Although IQ-CNN performs well in the same-domain evaluation, it performs poorly when tested on a dataset captured in a different location.  Specifically, the average testing accuracy when IQ-CNN is trained on location A and tested on locations B and C is 57.01\% and 24.96\%. This significant drop in performance is also seen in the other locations, as the testing accuracy on locations A and C when the model is trained on location B is 59\% and 24.7\%, whereas the average testing accuracy when the model is trained on location C and tested on locations A and B is 45.9\% and 61.4\%, respectively. Note the significant accuracy difference between that achieved by the proposed EPS-CNN and that achieved by the conventional IQ-CNN (i.e., a drop from 90+\% to as low as 25\%).
The confusion matrices in Figs.~\ref{IQ-loc1}, \ref{IQ-loc2}, \ref{IQ-loc3} show the struggle of the trained model to correctly classify the devices when the corresponding packets are captured in a different location. The results clearly demonstrate the superiority of the \proposed-based deep learning framework in device fingerprinting, particularly in cross-domain evaluations.

\begin{figure*}
\centering
\subfloat[Testing accuracy]{
   \includegraphics[width=.55\columnwidth]{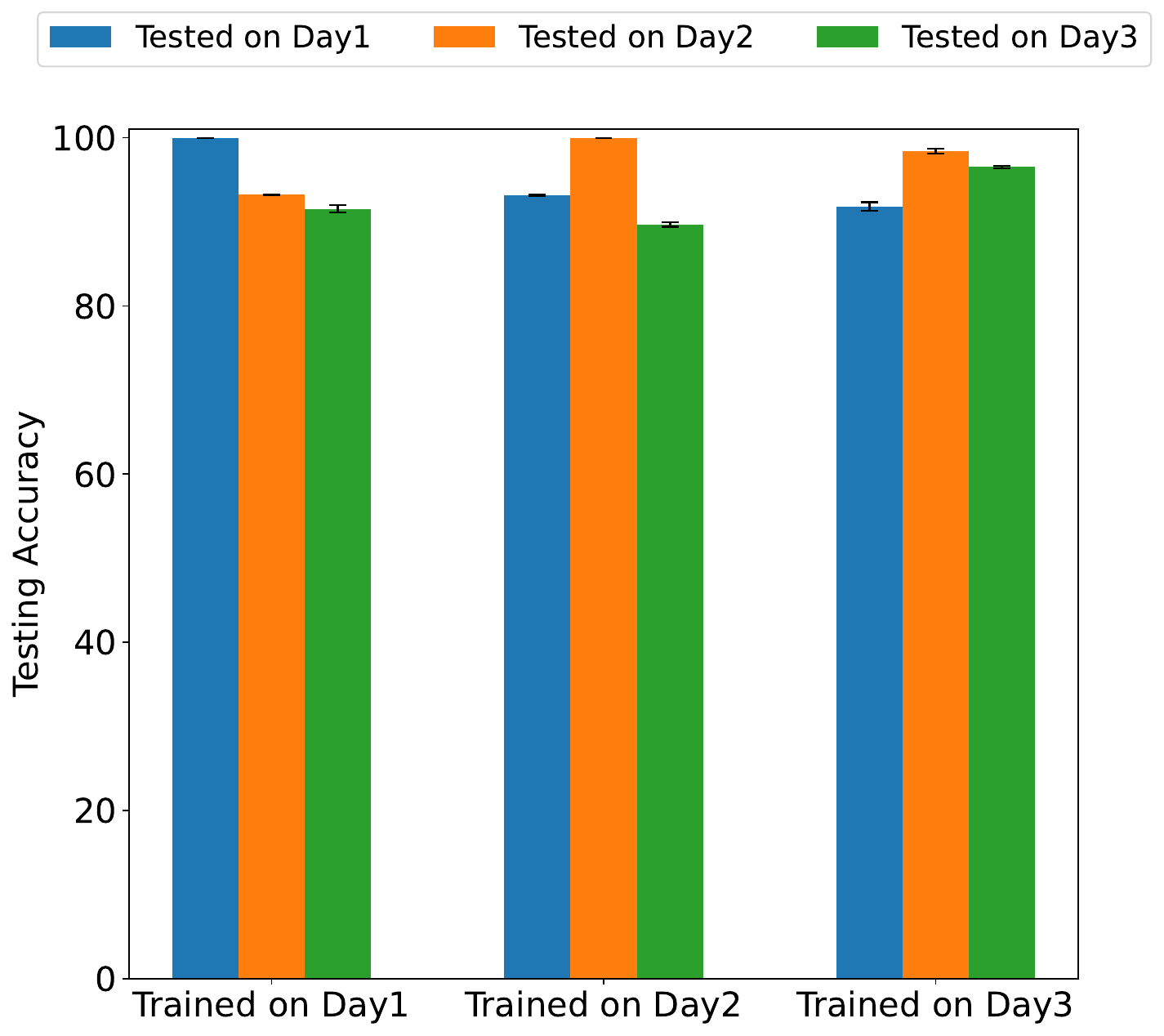}
   \label{eps-trained-days}}
\subfloat[Train-day1 / Test-day2]{
   \includegraphics[width=.47\columnwidth]{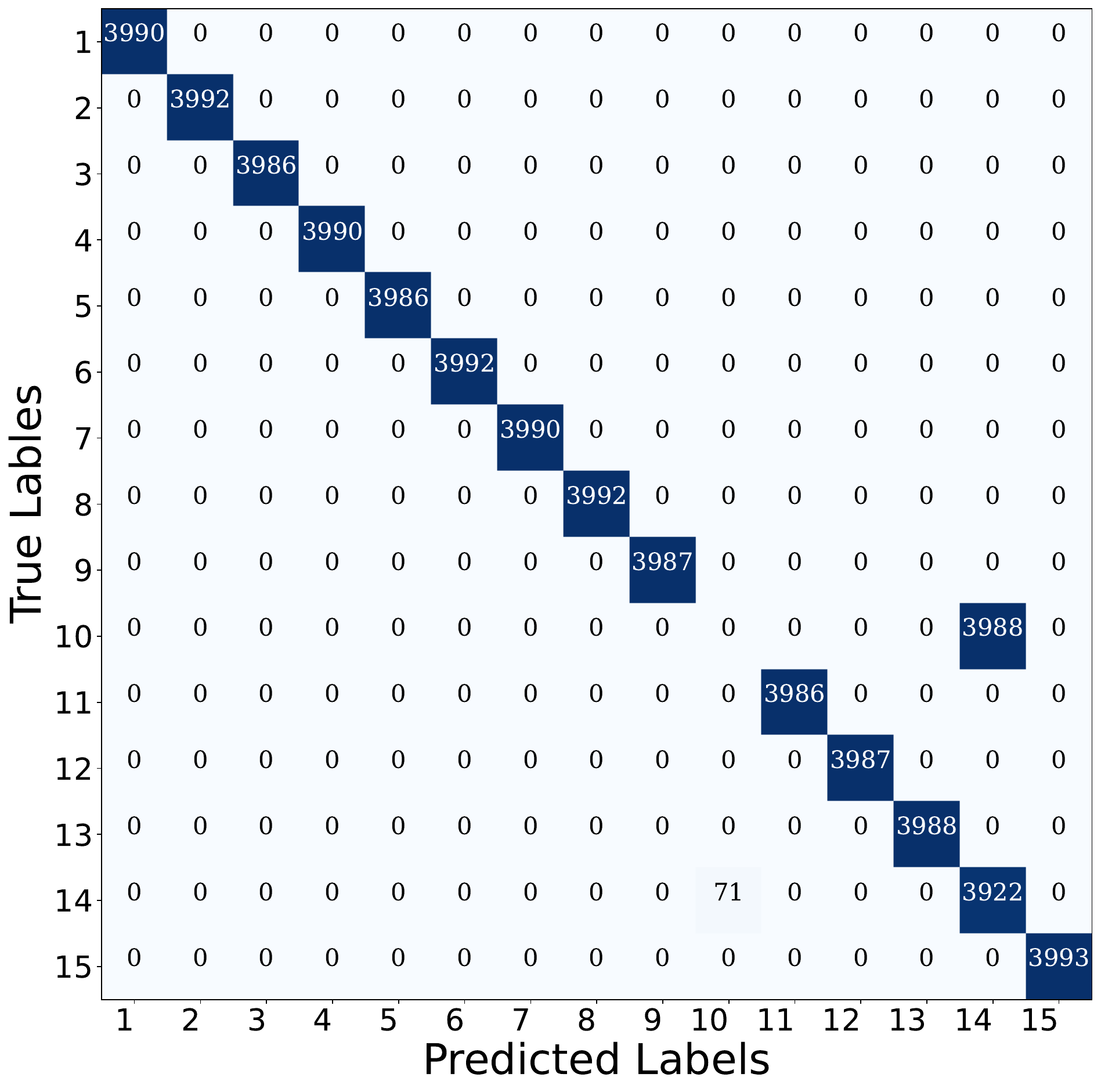}
   \label{eps-day1}}
\subfloat[Train-day2 / Test-day1]{
   \includegraphics[width=.47\columnwidth]{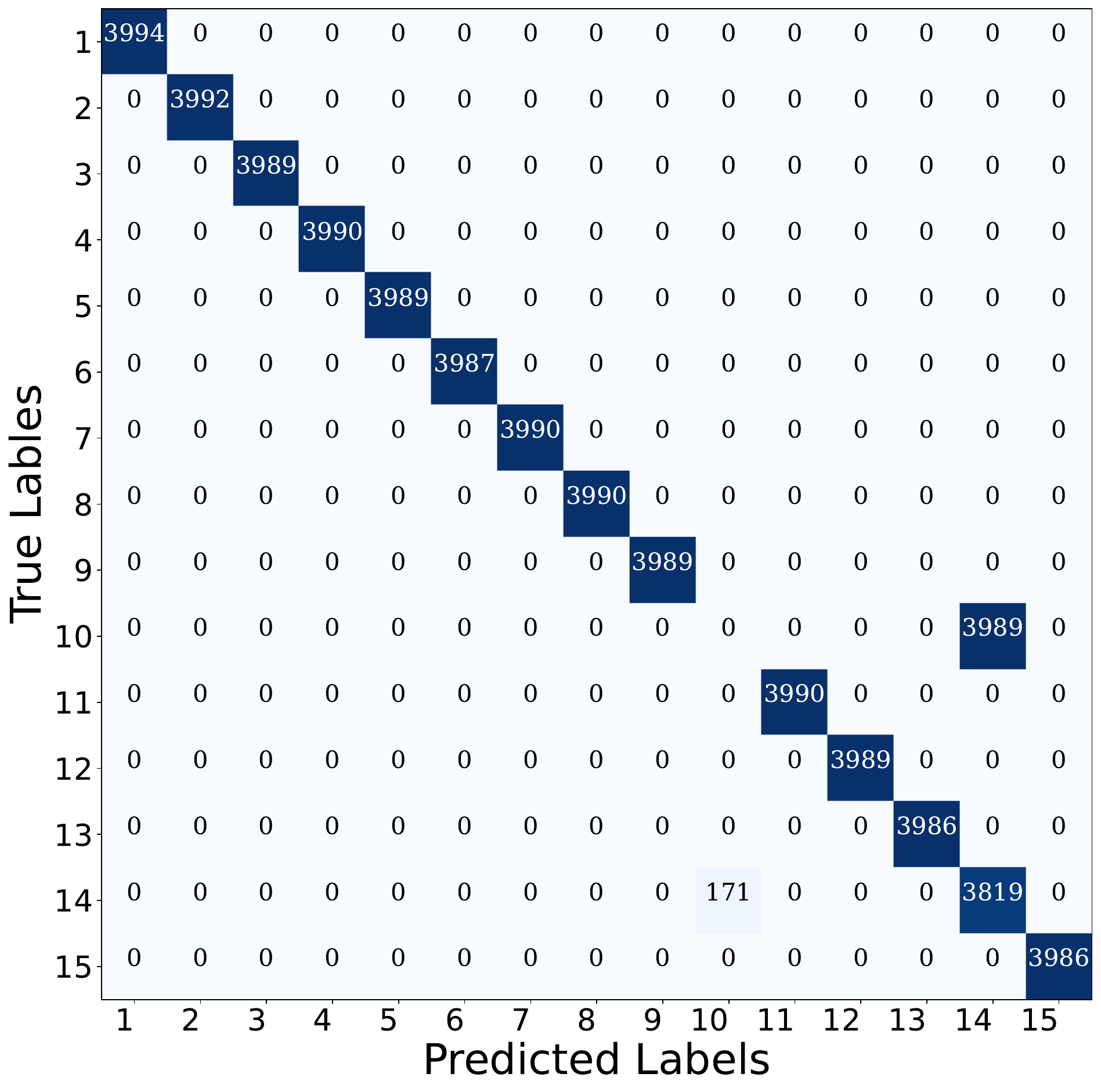}
   \label{eps-day2}}
   \subfloat[Train-day3 / Test-day2]{
   \includegraphics[width=.47\columnwidth]{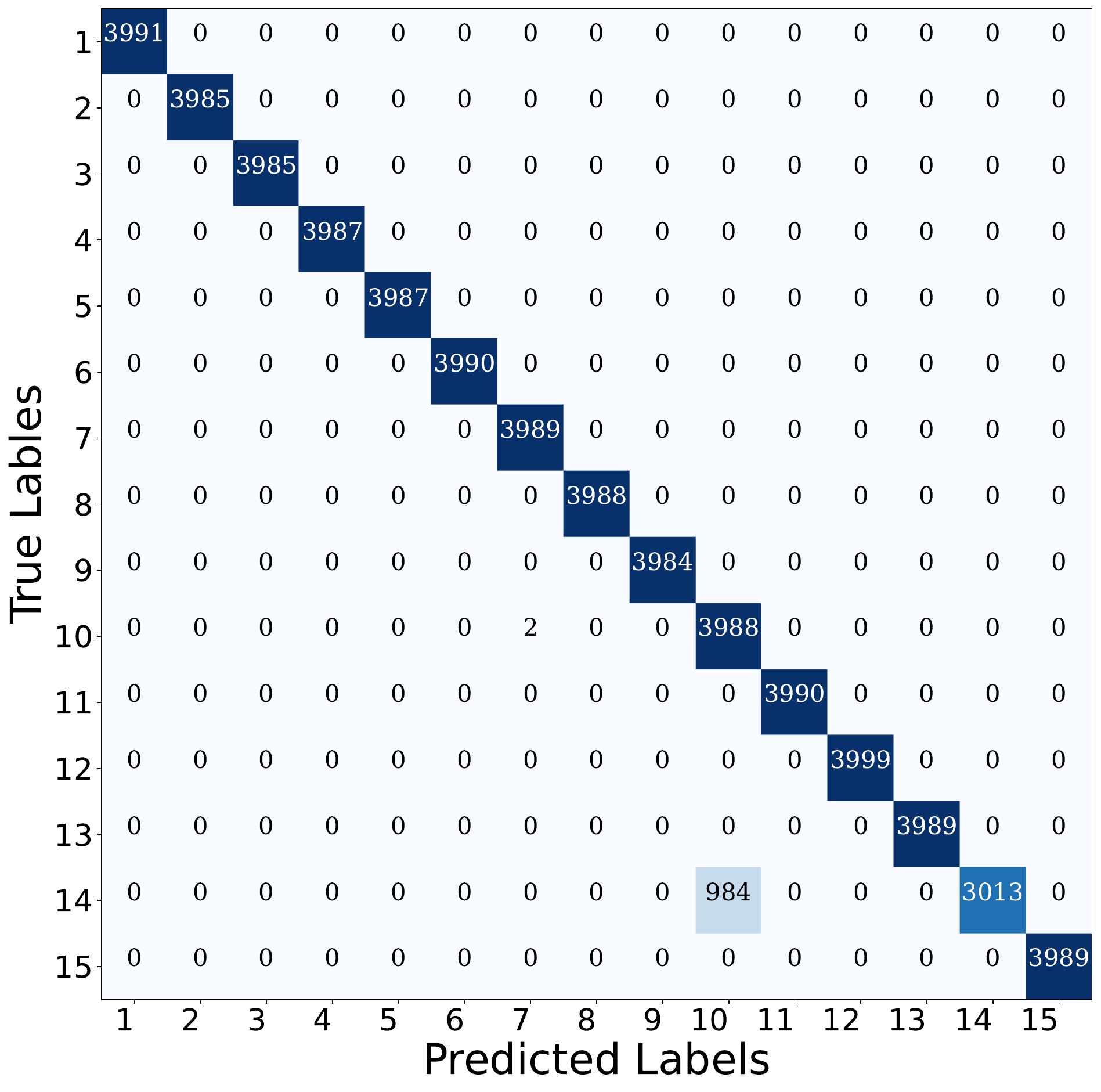}
   \label{eps-day3}}
\caption{EPS-CNN's performance across three different days: (a) Testing accuracy; (b)-(d) confusion matrices for different Train-day/Test-day combinations.}
\label{EPS-days}
\end{figure*}

\subsection{Robustness to Location Changes with Random Placement}\label{subsub:ran-loc}

We also considered evaluating the effectiveness of the proposed EPS-CNN framework under two random-location setups, as described in Sec.~\ref{random_Setup}. For each setup, during training (referred to as enrolment), all devices transmit from a fixed location, 1m away from the receiver; and during testing (referred to as deployment), the devices transmit from random locations all within 3m from the receiver; refer to Fig.~\ref{rand_loc} for visualization of this random deployment scenario.
The EPS-CNN framework exhibits strong performance in both same-domain and cross-domain evaluations under random-location setups. Fig.~\ref{eps-rand} shows that EPS-CNN achieves high average same-domain testing accuracies of 96.7\% and 98.3\% respectively under random-location setups 1 and 2. Furthermore, the figure demonstrates the robustness of the framework in cross-domain with testing accuracies of 93.1\% and 98.1\% under random-location setups 1 and 2, respectively. 
Notably, Fig.~\ref{eps-rand1}, showing the confusion matrix under random-location setup 2, showcases the framework's exceptional accuracy for most devices. 
In contrast, IQ-CNN experiences significant performance degradation on both random-location setups, with cross-domain testing accuracies of only 40.2\% and 58.2\% on setups 1 and 2, respectively. Through the confusion matrix, Fig.~\ref{iq-rand1} provides a clear depiction of the IQ-CNN framework's struggle in recognizing devices when randomly deployed around the receiver.

\subsection{Robustness to Time Changes}

\begin{figure*}
\centering
\subfloat[Testing accuracy]{
   \includegraphics[width=.55\columnwidth]{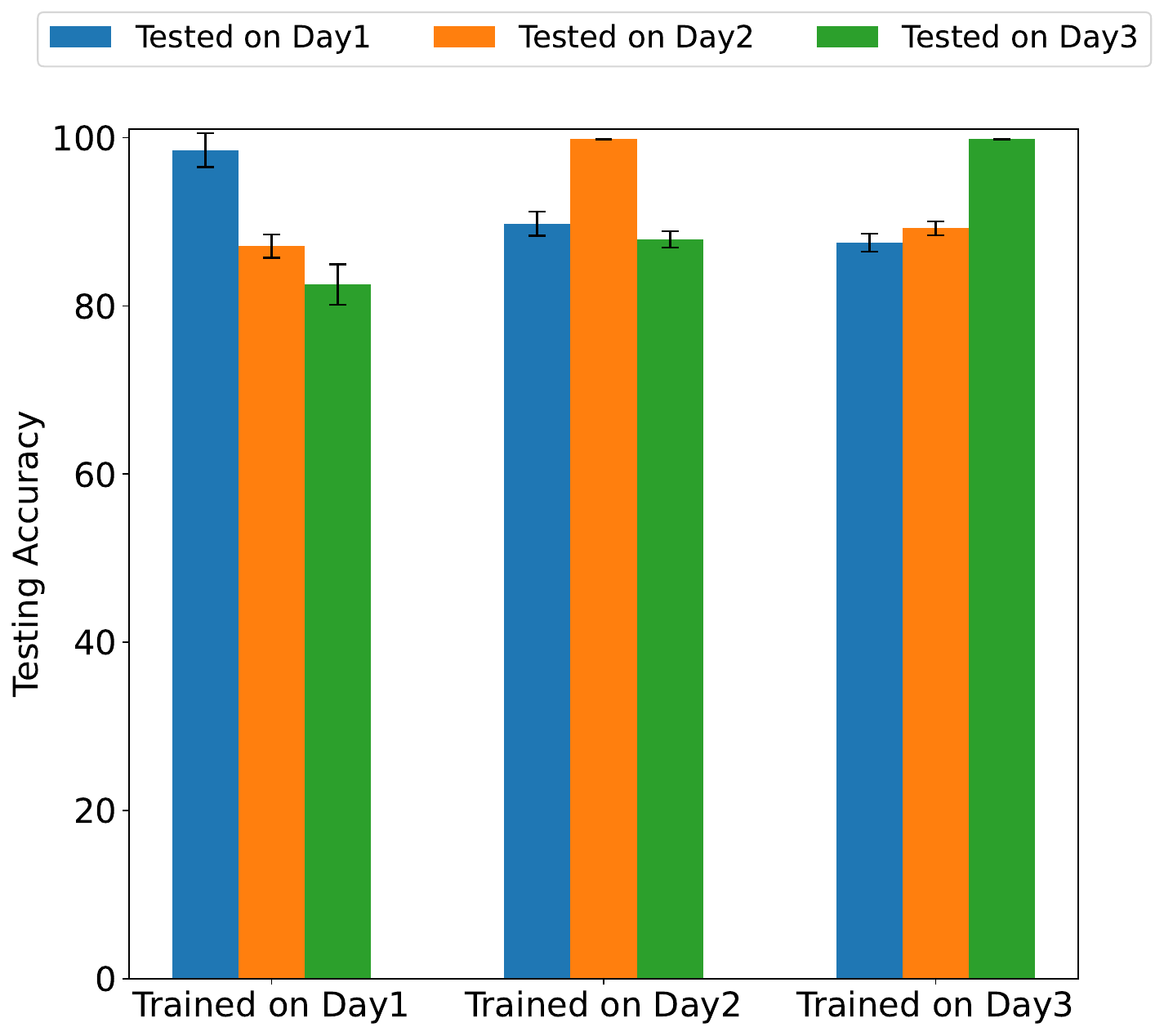}
   \label{IQ-trained-days}}
\subfloat[Train-day1 / Test-day3]{
   \includegraphics[width=.47\columnwidth]{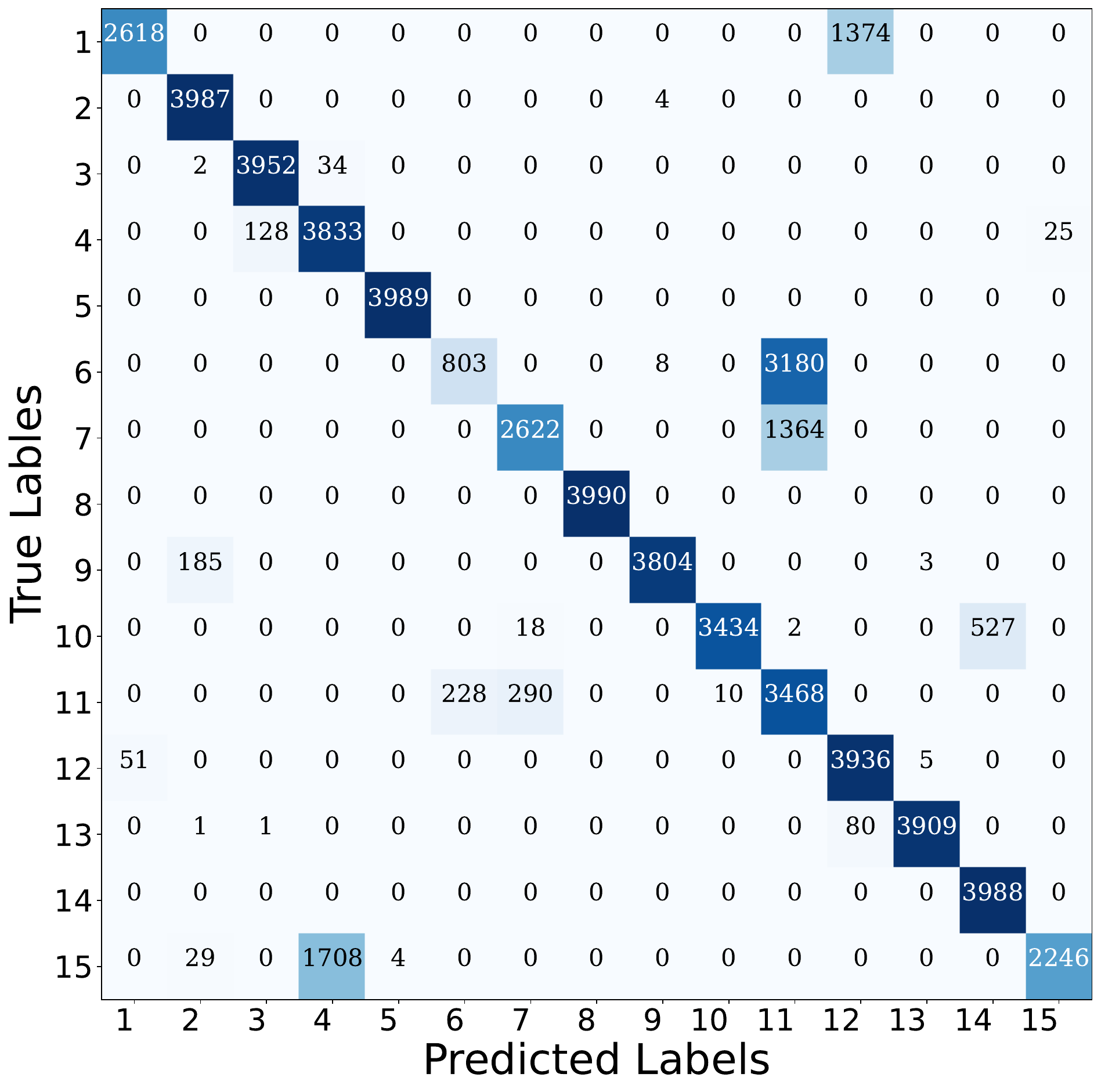}
   \label{IQ-day1}}
\subfloat[Train-day2 / Test-day1]{
   \includegraphics[width=.47\columnwidth]{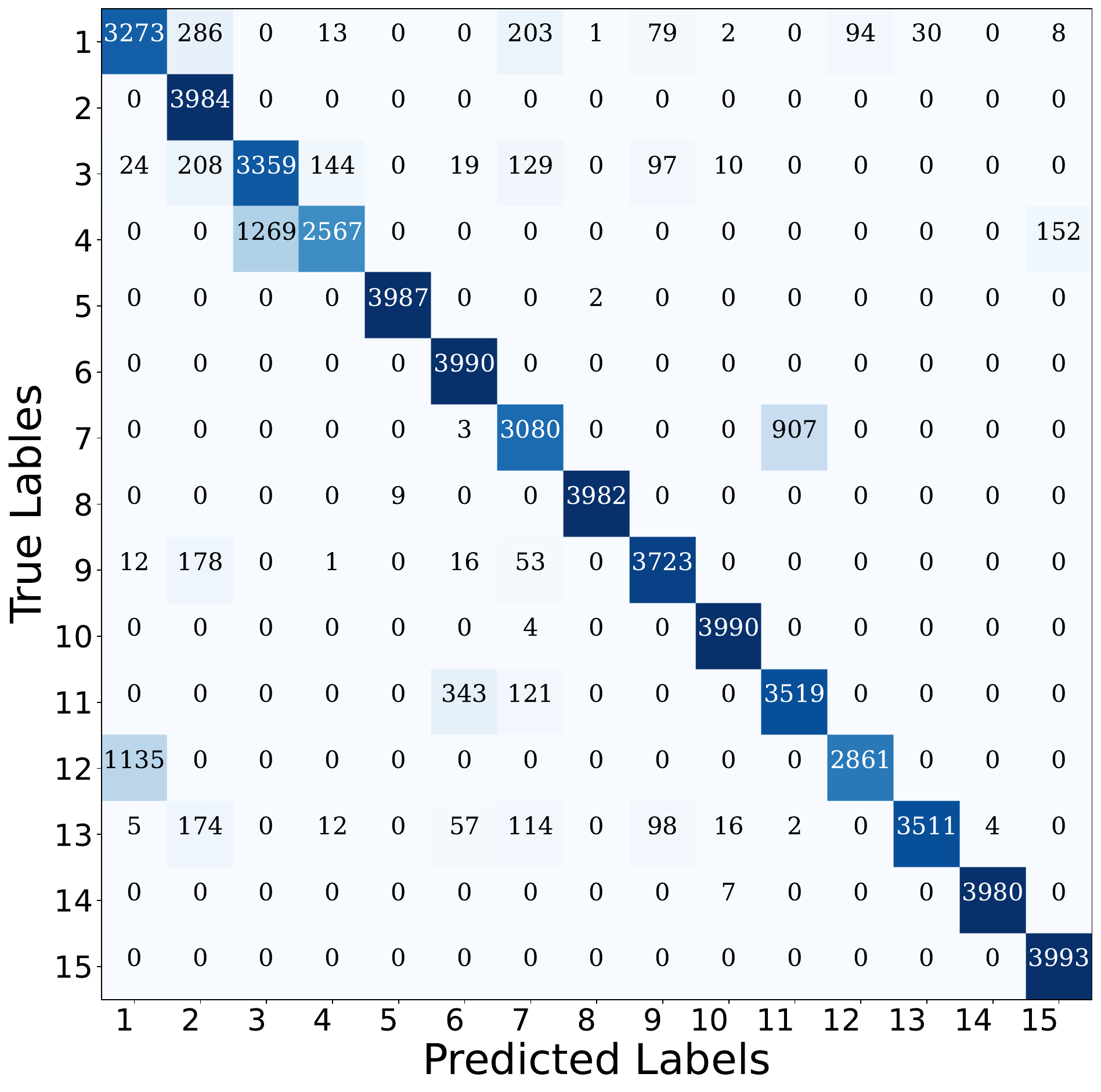}
   \label{IQ-day2}}
   \subfloat[Train-day3 / Test-day1]{
   \includegraphics[width=.47\columnwidth]{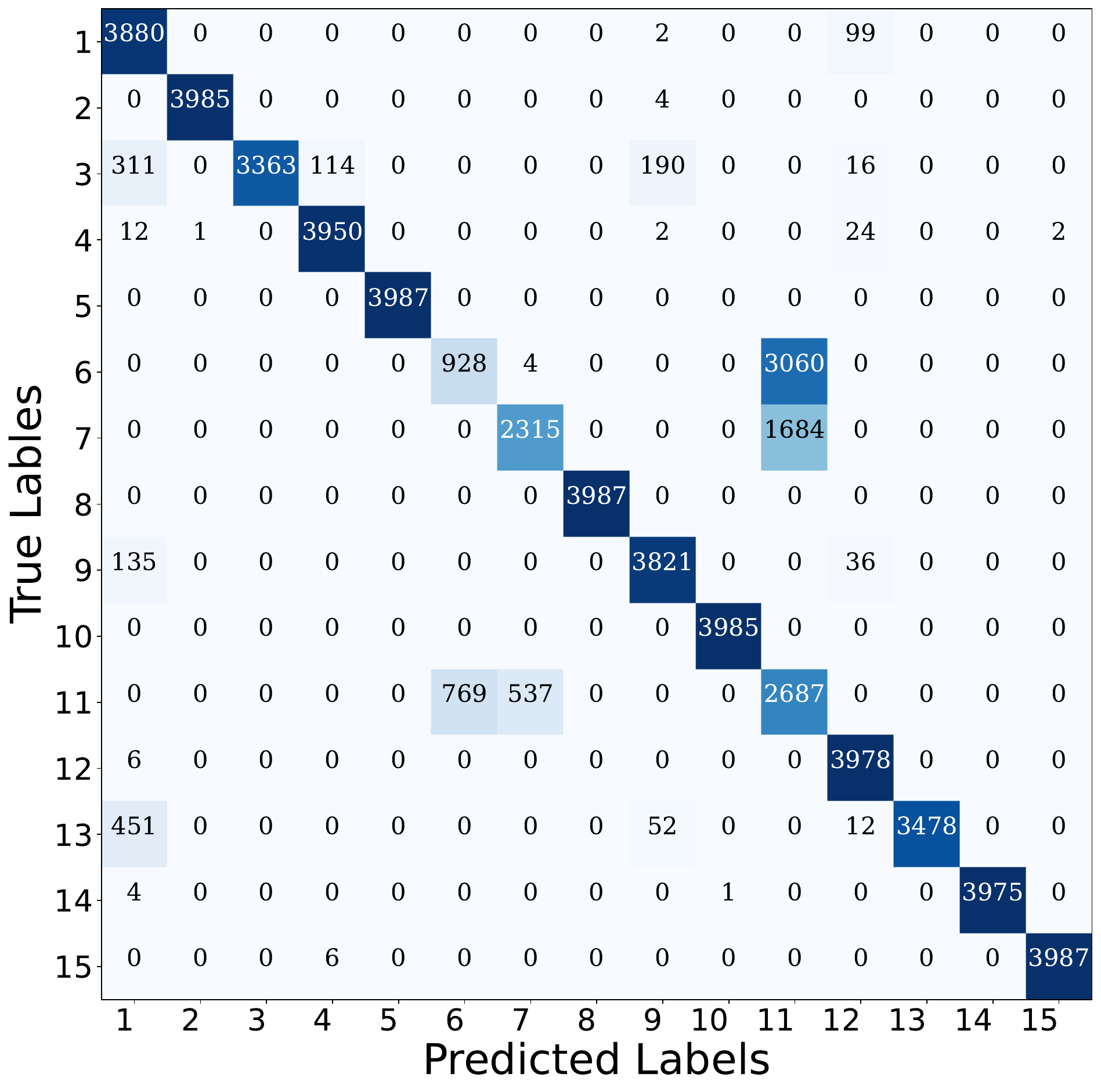}
   \label{IQ-day3}}
   \caption{IQ-CNN's performance across three different days: (a) Testing accuracy; (b)-(d) confusion matrices for different Train-day/Test-day combinations.}
\label{IQ-days}
\end{figure*}


The effectiveness of the proposed EPS-CNN framework was also evaluated on a cross-days scenario using the indoor wireless WiFi dataset as described in Sec.~\ref{wireless_Setup}. Fig.~\ref{eps-trained-days} presents the average testing accuracy of the proposed EPS-CNN framework when trained on one day and tested on one of the other three days. The same-domain testing accuracies (both training data and testing data are collected the same day) were found to be 100\%, 100\%, and 96.5\% for day 1, day 2, and day 3, respectively. 
These research findings demonstrate the distinguishability of the \proposed~feature representation, as the learning model was able to extract unique features from each device, achieving high performance on the same-domain performance metric.

More interestingly, for the cross-day evaluation, Fig.~\ref{eps-trained-days} shows that the proposed EPS-CNN framework maintains remarkable performance accuracy when tested on a different day. Specifically, when the learning model is trained on day 1 data, the average cross-domain testing accuracy is 93.2\% when the model is tested on day 2 data and 91.5\% when tested on day 3 data.
This achievable performance is consistent across other days, with an accuracy of 93.2\% or 89.7\% when training on day 2 data but testing on day 1 or day 3 data, respectively. Similarly, when the model is trained on day 3 and tested on day 1 data or day 2 data, the testing accuracies are 91.8\% or 98.4\%, respectively. The aggregate confusion matrices of the cross-day testing over the three days, shown in Figs.~\ref{eps-day1}, \ref{eps-day2}, \ref{eps-day3}, further indicate that most of the devices achieved perfect classification accuracies across the three tested days, with only one or two devices causing a small drop in performance.

In comparison, the performance of the IQ-CNN framework in cross-domain testing, shown in Fig.~\ref{IQ-days}, is inferior to that of the proposed EPS-CNN framework. Our results from Fig.~\ref{IQ-days} indicate that when the deep learning model is trained on day 1 data, the average cross-domain testing accuracy is 87.5\% when tested on day 2 data and 89.2\% when tested on day 3 data. When the model is trained on day 2 data, this average cross-domain testing accuracy is 89.8\% when tested on day 1 data or 87.9\% when tested on day 3 data. And the testing accuracy when the model is tested on day 1 or day 2 data but trained on day 3 data is 87.5\% or 89.2\%, respectively. The aggregate confusion matrices of the cross-domain testing over the three tested days are also shown in Figs.~\ref{IQ-day1}, \ref{IQ-day2}, \ref{IQ-day3}. Although both EPS-CNN and IQ-CNN frameworks achieved close-to-perfect performance in the same-domain testing accuracies, the proposed EPS-CNN outperforms the conventional IQ-CNN framework in the cross-domain performance metric on the three tested days. Our results show that the deep learning models when fed with our proposed \proposed~features are highly effective in addressing and mitigating the cross-day sensitivity of deep learning-based RF fingerprinting. 

The relatively good performance of conventional IQ-CNN fingerprinting in the cross-day metric suggests that the indoor wireless channel in this scenario did not change significantly over the three tested days of the experiment. This leads us to postulate that time itself is not a factor or domain that significantly affects the learning model's performance. Instead, time is simply a space in which various events can occur, leading to changes in the environment that can affect performance. Hence, we hypothesize that in a stable environment, cross-time evaluations may not be sufficient, and further tests are necessary to assess the model's performance under varying channel conditions, due to changing locations and distances. Our proposed EPS-CNN framework indeed maintains high performances even under varying locations as shown in our results presented earlier in Sections~\ref{subsub:fix-loc} and \ref{subsub:ran-loc}.

\section{On the Impact of Transceiver Hardware Warm-Up and Stabilization Time}
\label{sec:stability}
\begin{figure}
    \begin{minipage}[t]{\linewidth} 
        \centering 
        \subfloat[The in-phase (I) component of the WiFi signal]{
        \includegraphics[width=\linewidth]{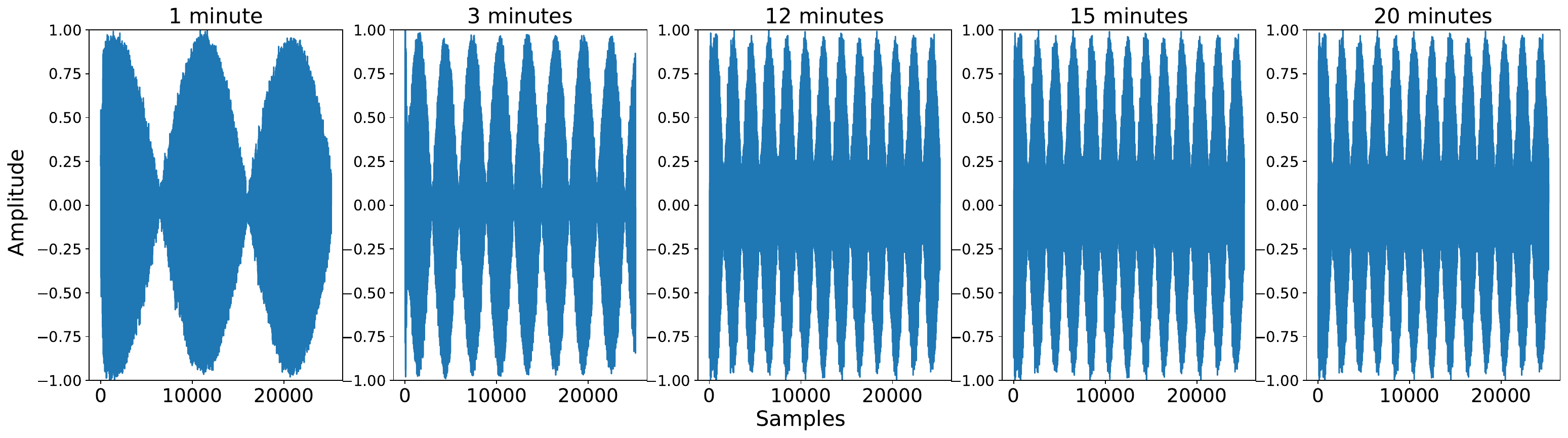}
        \label{dev7_I}}\\ 
        \subfloat[The quadrature (Q) component of the WiFi signal]{
        \includegraphics[width=\linewidth]{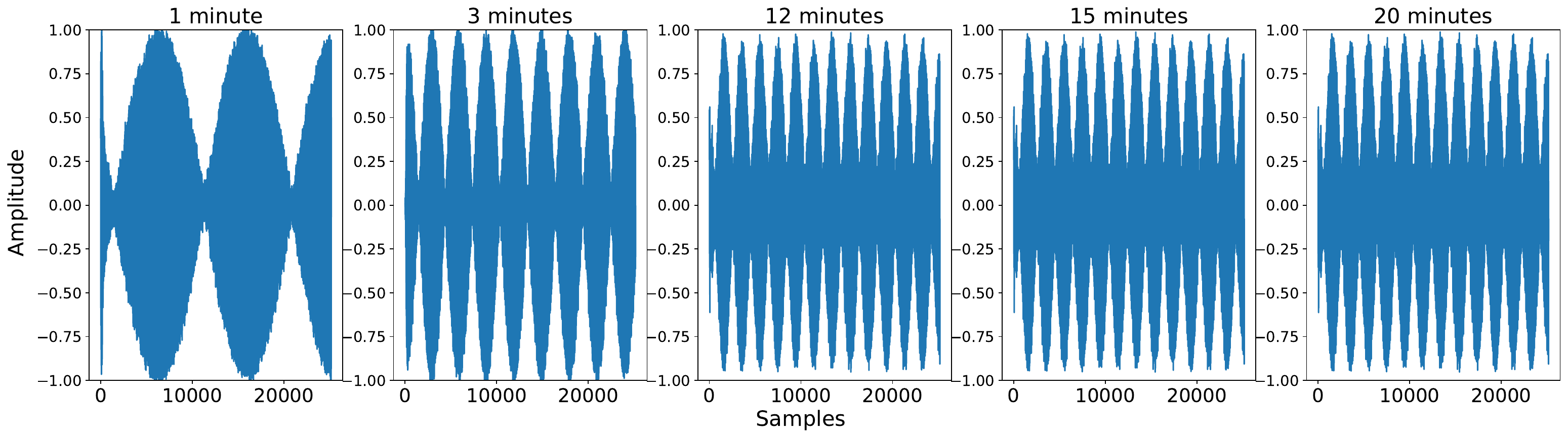}
        \label{dev7_Q}}\\
        \subfloat[The \proposed~feature representation of the WiFi signal]{
        \includegraphics[width=\linewidth]{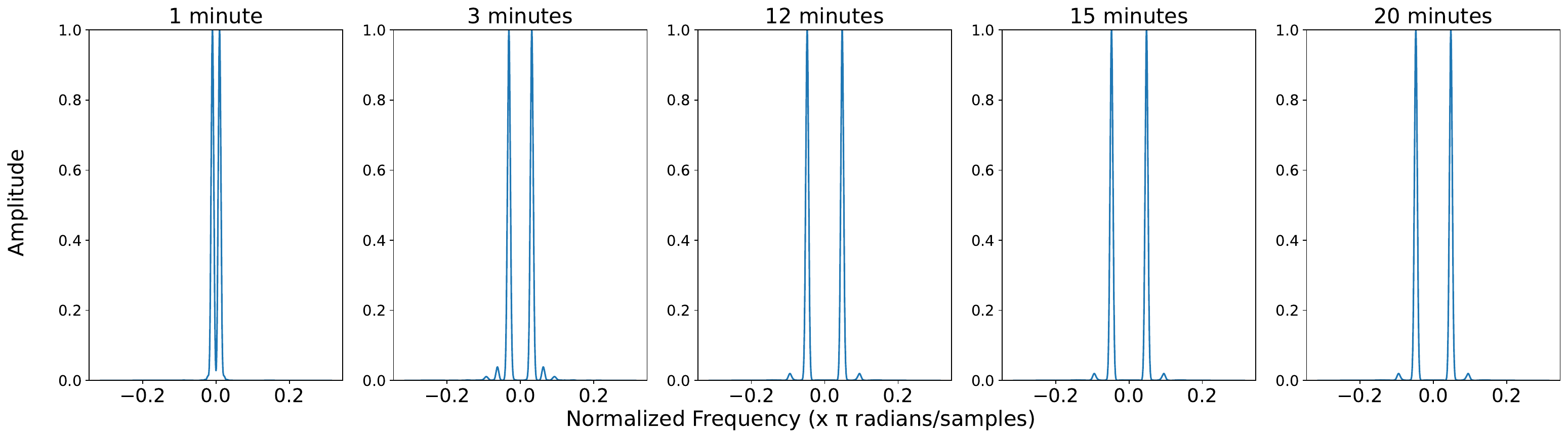}
        \label{dev7_env}}
        \caption{Representations of the RF signal captured from Device 7 and observed at different times during the warm-up period of the device.} 
        \label{dev7} 
        \end{minipage}
    \vfill \vspace{0.5in}
         \begin{minipage}[t]{\linewidth}
         \centering 
         \subfloat[The in-phase (I) component of the WiFi signal]{
        \includegraphics[width=\linewidth]{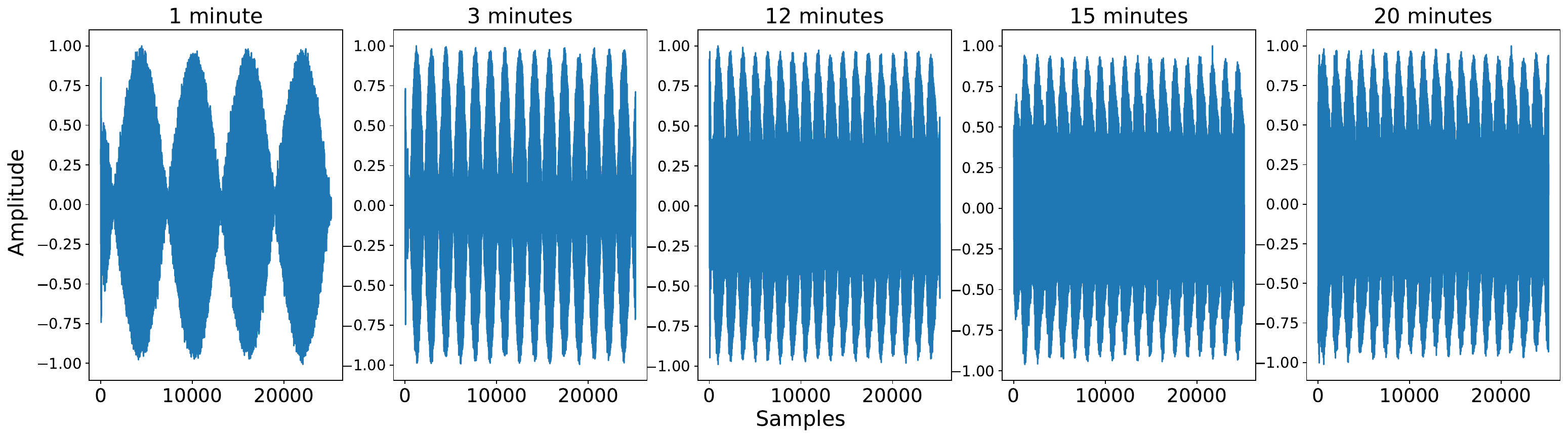}
        \label{dev10_I}}\\ 
        \subfloat[The quadrature (Q) component of the WiFi signal]{
        \includegraphics[width=\linewidth]{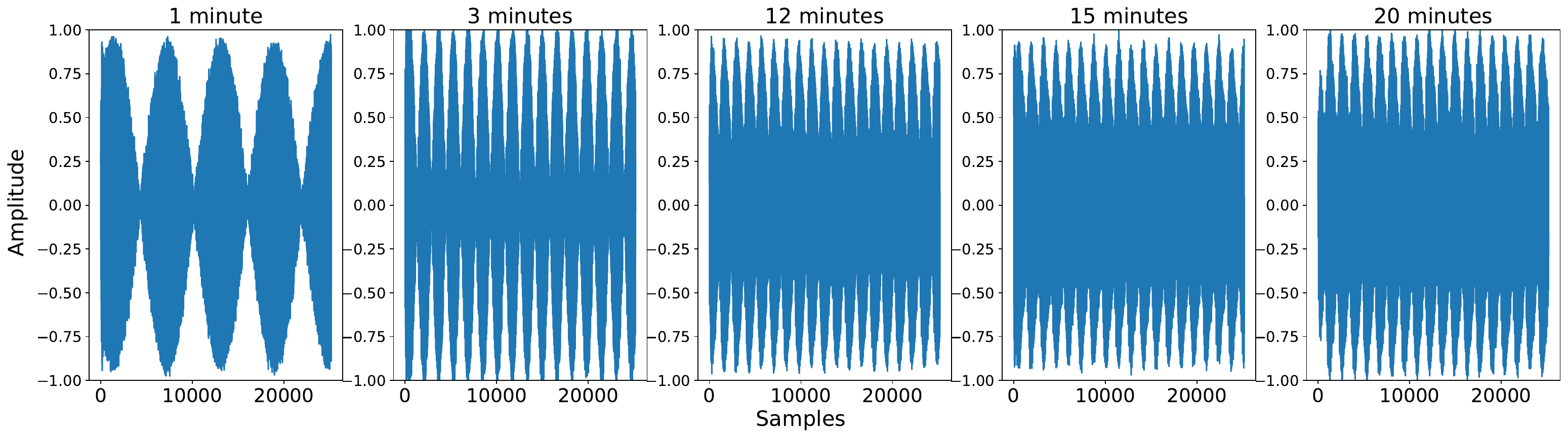}
        \label{dev10_Q}}\\
        \subfloat[The \proposed~feature representation of the WiFi signal]{
        \includegraphics[width=\linewidth]{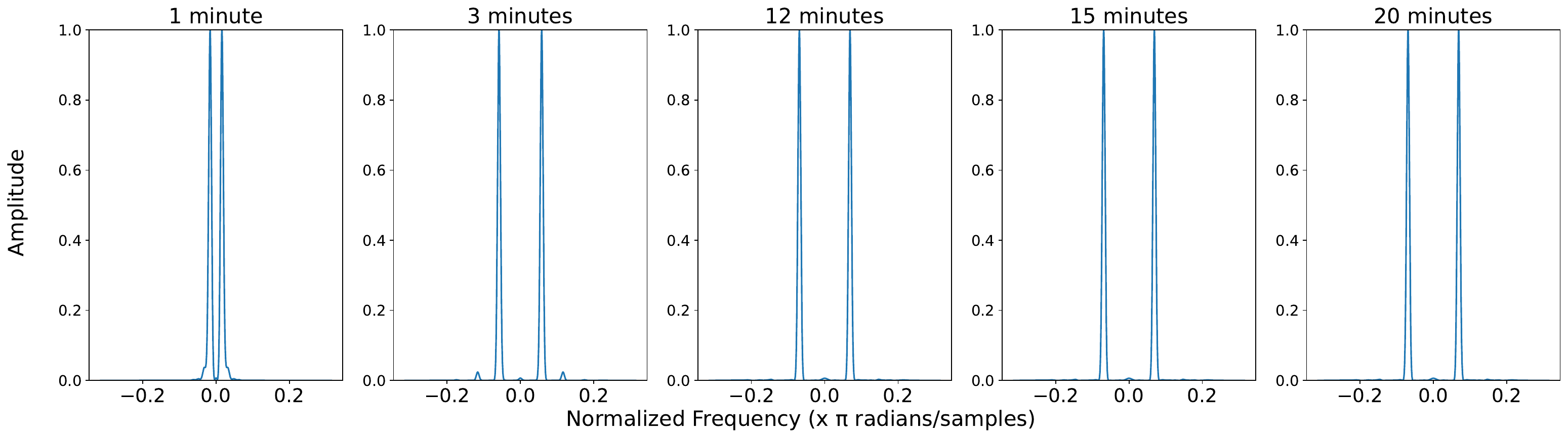}
        \label{dev10_env}}
        \caption{Representations of the RF signal captured from Device 10 and observed at different times during the warm-up period of the device.} 
        \label{dev10} 
    \end{minipage}
\end{figure}

We now turn our focus on investigating the effect of the warm-up time needed for the transceiver hardware to stabilize on the device fingerprinting performances. Despite the rich amount of literature available on this RF fingerprinting topic, the impact of hardware stabilization and warm-up time has not been carefully considered; see~\cite{elmaghbub2023impact} for our recent study on the topic. And for completeness, it is our goal here to shed some light on what could go wrong had such stabilization aspects not been carefully accounted for. More specifically, our objective in this section is to investigate and study the impact of the transceiver hardware warm-up on (i) the observed Envelope behavior of the time-domain IQ signals, (ii) the \proposed~features, and (iii) the overall \proposed-based device fingerprinting performance.

\subsection{Behavior of Received RF Signals During the Warm-Up and Stabilization Period of the Transmitting Device}\label{subsec:stab-beh}

We begin by studying the behavior of the I, Q and \proposed~representations during the hardware warm-up time. For this, we closely monitored the IQ signal behavior of two off-the-shelf (FiPy) devices (Devices 7 and 10) from our testbed during the initial 20 minutes following device activation. This involved capturing 802.11b WiFi packets transmitted by the devices using the USRP B210 at a sampling rate of 45MSps. The USRP receiver was clocked using an external 10 MHz OCXO (oven-controlled, high-performing crystal oscillator) reference signal to ensure measurement accuracy and stability.

We show in Fig.~\ref{dev7} the I, Q and \proposed~representations of the WiFi signal captured on Device 7 at different times during the device warm-up period; i.e., the figure on the far-left corresponds to the signal captured one minute from when the device was powered on, the figure on the far-right corresponds to the signal captured 20 minutes from when the device was powered on, and so on.
Four important observations can be drawn from this figure. First, the results confirm the presence of a CFO impairment, which is manifested in the observed sinusoidal shape of IQ signal's Envelope, as was illustrated and explained in Section~\ref{sec:motiv}. 
Second, observe that the I shape (Fig.~\ref{dev7_I}), Q shape (Fig.~\ref{dev7_Q}) and the \proposed~shape (Fig.~\ref{dev7_env}) all change over time as the device hardware warms up, with the frequency of humps in the Envelopes of the I and Q signals increasing over time until hardware stabilization. This increase indicates a varying CFO value during hardware warm-up time that is resulting from the instability of the local crystal oscillators; this finding is well aligned with the Envelope behavior observed and reported in Sec.~\ref{subsec:impact} and explained in Sec.~\ref{subsec:dem}.
Third, the I, Q and \proposed~shapes all seem to converge and stabilize after some time (i.e., around 12 minutes in the figure). Note that these shapes observed at minutes 12, 15 and 20 resemble one another, meaning that the shapes converge at around 12 minutes from device activation, which indicate that the local oscillator has reached a stable operating point by minute 12. 
Fourth, observe that the I and Q components vary on the opposite direction; i.e., shifted by 180 degrees, at any stage during the warm-up period; this also is well aligned with what was observed and reported in Sec.~\ref{sec:motiv}.

To assess the consistency of these trends across different devices, we also monitored these IQ data representations based on signals captured from several other devices also at different times during the device warm-up period. Our experimental results using other devices (we only show one more device, Device 10, here in Fig.~\ref{dev10}) confirm that the reported trends are also observed across all other devices, although each device exhibits slightly different initial and stable shapes.

\begin{figure}
\centering
   \includegraphics[width=0.8\columnwidth]{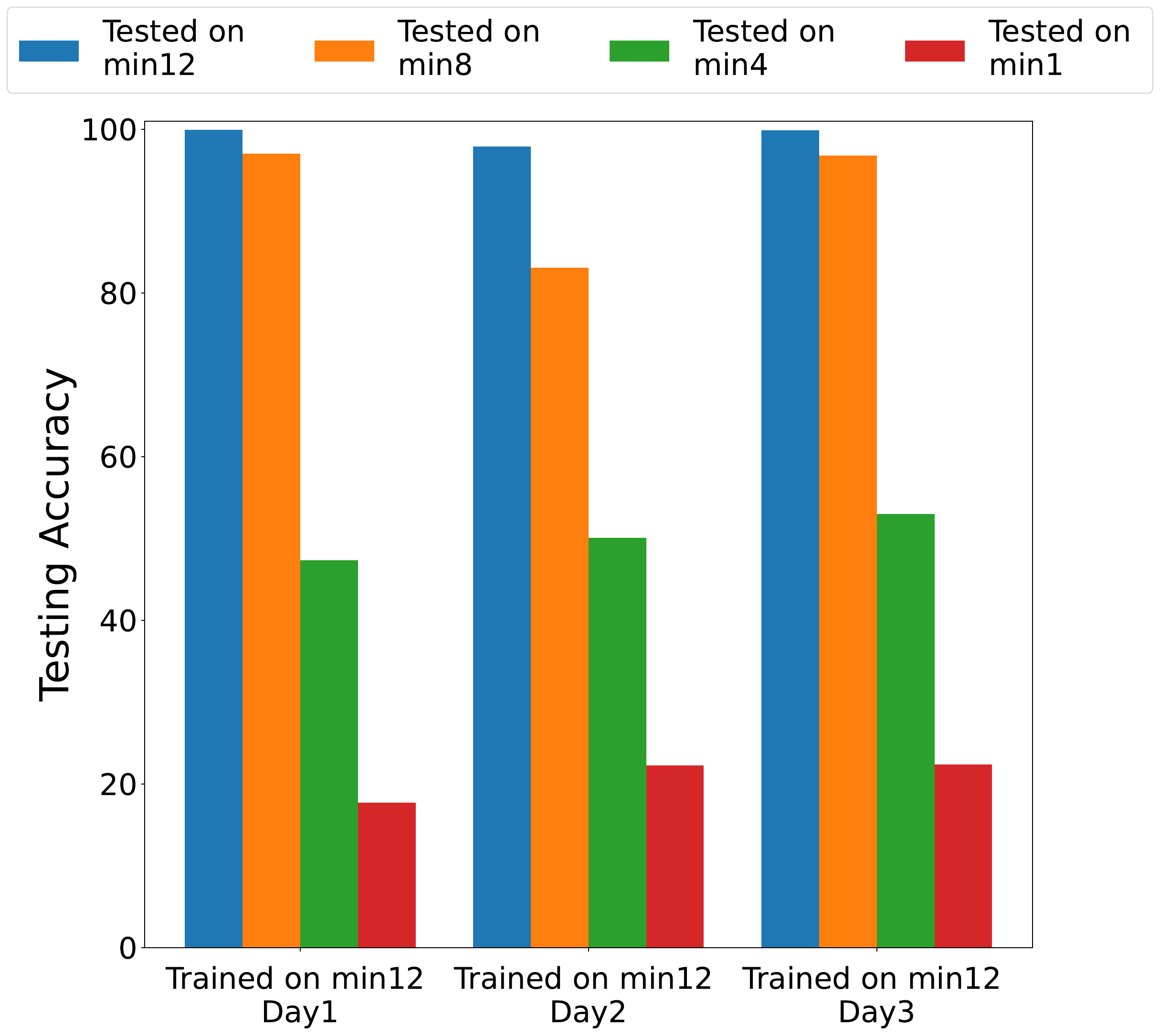}
   \label{cap1_days}
   
   \caption{Classification accuracy when training data is collected at minute 12 (after device stabilization) but testing data is collected at 1, 4, 8 and 12 minutes from device activation on the same day.}
\label{exp-all}
\end{figure}

\subsection{Sensitivity of RF Data-Driven Device Fingerprinting to Transceiver Hardware Warm-Up and Stabilization}

We now show the effect of hardware warm-up that is manifested in the observed IQ signal Envelope behavior on the device fingerprinting accuracy. For this, we run experiments whereby the proposed EPS-CNN framework is trained with data collected after stabilization (i.e., after 12 minutes from device activation) and tested with same-day data but collected at various different times during stabilization (i.e., during the initial 12 minutes from the activation of devices). To mask the impact of the wireless channel, we considered in this experiment the wired setting described in Sec.~\ref{wired_Setup}. 

Fig.~\ref{exp-all} shows the testing accuracy of EPS-CNN when testing data is collected before the device hardware is stabilized; that is, at 1, 4, and 8 minutes from when the devices are powered on. For ease of comparison, the figure also includes the case when the test data is collected after device stabilization; i.e., using data collected at minute 12. The figure clearly shows the dependency of the achieved accuracy on the time at which testing data is collected during stabilization time. Note that the closer to the stabilization time the testing data collection takes place, the higher the testing accuracy. The figure also confirms that this observed trend is consistent across different days.

One key observation that is worthy of note is that when both training and testing are done on data collected at about the same time from device activation even during warm-up time, the testing accuracy that the learning models achieve is not as low as what they achieved when testing and training data were done at different times during stabilization. For instance, we show in Fig.~\ref{cap1_days} the accuracy when the model is trained on data captured within the first minute after activation of one day and tested on data collected in the same time (i.e., within the first minute from activation) but of another day. The figure demonstrates that when the model is trained on minute 1 captures of day 1, an average testing accuracy of 70\% (resp. 62\%) is achieved when the model is tested on minute 1 captures of day 2 (resp. day 3), which is considerably higher than the testing accuracy when the model is tested on stabilized data (after 12 minutes) of the same day. 
These research findings indicate a systematic drift in the characteristics of the received IQ signals during the stabilization and warm-up period, with consistent behavior observed across different days, and highlight the challenges faced by the deep learning models in recognizing devices during the hardware warm-up period. 
These results thus underscore the importance of considering the stabilization aspects of the oscillator hardware (as well as the other transceiver hardware components) when developing hardware-impairment-driven RF fingerprinting techniques for robust device identification and classification.

\begin{figure}
\centering
   \includegraphics[width=0.7\columnwidth]{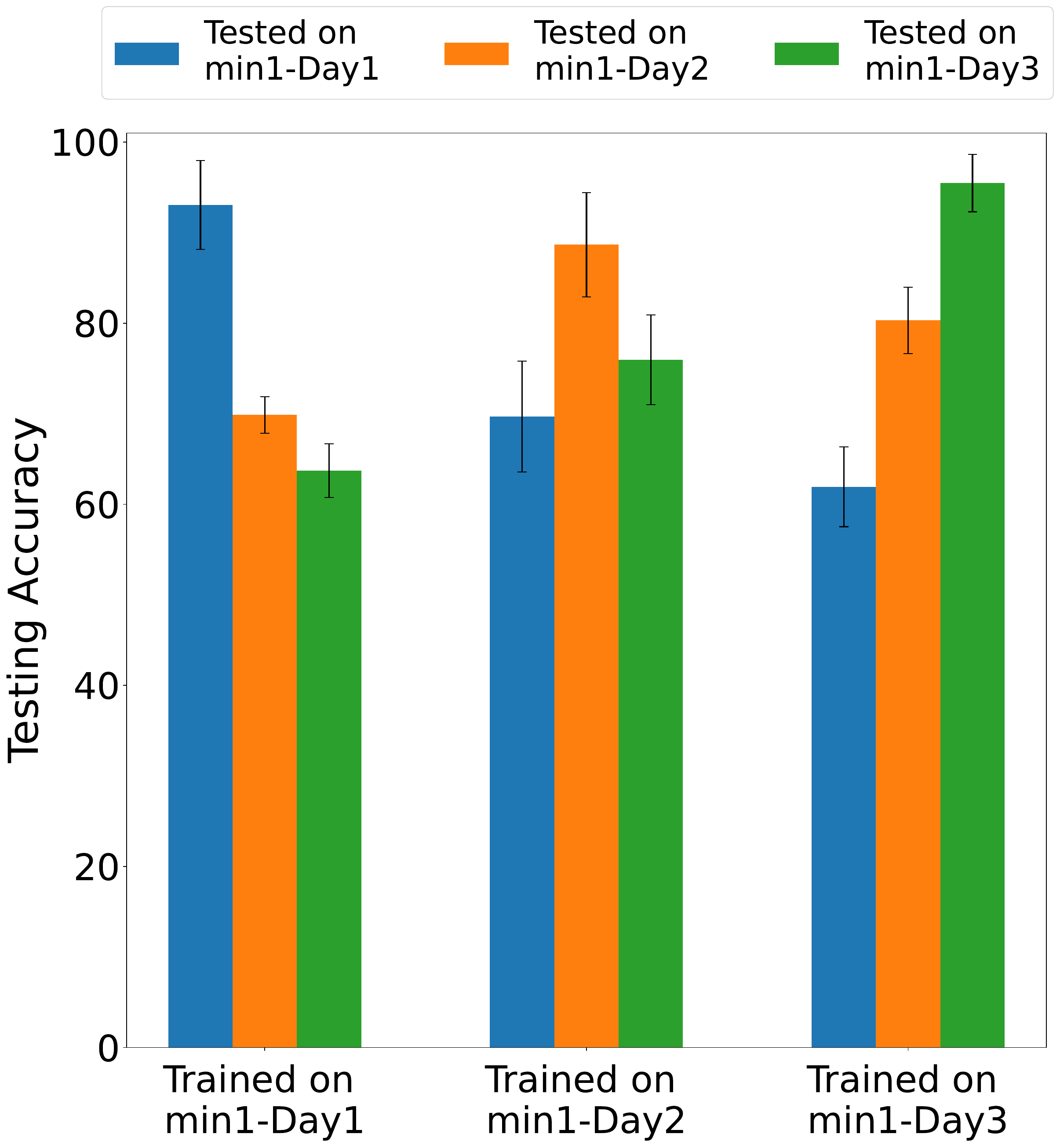}
   \label{cap1_days}
   
   \caption{Classification accuracy when training data is collected at minute 1 and testing data is collected also at minute 1 from device activation but on a different day.}
\label{cap1_days}
\end{figure}

 \section{Conclusion}
\label{sec:conc}


In conclusion, this paper addresses the limitations of conventional RF signal representations in deep learning-based RF fingerprinting methods. We propose the Double-Sided Envelope Power Spectrum (\proposed) as a novel RF signal representation that effectively captures device hardware impairments while eliminating irrelevant information. Experimental results demonstrate the superior performance of the \proposed~representation in terms of accuracy, robustness, and generalizability across various domains. By leveraging \proposed, DL-based RFFP methods can achieve unprecedented testing accuracy in same-domain evaluations and maintain high performance in cross-domain scenarios. The proposed representation offers a transformative solution for enhancing the security and privacy of wireless networks by advancing the accuracy and reliability of device identification through RF fingerprinting. Finally, we release large WiFi 802.11b datasets containing captures for different scenarios to allow others to further investigate these fingerprinting issues. 

\section{Acknowledgment}
\label{sec:ack}
We would like to thank Ms. Nora Basha for her help with the Matlab coding.
\bibliographystyle{IEEEtran}
\bibliography{IEEEexample}

\begin{IEEEbiography}[{\includegraphics[width=1in,height=1.25in,clip,keepaspectratio]{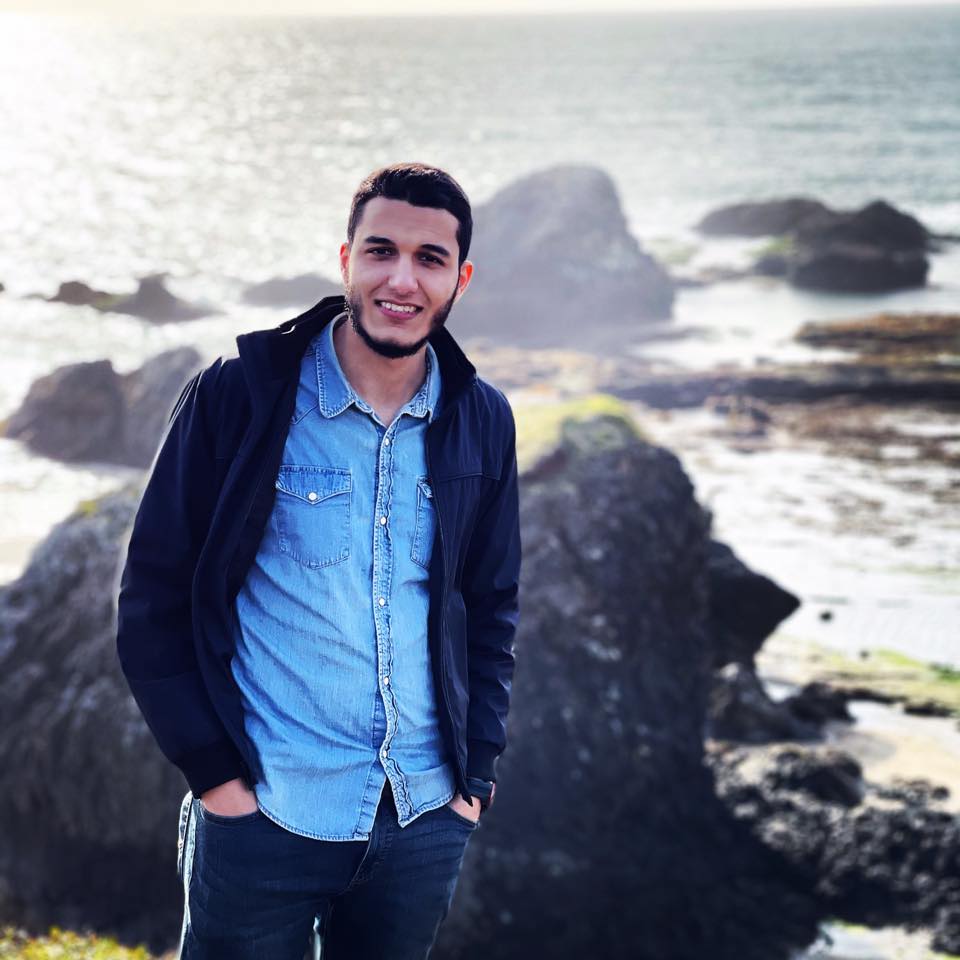}}]{Abdurrahman Elmaghbub} received the B.S. degree with summa cum laude, and MS in Electrical and Computer Engineering from Oregon State University in 2019, 2021, respectively, and is currently pursuing his Ph.D. degree in the School of Electrical Engineering and Computer Science at Oregon State University. His research interests are in the area of wireless communication and networking with a current focus on applying deep learning to wireless device classification.
\end{IEEEbiography}

\begin{IEEEbiography}[{\includegraphics[width=1in,height=1.25in,clip,keepaspectratio]{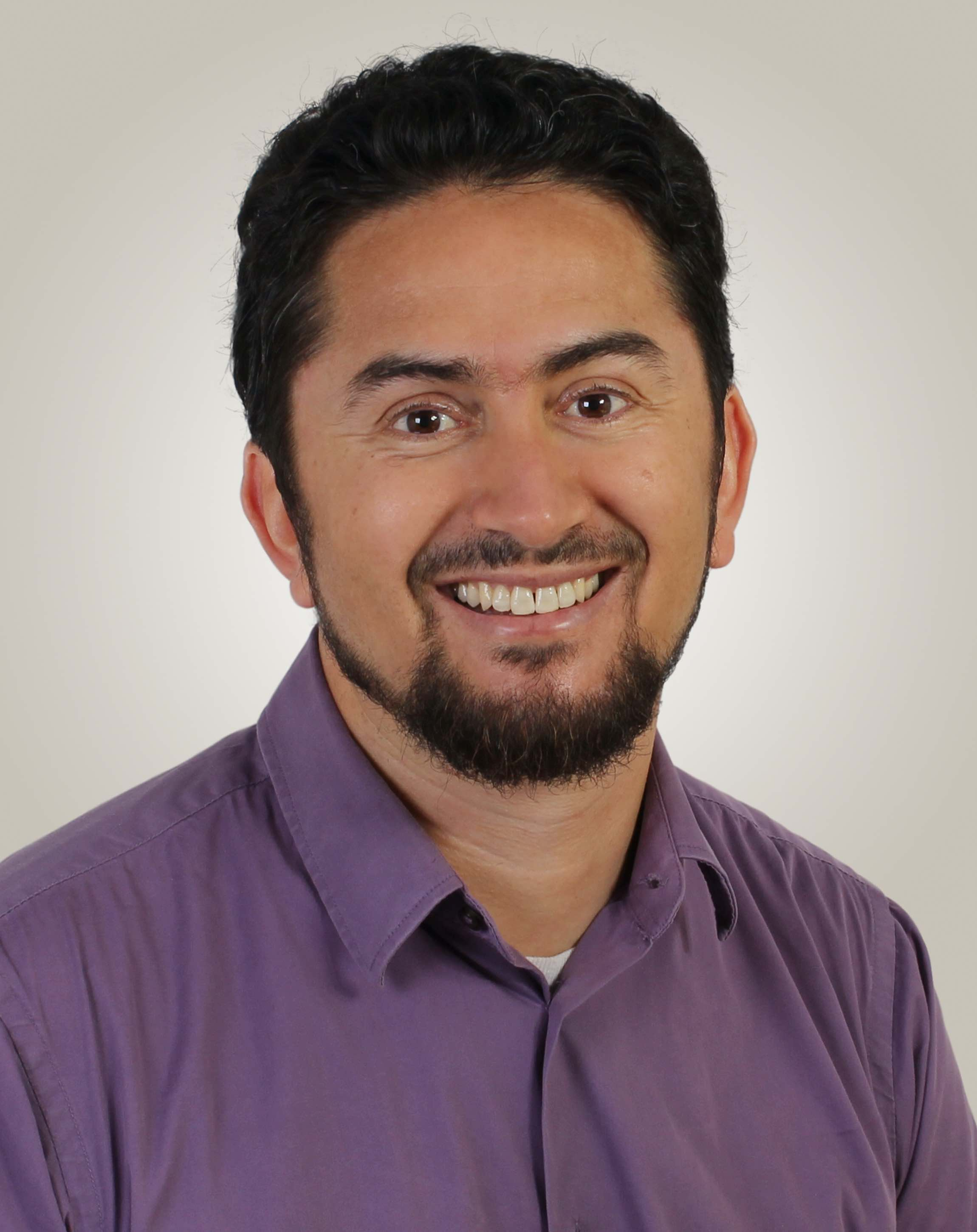}}]{Bechir Hamdaoui} is a Professor in the School of Electrical Engineering and Computer Science at Oregon State University. He received M.S. degrees in both ECE (2002) and CS (2004), and the Ph.D. degree in ECE (2005) all from the University of Wisconsin-Madison. 
His general interests are on theoretical and experimental research that enhances the cybersecurity \& resiliency of future intelligent networked systems, including connected \& autonomous vehicles, 5G/6G wireless, networked drones, smart cities, and cloud datacenters. He is the Founding Director of the NetSTAR Laboratory at Oregon State University.
Dr. Hamdaoui and his team have won several awards, including the ISSIP 2020 Distinguished Recognition Award, the 2009 NSF CAREER Award, the ICC 2017 Best Paper Award, and the 2016 EECS Outstanding Research Award. He serves/served as an Associate Editor for several IEEE journals and magazines and chaired \& organized many IEEE/ACM conference symposia \& workshop programs. He served as a Distinguished Lecturer for the IEEE Communication Society in 2016 and 2017 and served as the Chair \& Co-chair of the IEEE Communications Society's Wireless Technical Committee (WTC) from January 2019 until December 2022.
\end{IEEEbiography}

\end{document}